\documentclass[a4paper,12pt]{article}

\usepackage{amsmath,amssymb}
\usepackage{graphicx}
\usepackage{verbatim}
\usepackage[font=small,labelfont=bf,width=0.95\textwidth]{caption}
\usepackage{fixltx2e}             
\usepackage[amssymb]{SIunits}     
\usepackage{multirow}             
\usepackage{slashed}              
\usepackage{booktabs}             
\usepackage{color}                
\usepackage{subfigure}
\usepackage[a4paper,includehead,includefoot,ignorehead,left=3cm,right=3cm,top=2.4cm,bottom=2.4cm]{geometry}
\usepackage[pdftitle={Threshold Corrections in the Exceptional Supersymmetric Standard Model}, pdfauthor={Peter Athron,Dominik St\"ockinger,Alexander Voigt}, pdfkeywords={precision,threshold,correction,E6SSM,cE6SSM}, bookmarks=true, linktocpage]{hyperref}

\newcommand{\Qmatch}{T_{\ESSMmath}} 
\newcommand{\Qinter}{T_0}
\newcommand{\Qfix}{Q_\text{fix}}
\newcommand{\dd}{\mathrm{d}}
\newcommand{\mathi}{\mathrm{i}}

\newcommand{\Lagr}{\mathcal{L}}
\newcommand{\Order}[1]{O(#1)}
\newcommand{\cESSM}{CE\texorpdfstring{\textsubscript{6}}{6}SSM}
\newcommand{\ESSM}{E\texorpdfstring{\textsubscript{6}}{6}SSM}

\newcommand{\ESSMmath}{\text{E$_6$SSM}}
\newcommand{\MSSMmath}{\text{MSSM}}
\newcommand{\SMmath}{\text{SM}}
\newcommand{\UY}{\ensuremath{U(1)_{Y}}}
\newcommand{\Uem}{\ensuremath{U(1)_\text{em}}}
\newcommand{\UN}{\ensuremath{U(1)_{N}}}
\newcommand{\SUL}{\ensuremath{SU(2)_\text{L}}}
\newcommand{\SUc}{\ensuremath{SU(3)_\text{c}}}

\renewcommand{\figurename}{Fig.}
\renewcommand{\tablename}{Tab.}
\newcommand{\tabref}[1]{\tablename~\ref{#1}}
\newcommand{\figref}[1]{\figurename~\ref{#1}}
\newcommand{\figsref}[1]{Figs.~\ref{#1}}
\newcommand{\softsusy}{\texttt{SOFTSUSY}} 
\newcommand{\threelinebrace}{$\left. \begin{array}{c} \\ \\ \\ \end{array} \right\rbrace$}
\newcommand{\twolinebrace}{$\left. \begin{array}{c} \\ \\ \end{array} \right\rbrace$}
\newcommand{\elevenlinebrace}{$\left. \begin{array}{c} \\ \\ \\ \\ \\ \\ \\ \\ \\ \\ \\ \end{array} \right\rbrace$}
\newcommand{\SuperField}[1]{\hat{#1}}

\newcommand{\MSbar}{\ensuremath{\overline{\text{MS}}}}
\newcommand{\DRbar}{\ensuremath{\overline{\text{DR}}}}
\newcommand{\dimrep}[1]{\mathbf{#1}}
\newcommand{\dimrepadj}[1]{\mathbf{\bar{#1}}}

\DeclareMathOperator{\Tr}{Tr}

\DeclareMathOperator{\rre}{\Re e}
\DeclareMathOperator{\iim}{\Im m}
\DeclareMathOperator{\rretilde}{\widetilde{\Re e}}
\DeclareMathOperator{\CC}{\mathnormal{C}}  

\allowdisplaybreaks

\begin{document}

\begin{titlepage}
  \begin{flushright}
    ADP-12-16/T783
  \end{flushright}
  \vspace*{55mm}
  \begin{center}
    {\Large\bf Threshold Corrections in the 
      Exceptional Supersymmetric Standard Model}
    \\[8mm]
    Peter Athron$^{*}$,
    Dominik St\"ockinger$^{\dagger}$,
    Alexander Voigt$^{\dagger}$
    \\[3mm]
    {\small\it $^*$ ARC Centre of Excellence for Particle Physics at
      the Terascale, School of Chemistry and Physics, University of
      Adelaide, Adelaide, SA 5005, Australia
      \\[2mm]
      \small\it $^\dagger$ Institut f\"ur Kern und Teilchenphysik, TU
      Dresden, Dresden, D-01062, Germany.\\[2mm]}
  \end{center}
  \vspace*{0.75cm}
  \begin{abstract}
    \noindent
    We calculate threshold corrections to the running gauge and Yukawa
    couplings in the Exceptional Supersymmetric Standard Model (\ESSM)
    and analyse the more precise and reliable mass spectra in
    a constrained model (\cESSM). Full expressions for the corrections
    are provided and the implementation into a spectrum generator is
    described. We find a dramatic reduction in the matching scale
    dependency of the masses of many states and observe a significant
    adjustment of the correlation of low-scale physical masses and
    high-scale parameters.  Still, in substantial regions of parameter
    space the mass of the lightest Higgs is compatible with the new
    boson discovered at the LHC and the model satisfies limits from
    collider searches for squark, gluinos and $Z^\prime$ bosons.  We
    study the implications for gauge coupling unification from a new
    dependency of the spectrum on so-called survival Higgs fields
    which cannot be addressed without the inclusion of the threshold
    corrections.
  \end{abstract}
\end{titlepage}

\setcounter{page}{2}
\setcounter{footnote}{0}

\tableofcontents

\section{Introduction}

Supersymmetric (SUSY) extensions of the Standard Model (SM) with
TeV-scale SUSY breaking provide very attractive models for new physics
which could be discovered at the Large Hadron Collider (LHC). Such
models can solve the hierarchy problem of the SM due to the
cancellation of quadratic divergences, enable embedding into Grand
Unified Theories (GUTs), and thereby provide an explanation of the
$U(1)_Y$ rational charges, postulated \textit{ad hoc} in the SM.
These fundamental motivations do not imply any minimality condition on
the particle content or gauge structure.

Here we consider the Exceptional Supersymmetric Standard
Model\footnote{For a brief review see \cite{Athron:2010zz}.} (\ESSM)
\cite{King:2005jy,King:2005my}, which is a non-minimal SUSY extension
of the SM with an extra $U(1)_N$ gauge symmetry and new exotic matter
at the TeV scale.  A new Higgs singlet together with an extra $U(1)$
gauge symmetry solve the well-known $\mu$-problem \cite{Kim:1984} of
the Minimal Supersymmetric Standard Model (MSSM) without introducing
the tadpole or domain wall problems of the
Next-to-MSSM\footnote{Reviews are given in \cite{NMSSMreviews}.}
\cite{NMSSM}. These features also allow the lightest Higgs boson to be
substantially heavier than in the MSSM or NMSSM already at tree-level,
through new $U(1)_N$ $D$-term contributions and an additional $F$-term
contribution from the superpotential coupling between the singlet and
doublet Higgs bosons, which can be substantially larger than the
parallel NMSSM term due to different perturbative limits on the
coupling \cite{King:2005my}.

The model is inspired and motivated by $E_6$ GUTs, as the extra $U(1)$
appears from the breakdown of $E_6$ and the exotic new matter comes
from complete $E_6$ multiplets surviving to low energies, ensuring the
cancellation of gauge anomalies. The extra $U(1)$ gauge group of the
\ESSM\ is uniquely chosen such that right-handed neutrinos are
neutral, allowing large Majorana masses and a high-scale see-saw
mechanism.

The $E_6$ symmetry itself can arise from $E_8 \times E'_8$ Heterotic
string theory after the breakdown of $E_8$ \cite{Aguila:1986}.  The
$E'_8$ then plays the role of a hidden sector interacting with the
visible sector only through gravitational interactions and leads to
TeV-scale SUSY breaking, generating a set of soft SUSY-breaking
parameters.

In recent publications \cite{Athron:2009bs,Athron:2009ue} a
constrained version (\cESSM) was introduced where all the soft
SUSY-breaking masses are determined by a universal scalar mass, $m_0$,
gaugino mass $M_{1/2}$ and soft trilinear scalar coupling $A$. Owing
to the unification of matter and Higgs fields in complete GUT
multiplets the \cESSM\ is particularly well motivated. In
Refs.\ \cite{Athron:2009bs,Athron:2009ue} the low-energy mass spectra
were explored for the first time with benchmark scenarios,
representing the phenomenologically distinct possibilities in the
model.  This work has recently been updated to look at spectra which
are consistent with the $125$ GeV Higgs signal and recent searches for
squarks and gluinos \cite{Athron:2012sq}.  Other aspects of this model
and similar variants have also been discussed in
Refs.\ \cite{Howl:2007hq,Howl:2007zi,Howl:2008xz,Erler:2009jh,Howl:2009ds,Braam:2010sy,Hall:2011au,Belyaev:2012si,Rizzo:2012rf,Nevzorov:2012hs,Miller:2012he}.

The calculations in Refs.\
\cite{Athron:2009bs,Athron:2009ue,Athron:2012sq} neglect important
threshold corrections and involve a significant dependence on
unphysical threshold scales. This limits the accuracy of
phenomenological results and it was stated that the precision for the
masses was not better than $10\%$.

However to facilitate model discrimination from data, and ultimately
to reconstruct parameters in the event of a signal, precise
predictions of the TeV scale SUSY masses predicted from GUT scale
parameters are important. Indeed for the MSSM there are already a
number of state of the art spectrum generators publicly available
\cite{Porod:2003um,Djouadi:2002ze,Allanach:2001kg,Paige:2003mg,Chowdhury:2011zr}
which calculate the spectrum for scenarios, such as the constrained
MSSM (CMSSM).  Typically these employ two-loop renormalization group
equations, full one-loop matching conditions for gauge and Yukawa
couplings and one-loop shifts to pole masses and comparisons of their
predicted masses suggests an accuracy of about $1\%$
\cite{Belanger:2005jk}.

The present paper is devoted to the calculation of threshold
corrections to the running gauge and Yukawa couplings in the
Exceptional Supersymmetric Standard Model (\ESSM), which enable a more
precise evaluation of the mass spectrum from high-scale assumptions,
and to an extensive study of the phenomenological consequences in the
\cESSM. As in
Refs.\ \cite{Athron:2009bs,Athron:2009ue,Athron:2012sq} we employ
two-loop renormalization group equations for the running of the gauge
and Yukawa couplings between the weak scale and the GUT scale. As a
major improvement, we also compute and take into account the
corresponding one-loop threshold corrections arising in the transition
from the full \cESSM\ to the SM as a low-energy effective theory. We
show that the individual threshold corrections indeed drastically
reduce the dependence of our results on the unphysical threshold
scale. The resulting, more accurate and reliable mass spectra are
discussed. Particular attention is paid to the prediction for the
Higgs mass and the gluino mass. We confirm the recent result of
Ref.\ \cite{Athron:2012sq} that the lightest Higgs mass can easily be
in agreement with recent LHC discovery of a new boson
\cite{:2012gk,:2012gu}.  Finally we use our improved precision to
study gauge coupling unification\footnote{Previous studies
  \cite{King:2007uj,Athron:2009is} used trivial matching conditions,
  equivalent to assuming all new \ESSM\ states and all new MSSM
  states have common masses, $\Qmatch$ and $T_\text{MSSM}$ respectively.
  Such scenarios are clearly not realised in the constrained version
  of the model.} for the \cESSM\ and the impact of the survival Higgs
sector, which appears only via threshold corrections.

This paper is structured as follows.  In Sec.~\ref{sec:e6ssm} we
briefly outline the \ESSM\ model, in Sec.~\ref{sec:thresholdcorrections} we
describe the improved spectrum calculations, give analytical results
for the threshold corrections.  Sec.~\ref{sec:CESSM} describes the
application to an improved  \cESSM\ spectrum generator and explains the full numerical procedure
used.  In Sec.~\ref{sec:results} we illustrate the effect of the
thresholds and the improvement in accuracy of our results, and we give
an extensive discussion of the resulting model predictions.

\section{The \ESSM}
\label{sec:e6ssm}

The \ESSM\ is a supersymmetric gauge theory, inspired by $E_6$ GUTs in
which the $E_6$ is broken at the GUT scale $M_X$ via the Hosotani
mechanism \cite{Hosotani:1983xw} down to the low-energy gauge group of
the \ESSM,
\begin{align}
  \SUc\times \SUL\times \UY\times \UN.
  \label{eq:low-energy-gauge-group}
\end{align}
Here the $\UN$ is a special case of a $U(1)'$ symmetry arising from $E_6$ breaking,
\begin{align}
  U(1)' = U(1)_\chi \cos\theta + U(1)_\psi \sin\theta,
\end{align}
with $\tan\theta=\sqrt{15}$, where the gauge
groups $U(1)_\chi$ and $U(1)_\psi$ are defined by the breaking of $E_6
\rightarrow SO(10) \times U(1)_\psi$ and $SO(10) \rightarrow SU(5)
\times U(1)_\chi$ \cite{Hewett-Rizzo,Langacker:2008yv}.

The choice of $\UN$ makes the right-handed neutrino a pure gauge
singlet, thus allowing a gauge invariant Majorana mass term enabling a
high-scale seesaw mechanism for the generation of neutrino masses.
All other matter from complete $E_6$ matter multiplets survives to low
energies, thereby ensuring the cancellation of gauge anomalies.

The three families of \ESSM\  matter
particles fill complete $(\dimrep{27})_i$
representations of the $E_6$ group, ensuring that full low energy gauge group, including the new $U(1)_N$ gauge symmetry, is anomaly free. In addition to these, the model has
two Higgs-like doublets $\SuperField{H}'$ and $\SuperField{\bar{H'}}$,
the so-called survival Higgs doublets, both originating from extra
$(\dimrep{27})'$ and $(\dimrep{\overline{27}})'$ representations to
ensure gauge coupling unification at a high scale $M_X$.  The
decomposition of the fundamental $(\dimrep{27})$ representation under
$SU(5)\times U(1)_N$ and \eqref{eq:low-energy-gauge-group} is listed
in \tabref{tab:ESSMparticles}.
\begin{table}[tbh]
  \centering
  \begin{tabular}{llllll}
    \toprule
    Field & $G_\SMmath\times\UN$ & & $SU(5)\times\UN$ & & $E_6$ \\
    \midrule
    $\SuperField{Q}_i=(\SuperField{Q}_{u_i}\ \ \SuperField{Q}_{d_i})$ & 
    $(\dimrep{3}, \dimrep{2},{\textstyle\frac{1}{6}},1)_i$ & 
    \multirow{3}{*}{\threelinebrace} &
    \multirow{3}{*}{$(\dimrep{10},1)_i$} &
    \multirow{11}{*}{\elevenlinebrace} &
    \multirow{11}{*}{$(\dimrep{27})_i$}\\
    $\SuperField{u}_i^C$ & 
    $(\dimrepadj{3}, \dimrep{1}, {\textstyle -\frac{2}{3}}, 1)_i$ & \\
    $\SuperField{e}_i^C$ & 
    $(\dimrep{1}, \dimrep{1}, 1, 1)_i$ & \\
    $\SuperField{d}_i^C$ & 
    $(\dimrepadj{3}, \dimrep{1}, {\textstyle \frac{1}{3}}, 2)_i$ & 
    \multirow{2}{*}{\twolinebrace} &
    \multirow{2}{*}{$(\dimrepadj{5},2)_i$}\\ 
    $\SuperField{L}_i=(\SuperField{L}_{\nu_i}\ \ \SuperField{L}_{e_i})$ & 
    $(\dimrep{1}, \dimrep{2}, {\textstyle -\frac{1}{2}}, 2)_i$ & \\
    $\SuperField{\bar{D}}_i$ & 
    $(\dimrepadj{3},\dimrep{1},{\textstyle \frac{1}{3}},-3)_i$ &
    \multirow{2}{*}{\twolinebrace} &
    \multirow{2}{*}{$(\dimrepadj{5},-3)_i$} \\
    $\SuperField{H}_{1i} = (\SuperField{H}^0_{1i} \ \ \SuperField{H}^-_{1i})$ & 
    $(\dimrep{1},\dimrep{2},{\textstyle -\frac{1}{2}},-3)_i$ & \\
    $\SuperField{D}_i$ & 
    $(\dimrep{3},\dimrep{1},{\textstyle -\frac{1}{3}},-2)_i$ &
    \multirow{2}{*}{\twolinebrace} &
    \multirow{2}{*}{$(\dimrep{5},-2)_i$} \\
    $\SuperField{H}_{2i} = (\SuperField{H}^+_{2i} \ \ \SuperField{H}^0_{2i})$ & 
    $(\dimrep{1},\dimrep{2},{\textstyle \frac{1}{2}},-2)_i$ & \\
    $\SuperField{S}_i$ & $(\dimrep{1},\dimrep{1},0,5)_i$ & & $(\dimrep{1},5)_i$ \\
    $\SuperField{N}_i^C$ & $(\dimrep{1},\dimrep{1},0,0)_i$ & & $(\dimrep{1},0)_i$ \\
    \midrule
    $\SuperField{H}' = (\SuperField{H'}^0 \ \ \SuperField{H'}^-)$ & 
    $(\dimrep{1},\dimrep{2},{\textstyle -\frac{1}{2}},2)$ & & 
    $\ni (\dimrepadj{5},2)'$ & &
    $\ni (\dimrep{27})'$\\
    $\SuperField{\bar{H'}} = ({\SuperField{\bar{H'}}}^+ \ \ {\SuperField{\bar{H'}}}^0)$ & 
    $(\dimrep{1},\dimrep{2},{\textstyle \frac{1}{2}},-2)$ & & 
    $\ni (\dimrep{5},-2)'$ & &
    $\ni (\dimrep{\overline{27}})'$\\
    \midrule
    $\SuperField{V}^a_g$ & $(\dimrep{8}, \dimrep{1}, 0, 0)$ & & $\ni (\dimrep{24}, 0)$ & &
    $\ni (\dimrep{78})$ \\
    $\SuperField{V}_W^i$ & $(\dimrep{1}, \dimrep{3}, 0, 0)$ & & $\ni (\dimrep{24}, 0)$ & &
    $\ni (\dimrep{78})$ \\
    $\SuperField{V}_Y$ & $(\dimrep{1}, \dimrep{1}, 0, 0)$ & & $\ni (\dimrep{24}, 0)$ & &
    $\ni (\dimrep{78})$ \\
    $\SuperField{V}_N$ & $(\dimrep{1}, \dimrep{1}, 0, 0)$ & & $\ni (\dimrep{1}, 0)$ & &
    $\ni (\dimrep{78})$ \\
    \bottomrule
  \end{tabular}
  \caption{\ESSM\ SUSY multiplets and their gauge quantum numbers
    (generation index $i=1, 2, 3$), where
    $G_\SMmath\equiv\SUc\times\SUL\times\UY$.  For the abelian groups
    \UY\ and \UN\ the charges $Y/2$ and $N/2$ are listed.}
  \label{tab:ESSMparticles}
\end{table}
\begin{table}[tbh]
  \centering
  \begin{tabular}{llll}
    \toprule
    Superfield & \multicolumn{3}{c}{Component fields} \\
     & Spin $0$ & Spin $1/2$ & Spin $1$ \\
    \midrule
    $\SuperField{Q}_i = ( \SuperField{Q}_{u_i} \ \ \SuperField{Q}_{d_i} )^T$ &
    $\tilde{q}_{iL} = ( \tilde{u}_{iL} \ \ \tilde{d}_{iL} )^T$ &
    $q_{iL} = ( u_{iL} \ \ d_{iL} )^T$ \\
    $\SuperField{u}^C_i$ & $\tilde{u}^*_{iR}$ & $u^C_{iR}$ \\
    $\SuperField{d}^C_i$ & $\tilde{d}^*_{iR}$ & $d^C_{iR}$ \\ 
    $\SuperField{L}_i = ( \SuperField{L}_{\nu_i} \ \ \SuperField{L}_{e_i} )^T$ &
    $\tilde{\ell}_{iL} = ( \tilde{\nu}_{iL} \ \ \tilde{e}_{iL} )^T$ &
    $\ell_{iL} = ( \nu_{iL} \ \ e_{iL} )^T$ \\
    $\SuperField{e}^C_i$ & $\tilde{e}^*_{iR}$ & $e^C_{iR}$ \\
    $\SuperField{N}^C_i$ & $\tilde{\nu}^*_{iR}$ & $\nu^C_{iR}$ \\
    $\SuperField{D}_i$       & $\tilde{D}_{iL}$ & $D_{iL}$ \\
    $\SuperField{\bar{D}}_i$ & $\tilde{D}^*_{iR}$ & $D_{iR}^C$ \\
    $\SuperField{H}_{1i} = ( \SuperField{H}^0_{1i} \ \ \SuperField{H}^-_{1i} )^T$ &
    $H_{1i} = ( H^0_{1i} \ \ H^-_{1i} )^T$ &
    $\tilde{H}_{1iL} = ( \tilde{H}^0_{1iL} \ \ \tilde{H}^-_{1iL} )^T$ \\
    $\SuperField{H}_{2i} = ( \SuperField{H}^+_{2i} \ \ \SuperField{H}^0_{2i} )^T$ &
    $H_{2i} = ( H^+_{2i} \ \ H^0_{2i} )^T$ &
    $\tilde{H}_{2iL} = ( \tilde{H}^+_{2iL} \ \ \tilde{H}^0_{2iL} )^T$ \\
    $\SuperField{S}_i$ & $S_i$ & $\tilde{S}_i$ \\
    $\SuperField{H'} = ( {\SuperField{H'}}^0 \ \ {\SuperField{H'}}^- )^T$ &
    $H' = ( {H'}^0 \ \ {H'}^- )^T$ &
    $\tilde{H'}_L = ( \tilde{H'}^0_L \ \ \tilde{H'}^-_L )^T$ \\
    $\SuperField{\bar{H'}} = ( {\SuperField{\bar{H'}}}^+ \ \ {\SuperField{\bar{H'}}}^0 )^T$ &
    $\bar{H'} = ( {\bar{H'}}^+ \ \ {\bar{H'}}^0 )^T$ &
    $\tilde{\bar{H'}}_L = ( {\tilde{\bar{H'}}}^+_L \ \ {\tilde{\bar{H'}}}^0_L )^T$
    \\
    \midrule
    $\SuperField{V}^a_g$ & & $\tilde{g}^a$ & $G^a_\mu$ \\
    $\SuperField{V}_W^i$ & & $\tilde{W}^i$ & $W_\mu^i$ \\
    $\SuperField{V}_Y$ & & $\tilde{B}$ & $B_\mu$ \\
    $\SuperField{V}_N$ & & $\tilde{Z}'$ & $Z'_\mu$ \\
    \bottomrule
  \end{tabular}
  \caption{Component fields of the \ESSM\ superfields
    (generation index $i=1, 2, 3$).  The charge conjugation of a spinor $\psi$ 
    is defined as $\psi^C:=\CC\bar{\psi}^T$, where $\CC=i\gamma^2\gamma^0$.}
  \label{tab:ESSMcomponents}
\end{table}
All Standard Model fermions as well as their superpartners fit into
the multiplets $(\dimrep{10},1)_i$ and $(\dimrepadj{5},2)_i$.  The
$(\dimrepadj{5},-3)$ and $(\dimrep{5},-2)$ representations contain
Higgs-like doublets $\SuperField{H}_{1i}$, $\SuperField{H}_{2i}$ and exotic colored
matter fields $\SuperField{D}_i$, $\SuperField{\bar{D}}_i$.  The remaining $SU(5)$
singlets $(\dimrep{1},0)_i$ and $(\dimrep{1},5)_i$ equate to
right-handed neutrinos and fields $\SuperField{S}_i$, respectively.  A
complete particle listing can be found in \tabref{tab:ESSMcomponents}.

In an $E_6$ GUT the gauge bosons and their superpartners fit into the
adjoint $(\dimrep{78})$ representation of the $E_6$, which is
decomposed under the low-energy gauge group
$\SUc\times\SUL\times\UY\times\UN$ into
\begin{align}
  (\dimrep{78})
  &\rightarrow 
  (\dimrep{8}, \dimrep{1}, 0, 0)
  + (\dimrep{1}, \dimrep{3}, 0, 0)
  + (\dimrep{1}, \dimrep{1}, 0, 0)
  + (\dimrep{1}, \dimrep{1}, 0, 0)
  + \dotsb
\end{align}
The gluons present in a low-energy model are associated to
$(\dimrep{8}, \dimrep{1}, 0, 0)$.  The $(\dimrep{1}, \dimrep{3}, 0,
0)$ multiplet contains the weak gauge bosons and the two $U(1)$ gauge
fields belong to the two $(\dimrep{1}, \dimrep{1}, 0, 0)$
representations.  When the $E_6$ is broken at the GUT scale $M_X$, the
other gauge bosons are expected to have masses of the order of $M_X$.
However, the \ESSM\ is a low-energy model by construction and does not
include these heavy GUT gauge bosons.

As with the MSSM embedded into GUT models, the new bosons absent from
the low-energy theory could give rise to significant threshold
corrections to gauge coupling unification at $M_X$.  This should be
borne in mind when studying the success or failure of gauge coupling
unification within a low-energy SUSY model.  However the purpose of
this paper is to consider the low-energy threshold effects from
sparticle masses and to test the significance of these on the physical
predictions made by postulates about the high scale
parameters.

For $(\dimrep{27})_i$ representations of $E_6$ the most general
renormalizable superpotential with full $E_6$ invariance is given by
the trace of the $(\dimrep{27})_i\times (\dimrep{27})_j\times
(\dimrep{27})_k$. Invariance under the low-energy gauge group allows
further terms.  But as in the case of the MSSM, the most general gauge
invariant superpotential is not phenomenologically viable
\cite{King:2005jy,Athron:2009bs}, as it contains ba\-ry\-on number
violation and unacceptably large contributions to non-diagonal flavor
transitions. To conserve baryon number and avoid flavor changing
neutral currents, one first imposes a $Z_2^H$ symmetry, under which
all chiral superfields transform as odd, except $\SuperField{H}_{13}$,
$\SuperField{H}_{23}$ and $\SuperField{S}_3$.\footnote{Although a
  high-scale family structure is not a part of our model by
  construction, such a $Z_2^H$ symmetry could be the result of a
  $\Delta_{27}$ family symmetry at the GUT scale \cite{Howl:2009ds}.}
The remaining $\SUc\times\SUL\times\UY\times\UN$ and $Z_2^H$ invariant
superpotential reads
\begin{align}
  \begin{split}
    W_{\ESSMmath} &=
    - y^e_{ij} (\SuperField{H}_{13} \SuperField{L}_i) \SuperField{e}^C_j
    - y^d_{ij} (\SuperField{H}_{13} \SuperField{Q}_i) \SuperField{d}^C_j
    - y^u_{ij} (\SuperField{Q}_i \SuperField{H}_{23}) \SuperField{u}^C_j \\
    &\phantom{{}=}
    + \frac{1}{2} M_{ij} \SuperField{N}^C_i \SuperField{N}^C_j
    + h_{4j}^E (\SuperField{H}_{13} \SuperField{H}') \SuperField{e}^C_j
    + h_{4j}^N (\SuperField{H}_{23} \SuperField{H}') \SuperField{N}^C_j
    + \mu' (\SuperField{H}' \SuperField{\bar{H'}})   
    \\
    &\phantom{{}=}
    + \lambda_i \SuperField{S}_3 (\SuperField{H}_{1i} \SuperField{H}_{2i})
    + \kappa_i  \SuperField{S}_3 \SuperField{D}_i \SuperField{\bar{D}}_i
    + f_{\alpha\beta} \SuperField{S}_\alpha (\SuperField{H}_{13}  \SuperField{H}_{2\beta})
    + \tilde{f}_{\alpha\beta} \SuperField{S}_\alpha (\SuperField{H}_{1\beta}  \SuperField{H}_{23}),
  \end{split}
  \label{eq:ESSM-ZH-superpotential}
\end{align}
where the $SU(2)$ superfield spinor product is defined as $(A B) :=
A^2B^1 - A^1B^2$.  A problematic consequence of this $Z_2^H$ symmetry
would be that the exotic quarks would only have gauge interactions and
interactions with the singlet field, leading to stable charged matter
in violation of experimental constraints
\cite{Rich:1987jd,Smith:1988ni,Hemmick:1989ns}.  Therefore the $Z_2^H$
symmetry can only be approximate.

The $Z_2^H$-violating terms should not lead to rapid proton decay.
Hence, another discrete symmetry, analogous to $R$-parity, must be
required.  This can be done in two ways: either a $Z_2^L$ symmetry is
implemented, under which all superfields except the leptons are even
(Model I) or one imposes a $Z_2^B$ symmetry, under which the exotic
quarks and leptons are odd whereas the others remain even (Model II).
The superpotential \eqref{eq:ESSM-ZH-superpotential} is then enlarged
by one of the following $Z_2^H$-violating but $Z_2^{L,B}$-conserving
terms
\begin{align}
  W_\text{Model I} &= 
  g_{ijk}^Q \SuperField{D}_i (\SuperField{Q}_j \SuperField{Q}_k)
  + g_{ijk}^q \SuperField{\bar{D}}_i \SuperField{d}^C_j \SuperField{u}^C_k ,
  \label{eq:ModelI}
  \\
  W_\text{Model II} &= 
  g_{ijk}^N \SuperField{N}^C_i \SuperField{D}_j \SuperField{u}^C_k
  + g_{ijk}^E \SuperField{e}^C_i \SuperField{D}_j \SuperField{u}^C_k
  + g_{ijk}^D (\SuperField{Q}_i \SuperField{L}_j) \SuperField{\bar{D}}_k.
  \label{eq:ModelII}
\end{align}
In Model I the scalar exotic quarks can decay into two quarks (they are
diquarks) and in Model II they are leptoquarks since they decay into a
lepton and a quark.

For correct electroweak symmetry breaking only the scalar components
of $\SuperField{H}_{13}$, $\SuperField{H}_{23}$ and $\SuperField{S}_3$ get a non-zero VEV.
To ensure this, we impose that a certain hierarchy between the Yukawa
couplings must exist
\begin{align}
  \kappa_i \sim \lambda_3 \gtrsim \lambda_{1,2} \gg 
  f_{\alpha\beta}, \tilde{f}_{\alpha\beta}, h_{4j}^E, h_{4j}^N.
\end{align}
This hierarchical structure allows a simplification of the
superpotential \eqref{eq:ESSM-ZH-superpotential}.  Integrating out the
right-handed neutrinos, which are assumed to be very heavy, and
keeping only the dominant terms, one arrives at
\begin{align}
  \begin{split}
    W_{\ESSMmath} &\approx - y_{\tau} (\SuperField{H}_d
    \SuperField{L}_3) \SuperField{e}^C_3 - y_b (\SuperField{H}_d
    \SuperField{Q}_3) \SuperField{d}_3^C - y_t (\SuperField{Q}_3
    \SuperField{H}_u) \SuperField{u}_3^C\\
    &\phantom{{}\approx}
    + \lambda_i \SuperField{S} (\SuperField{H}_{1i}
    \SuperField{H}_{2i}) + \kappa_i \SuperField{S} \SuperField{D}_i
    \SuperField{\bar{D}}_i + \mu' (\SuperField{H}'
    \SuperField{\bar{H'}}),
  \end{split}
  \label{SuPot_RGE}
\end{align}
where the scalar components of $\SuperField{H}_u:=\SuperField{H}_{23}$
and $\SuperField{H}_d:=\SuperField{H}_{13}$ and
$\SuperField{S}:=\SuperField{S}_3$ develop VEVs, $\langle H^0_u
\rangle = v_u/\sqrt{2}$, $\langle H^0_d \rangle = v_d/\sqrt{2}$,
giving mass to ordinary matter while $\langle S \rangle = s/\sqrt{2}$
gives exotic quark masses, $\kappa_i S \rightarrow \kappa_i s/\sqrt{2}
=: \mu_{D_i}$; masses for the fermion components of the first two generations of ``inert'' Higgs-like doublets (the ones which don't get VEVs) $\mu_{\tilde{H}_\alpha}:= \lambda_\alpha s/\sqrt{2}$ and an effective $\mu$-term for the Higgs doublets $\mu_\text{eff} :=
\lambda_3 s/\sqrt{2}$.  Eq.~\eqref{SuPot_RGE} is the superpotential
which will be used in the following analysis to determine the
particle spectrum from high scale assumptions, inspired by minimal
Supergravity and $E_6$ GUTs.

The Higgs potential, electroweak symmetry breaking conditions and mass
eigenstates of all the particles in the model have been presented in
Ref.~\cite{Athron:2009bs}.

\section{Threshold corrections in the \ESSM}
\label{sec:thresholdcorrections}

The \ESSM\ is a low-energy model which is motivated by a particular
high-scale structure. The high and low scales are connected by
renormalization group equations which have already been given at the
two-loop level in Ref.~\cite{Athron:2009bs}. In the following we
present the results for the threshold corrections which are required
for a consistent inclusion of subleading effects. Before presenting
our computation and the results we give a brief summary of relevant
general features of threshold corrections.

\subsection{Threshold corrections for matching full and effective theories}
\label{sec:visualisation-of-TC}

\begin{figure}[tbh]
  \centering
  \includegraphics[width=0.9\textwidth]{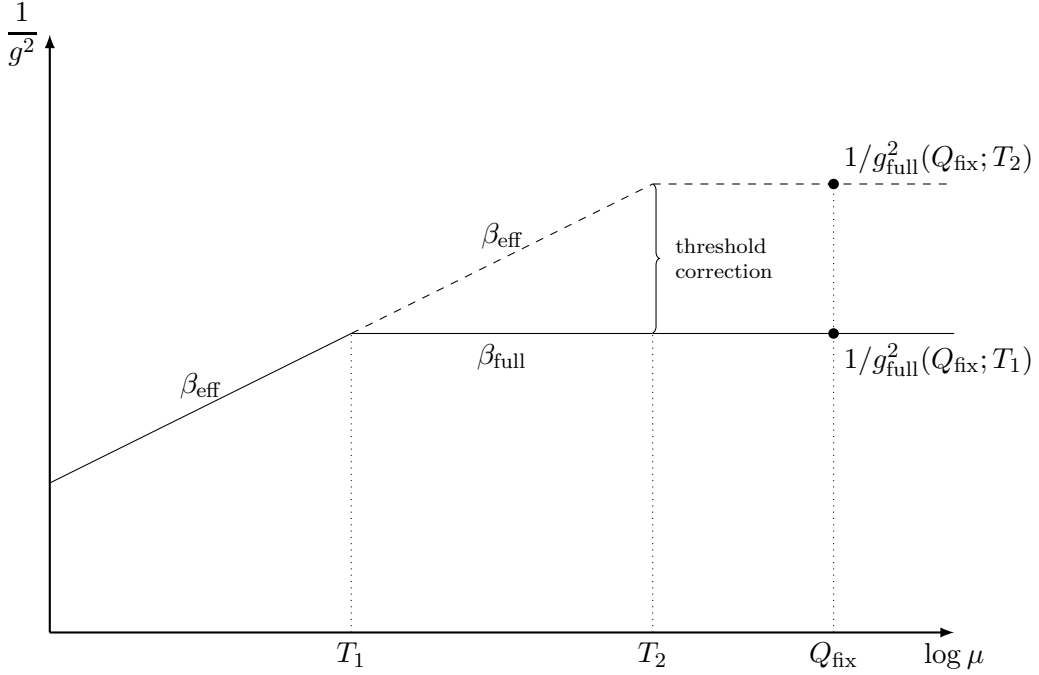}
  \caption{Visualization of gauge coupling threshold corrections}
  \label{fig:visualisation-of-gauge-tc}
\end{figure}

To connect running fundamental \ESSM\ parameters with SM
quantities we consider the SM as an effective theory of the \ESSM\ and
match the two theories at a threshold scale $\Qmatch$, which should be
of the order of the heavy \ESSM\ particles \cite{Ovrut:1980bk}.
To see the need for and the properties of threshold corrections most clearly, we slightly
generalize and suppose  we have a full and an effective
gauge theory and we want to calculate the \DRbar\ gauge coupling
$g_\text{full}(\Qfix)$ at a fixed scale $\Qfix$ in the
full theory.  As input we know $g_\text{eff}(Q_\text{low})$ in the
effective theory at some low scale $Q_\text{low} < \Qfix$.  We
start with our effective theory at $Q_\text{low}$ and run the gauge
coupling $g_\text{eff}$ from $Q_\text{low}$ to the matching scale
$T_1$, using the beta function $\beta_\text{eff}$ of the effective
theory, see \figref{fig:visualisation-of-gauge-tc}.
At $T_1$ we match the full and the effective theory and
calculate $g_\text{full}(T_1)$ without the use of threshold
corrections,
\begin{align}
  g_\text{full}(T_1) := g_\text{eff}(T_1) .
  \label{eq:toymodel-trivial-threshold-correction}
\end{align}
Now we use the beta functions of the full theory $\beta_\text{full}$
to evolve $g_\text{full}(T_1)$ to the desired scale $\Qfix$ and
obtain $g_\text{full}(\Qfix)$.

The problem with this approach is that the resulting coupling
$g_\text{full}(\Qfix)$ depends on the choice of the unphysical
matching scale $T_1$, i.e.\ $g_\text{full}(\Qfix;T_1)$.  More
precisely, if one would do the same calculation using another matching
scale $T_2$ the difference would be
\begin{align}
  \frac{1}{g_\text{full}^2(\Qfix;T_1)} -
  \frac{1}{g_\text{full}^2(\Qfix;T_2)} =
  (T_1 - T_2) \frac{2}{(4\pi)^2} \left(
    \beta_\text{full} - \beta_\text{eff}
  \right),\label{eq:diff-one-over-gsqr}
\end{align}
because the solution of the RGE is given by
\begin{align}
  \frac{1}{g_\text{full/eff}^2(\mu)} =
  -\frac{2\beta_\text{full/eff}}{(4\pi)^2}\log\mu + C
  .
\end{align}
In the limit $T_1\rightarrow T_2$ Eq.~\eqref{eq:diff-one-over-gsqr}
becomes
\begin{align}
  \frac{\dd}{\dd t} \frac{(4\pi)^2}{2 g_\text{full}^2(\Qfix;t)} =
  \beta_\text{full} - \beta_\text{eff} \neq 0
  ,\label{eq:toymodel-threshold-test}
\end{align}
which means that the gauge coupling in the full theory at the scale
$\Qfix$ depends on the unphysical matching scale $t$ and the
slope of $g_\text{full}^{-2}(\Qfix;t)$ with respect to $t$ is
proportional to the difference of the beta functions of the full and
the effective theory.

The way out is to use threshold corrections $\Delta g$ when
matching the full and the effective theory, i.e.\ replace
Eq.~\eqref{eq:toymodel-trivial-threshold-correction} by
\begin{align}
  g_\text{full}(T_i) := g_\text{eff}(T_i) + \Delta g(T_i)
  \label{eq:toymodel-threshold-correction},
\end{align}
where $\Delta g(T_i)$ can be obtained by matching Green
functions of the full and the effective theory, see below.  As a
result the right-hand side of Eq.~\eqref{eq:toymodel-threshold-test}
will vanish, i.e.\ the gauge coupling $g_\text{full}(\Qfix)$
will not depend on the matching scale.  This property will be used as
a test in Sec.~\ref{sec:matching-scale-dependency}.

To calculate the threshold corrections we consider a full theory
consisting of parameters $\rho_1,\dotsc, \rho_p$, light fields $l_1,
\dotsc, l_q$ and heavy fields $H_1,\dotsc, H_r$
\begin{align}
  \Lagr^\text{full} = \Lagr^\text{full}(\rho_1,\dotsc, \rho_p;
  l_1,\dotsc, l_q, H_1,\dotsc, H_r)
  .
\end{align}
The effective theory might contain effective parameters $\hat{\rho}_1,
\dotsc, \hat{\rho}_k$ and effective light fields $\hat{l}_1,\dotsc,
\hat{l}_q$, where the heavy fields were integrated out
\begin{align}
  \Lagr^\text{eff} = \Lagr^\text{eff}(\hat{\rho}_1,\dotsc, \hat{\rho}_k;
  \hat{l}_1,\dotsc, \hat{l}_q)
  .
\end{align}
The requirement that both the full and the effective theory describe the
same physics in the limit $p\rightarrow 0$ can be achieved by equating
all connected Green functions with light external fields $l_i$ in the
full and in the effective theory at zero external momenta.  This
condition leads to the equality of all one-particle irreducible
correlation functions $\Gamma$ that are one-particle irreducible with
respect to the light fields $l_i$ (1LPI)
\begin{align}
  \left.
    \Gamma^\text{full}_{l_{i_1} l_{i_2} \dotsm l_{i_n}}(\rho_1,\dotsc, \rho_p)
  \right|_{k_i=0}
  &= 
  \left.
    \Gamma^\text{eff}_{l_{i_1} l_{i_2} \dotsm l_{i_n}}(\hat{\rho}_1,\dotsc, \hat{\rho}_k)
  \right|_{k_i=0} ,
  \label{eq:matching-cond}  
\end{align}
where $k_i$ are the momenta of the external fields.
The next step is to decompose the renormalized 1LPI correlation
functions into a tree-level part and a part which contains one-loop
contributions
\begin{align}
  \Gamma^\text{full}_{l_{i_1} l_{i_2} \dotsm l_{i_n}} 
  &= \Gamma^\text{\text{full},tree}_{l_{i_1} l_{i_2} \dotsm l_{i_n}}
  + \Gamma^\text{\text{full},1L}_{l_{i_1} l_{i_2} \dotsm l_{i_n}} 
  ,
  \label{eq:dec-full}
  \\
  \Gamma^\text{eff}_{l_{i_1} l_{i_2} \dotsm l_{i_n}} 
  &= \Gamma^\text{eff,tree}_{l_{i_1} l_{i_2} \dotsm l_{i_n}}
  + \Gamma^\text{eff,1L}_{l_{i_1} l_{i_2} \dotsm l_{i_n}}
  \label{eq:dec-eff}
  .
\end{align}
Imposing a relative field renormalization for the renormalized fields
in the full and effective theory
\begin{align}
  \hat{l}_i &= \left( 1 + \frac{1}{2} K_{l_i} \right) l_i  
  , \qquad \qquad
  (i = 1,\dotsc, q)
  \label{eq:field-ren}
\end{align}
and inserting Eq.~\eqref{eq:dec-full}--\eqref{eq:field-ren} into
\eqref{eq:matching-cond} yields the matching condition
\begin{align}
    \Gamma^\text{full,tree}_{l_{i_1} l_{i_2} \dotsm l_{i_n}}
    + \Gamma^\text{full,1L}_{l_{i_1} l_{i_2} \dotsm l_{i_n}}
    =
    \left( 1 + \frac{1}{2}\sum_{n=1}^q K_{l_{i_n}} \right)
    \Gamma^\text{eff,tree}_{\hat{l}_{i_1} \hat{l}_{i_2} \dotsm \hat{l}_{i_n}}
    + \Gamma^\text{eff,1L}_{\hat{l}_{i_1} \hat{l}_{i_2} \dotsm \hat{l}_{i_n}}
    + \Order{\hbar^2}.
  \label{eq:matching-formula}
\end{align}
This equation is evaluated at zero external momenta.  The analogous
equation holds for the derivatives of the 1LPI correlation functions.
Imposing all matching conditions yields the definitions for the
$K_{l_i}$ in terms of 1LPI correlation functions and at the same time
the desired relations between the parameters of the effective and full
theory (threshold corrections)
\begin{align}
  \hat{\rho}_i = \rho_i 
  + \Delta\rho_i^\text{(1-loop)}(\rho_1,\dotsc, \rho_p, K_{l_1},\dotsc, K_{l_q})
  , \qquad \qquad (i = 1,\dotsc, k) .
\end{align}

\subsection{Gauge coupling threshold corrections in the \ESSM}
\label{sec:calculation-of-gauge-coupling-TCs}
\paragraph{Strong gauge coupling}

To calculate the threshold correction for the gauge coupling $g_3$ of
the unbroken \SUc\ one needs to apply the matching procedure of
Sec.~\ref{sec:visualisation-of-TC} to the following 1LPI correlation
functions
\begin{align}
  \partial_{p} \left.
  \Gamma^\ESSMmath_{q_i\bar{q}_i}(p,-p)
  \right|_{p=0}
  &= 
  \partial_{p} \left.
  \Gamma^\SMmath_{q_i\bar{q}_i}(p,-p) 
  \right|_{p=0}
  ,\label{eq:matching-fermions}
  \\
  \partial_{k^2} \left.
  \Gamma^\ESSMmath_{G^a_\mu G^b_\nu}(k,-k) 
  \right|_{k^2=0}
  &= 
  \partial_{k^2} \left.
  \Gamma^\SMmath_{G^a_\mu G^b_\nu}(k,-k) 
  \right|_{k^2=0}
  ,\label{eq:matching-G}
  \\
  \left.
  \Gamma^\ESSMmath_{G^a_\mu q_i\bar{q}_i}(k,p,-(p+k))
  \right|_{p=k=0}
  &= 
  \left.
  \Gamma^\SMmath_{G^a_\mu q_i\bar{q}_i}(k,p,-(p+k)) 
  \right|_{p=k=0}
  ,\label{eq:matching-g3}
\end{align}
where $q_i$ are the light colored fields that remain in the SM, i.e.\
the SM quarks.  After imposing the relative field 
renormalizations \eqref{eq:field-ren} and decomposing the 1LPI
functions into a tree-level and a loop part one gets the threshold
correction for $g_3$
\begin{align}
  g_3^{\MSbar,\SMmath} &= g_3^{\MSbar,\ESSMmath} 
  + \frac{g_3^3}{(4\pi)^2} \left(
    \sum_f \frac{4}{3} F_f C(r_f) \log\frac{m_f}{\mu}
    + \sum_s \frac{1}{3} F_s C(r_s) \log\frac{m_s}{\mu}
  \right).
  \label{eq:g3-general-threshold-correction}
\end{align}
The sums run over all heavy fermions $f$ and scalars $s$ that are
integrated out.  Eq.~\eqref{eq:g3-general-threshold-correction} is
also in agreement with the general result in \cite{Hall:1980kf}.  The
constants $C(r)$ are invariants of the representations $r$ of \SUc\
and are given by
\begin{align}
  C(N) &= \frac{1}{2} & &\text{(fundamental representation $N$)},\\
  C(G) &= 3           & &\text{(adjoint representation $G$)},
\end{align}
and $F_f$, $F_s$ account for the different number of field degrees of
freedom
\begin{align}
  F_f &=
  \begin{cases}
    1   & \text{if $f$ is a Dirac fermion}, \\
    1/2 & \text{if $f$ is a Majorana fermion},
  \end{cases} \\
  F_s &=
  \begin{cases}
    1   & \text{if $s$ is a complex scalar}, \\
    1/2 & \text{if $s$ is a real scalar}.
  \end{cases}
\end{align}
When matching the \ESSM\ to the SM, only the gluino, the
squarks and the exotics contribute in
Eq.~\eqref{eq:g3-general-threshold-correction} and we have
\begin{align}
  C(r_{\tilde{g}}) &= C(G) = 3,
  & C(r_{\tilde{q}_{ik}}) &= C(r_{\tilde{D}_{ik}}) = C(r_{D_i}) = C(N) = 1/2, \\
  F_{\tilde{g}} &= 1/2,
  & F_{\tilde{q}_{ik}} &= F_{\tilde{D}_{ik}} = F_{D_i} = 1.
\end{align}
Then the threshold correction
\eqref{eq:g3-general-threshold-correction} reduces to
\begin{align}
  \begin{split}
    g_3^{\DRbar,\ESSMmath} =
    g_3^{\MSbar,\SMmath} 
    + \frac{g_3^3}{(4\pi)^2}
    \Bigg[
    &\frac{1}{2}
    - 2 \log\frac{m_{\tilde{g}}}{\mu}
    - \frac{1}{6} \sum_{\tilde{q}\in\{\tilde{u},\tilde{d}\}} \sum_{i=1}^3 \sum_{k=1}^2 
    \log\frac{m_{\tilde{q}_{ik}}}{\mu}
    \\
    &- \frac{2}{3} \sum_{i=1}^3 \log\frac{m_{D_i}}{\mu}
    - \frac{1}{6} \sum_{i=1}^3 \sum_{k=1}^2 \log\frac{m_{\tilde{D}_{ik}}}{\mu}
    \Bigg],
  \end{split}
  \label{eq:g3-threshold-program}
\end{align}
where we have added a finite counterterm which converts $g_3$ from
the \MSbar\ to the \DRbar\ scheme \cite{Martin:1993yx}.

\paragraph{Electroweak gauge couplings}
The calculation of threshold corrections for the couplings $g_Y$ and
$g_2$ is more involved, because the gauge group $\SUL\times\UY$ is
spontaneously broken to $\Uem$.  From the relations
\begin{align}
  g_Y &= \frac{e}{c_W}, & g_2 &= \frac{e}{s_W}, & c_W &= \frac{m_W}{m_Z}
\end{align}
one can see that the threshold corrections for $g_Y$ and $g_2$ are
related to those of the $W^\pm$ and $Z$ boson masses as well as to the
gauge coupling $e$ of the remaining \Uem\ gauge symmetry.  Therefore
the following matching conditions are imposed to obtain threshold
corrections for $e$, $m_Z$ and $m_W$
\begin{align} 
  \partial_{k^2}^n \left.
  \Gamma^\ESSMmath_{W^+_\mu W^-_\nu}(k,-k)
  \right|_{k^2=0}
  &=   
  \partial_{k^2}^n \left.
  \Gamma^\SMmath_{W^+_\mu W^-_\nu}(k,-k) 
  \right|_{k^2=0}
  \qquad (n=0,1)
  \label{eq:matching-W}
  ,\\
  \partial_{k^2}^n \left.
  \Gamma^\ESSMmath_{Z_\mu Z_\nu}(k,-k)
  \right|_{k^2=0}
  &= 
  \partial_{k^2}^n \left.
  \Gamma^\SMmath_{Z_\mu Z_\nu}(k,-k) 
  \right|_{k^2=0}
  \quad\qquad (n=0,1)
  \label{eq:matching-Z}
  ,\\
  \partial_{k^2} \left.
  \Gamma^\ESSMmath_{A_\mu A_\nu}(k,-k) 
  \right|_{k^2=0}
  &= 
  \partial_{k^2} \left.
  \Gamma^\SMmath_{A_\mu A_\nu}(k,-k) 
  \right|_{k^2=0}
  \label{eq:matching-Aem}
  ,\\
  \left.
  \Gamma^\ESSMmath_{Z_\mu A_\nu}(k,-k) 
  \right|_{k^2=0}
  &= 
  \left.
  \Gamma^\SMmath_{Z_\mu A_\nu}(k,-k) 
  \right|_{k^2=0}
  \label{eq:matching-ZA}
  ,\\
  \partial_{p} \left.
  \Gamma^\ESSMmath_{\psi_{Li} \bar{\psi}_{Lj}}(p,-p)
  \right|_{p=0}
  &= 
  \partial_{p} \left.
  \Gamma^\SMmath_{\psi_{Li} \bar{\psi}_{Lj}}(p,-p) 
  \right|_{p=0}
  \label{eq:matching-psiL}
  ,\\
  \partial_{p} \left.
  \Gamma^\ESSMmath_{\psi_{Ri} \bar{\psi}_{Rj}}(p,-p)
  \right|_{p=0}
  &= 
  \partial_{p} \left.
  \Gamma^\SMmath_{\psi_{Ri} \bar{\psi}_{Rj}}(p,-p) 
  \right|_{p=0}
  \label{eq:matching-psiR}
  ,\\
  \left.
  \Gamma^\ESSMmath_{A_\mu \psi_i \bar{\psi}_j}(k,p,-(p+k))
  \right|_{k=p=0}
  &= 
  \left.
  \Gamma^\SMmath_{A_\mu \psi_i \bar{\psi}_j}(k,p,-(p+k)) 
  \right|_{k=p=0}
  \label{eq:matching-e}
  .
\end{align}
The additional matching condition for $\Gamma_{Z_\mu A_\nu}$ is
necessary, because the gauge fields $B_\mu$, $\vec{W}_\mu$ mix to
$A_\mu$, $Z_\mu$ and $W^\pm_\mu$.  Introducing relative field
renormalizations for $W^\pm_\mu$, $Z_\mu$, $A_\mu$, $\psi_{iL}$ and
$\psi_{iR}$ and inserting them into the matching conditions
\eqref{eq:matching-W}--\eqref{eq:matching-e} leads to the threshold
corrections for the $W$ and $Z$ boson masses and the electromagnetic
coupling%
\footnote{The Ward identity which reflects the \Uem\ gauge invariance
  in the Standard Model was used to simplify the result.}
\begin{align}
  (m_V^\SMmath)^2 &= (m_V^\ESSMmath)^2 + \left.
    \Gamma^\text{\ESSM,1L,heavy}_{V_\mu V_\nu,T} 
  \right|_{k^2=0}
  - m_V^2 K_{VV}
  , \qquad V\in \{W,Z\}
  \label{eq:thredhold-mV} ,\\
  &\equiv (m_V^\ESSMmath)^2 + \Delta m_V^2 ,\\
  e^\SMmath &= e^\ESSMmath \left(1 - \frac{1}{2} K_{AA} - \frac{s_W}{2 c_W} K_{ZA} \right)
  \equiv e^\ESSMmath + \Delta e
  \label{eq:threshold-simple-e}
  .
\end{align}
The relative field renormalization constants in
Eq.~\eqref{eq:thredhold-mV} and \eqref{eq:threshold-simple-e} are
given by
\begin{align}
  K_{AA} &= - \frac{\partial}{\partial k^2} \left. 
    \Gamma^\text{\ESSM,1L,heavy}_{A_\mu A_\nu,T} 
  \right|_{k^2=0} , &
  K_{ZA} &= \frac{2}{m^2_Z} \left. 
    \Gamma^\text{\ESSM,1L,heavy}_{A_\mu Z_\nu,T} 
  \right|_{k^2=0} \label{eq:def-KZA} ,\\
  K_{WW} &= - \frac{\partial}{\partial k^2} \left. 
    \Gamma^\text{\ESSM,1L,heavy}_{W^+_\mu W^-_\nu,T} 
  \right|_{k^2=0} , &
  K_{ZZ} &= - \frac{\partial}{\partial k^2} \left. 
    \Gamma^\text{\ESSM,1L,heavy}_{Z_\mu Z_\nu,T} 
  \right|_{k^2=0}
\end{align}
and by $\Gamma^\text{\ESSM,1L,heavy}_{V_\mu V'_\nu,T}$ we denote the
one-loop parts of the gauge boson 2-point functions that contain all
particles that we integrate out, i.e.\ all non-SM particles.  Using
Eq.~\eqref{eq:thredhold-mV}--\eqref{eq:threshold-simple-e} one can
write the threshold corrections for $g_Y$ and $g_2$ in the form
\begin{align}
  g_Y^\SMmath &= 
  \frac{e^\SMmath}{c_W^\SMmath}
  = \frac{e^\SMmath\,m_Z^\SMmath}{m_W^\SMmath}
  = g_Y^\ESSMmath \left( 
    1 
    + \frac{\Delta e}{e} 
    + \frac{1}{2}\frac{\Delta m_Z^2}{m_Z^2} 
    - \frac{1}{2}\frac{\Delta m_W^2}{m_W^2} 
  \right) ,\label{eq:general-tc-gY}\\
  g_2^\SMmath &= 
  \frac{e^\SMmath}{s_W^\SMmath}
  = \frac{e^\SMmath}{\sqrt{1-\left(\frac{m_W^\SMmath}{m_Z^\SMmath}\right)^2}}
  = g_2^\ESSMmath \left( 
    1 
    + \frac{\Delta e}{e} 
    - \frac{c_W^2}{s_W^2}\frac{\Delta m_Z^2}{2 m_Z^2} 
    + \frac{c_W^2}{s_W^2}\frac{\Delta m_W^2}{2 m_W^2} 
  \right) \label{eq:general-tc-g2}
  .
\end{align}
Inserting the explicit form of $\Delta e$, $\Delta m_W^2$ and
$\Delta m_Z^2$ yields
\begin{align}
    g_Y^{\DRbar,\ESSMmath} &= g_Y^{\MSbar,\SMmath}
    + \frac{g_Y^3}{(4\pi)^2}
    \Bigg[ 
      - \frac{4}{3} N_c \sum_{i=1}^3 \left(\frac{Y_{D_i}}{2}\right)^2
      \log\frac{m_{D_i}}{\mu}
      \notag\\
      &\phantom{{}=}
      - \frac{1}{3} N_c \sum_{i=1}^3 \sum_{k=1}^2 
      \left(\frac{Y_{\tilde{D}_{ik}}}{2}\right)^2
      \log\frac{m_{\tilde{D}_{ik}}}{\mu}
      \notag\\
      &\phantom{{}=}
      - \frac{1}{3} N_c \sum_{i=1}^3 \sum_{k=L,R}
      \left\{
        \left(\frac{Y_{\tilde{u}_{ik}}}{2}\right)^2
        \log\frac{m_{\tilde{u}_{ik}}}{\mu}
        + \left(\frac{Y_{\tilde{d}_{ik}}}{2}\right)^2
        \log\frac{m_{\tilde{d}_{ik}}}{\mu}
      \right\}
      \notag\\
      &\phantom{{}=}
      - \frac{1}{3} \sum_{i=1}^3 \sum_{k=L,R}
      \left(\frac{Y_{\tilde{e}_{ik}}}{2}\right)^2
      \log\frac{m_{\tilde{e}_{ik}}}{\mu}
      - \frac{1}{3} \sum_{i=1}^3
      \left(\frac{Y_{\tilde{\nu}_{iL}}}{2}\right)^2
      \log\frac{m_{\tilde{\nu}_{iL}}}{\mu}
      \notag\\
      &\phantom{{}=}
      - \frac{1}{3} \sum_{i=1}^2 \sum_{p=1}^2 \sum_{j=1}^2
      \left(\frac{Y_{H_{pi}^j}}{2}\right)^2
      \log\frac{m_{H_{pi}^j}}{\mu}
      - \frac{1}{3} \sum_{j=1}^2
      \left(\frac{Y_{H_{23}^j}}{2}\right)^2
      \log\frac{m_H}{\mu}
      \notag\\
      &\phantom{{}=}
      - \frac{2}{3} \sum_{i=1}^3 \sum_{p=1}^2 \sum_{j=1}^2
      \left(\frac{Y_{\tilde{H}_{piL}^j}}{2}\right)^2
      \log\frac{m_{\tilde{H}_{piL}^j}}{\mu}
      \notag\\
      &\phantom{{}=}
      - \frac{1}{3} \sum_{j=1}^2
      \left(\frac{Y_{{H'}^j}}{2}\right)^2
      \log\frac{m_{{H'}^j}}{\mu}
      - \frac{2}{3} \sum_{j=1}^2
      \left(\frac{Y_{\tilde{H'}_{L}^j}}{2}\right)^2
      \log\frac{m_{\tilde{H'}_{L}^j}}{\mu}
      \notag\\
      &\phantom{{}=}
      - \frac{1}{3} \sum_{j=1}^2
      \left(\frac{Y_{\bar{H'}^j}}{2}\right)^2
      \log\frac{m_{\bar{H'}^j}}{\mu}
      - \frac{2}{3} \sum_{j=1}^2
      \left(\frac{Y_{{\tilde{\bar{H'}}}_{L}^j}}{2}\right)^2
      \log\frac{m_{{\tilde{\bar{H'}}}_{L}^j}}{\mu}
    \Bigg]
    ,
  \label{eq:g1-threshold-program} 
\end{align}
\begin{align}
    g_2^{\DRbar,\ESSMmath} = g_2^{\MSbar,\SMmath}
    + \frac{g_2^3}{(4\pi)^2}
    \Bigg[ 
    &\frac{1}{3}
    - \frac{1}{6} N_c \sum_{i=1}^3 
    \log\frac{m_{\tilde{q}_{iL}}}{\mu} 
    - \frac{1}{6} \sum_{i=1}^3
    \log\frac{m_{\tilde{\ell}_{iL}}}{\mu} 
    \notag\\
    &- \frac{4}{3}  \log\frac{m_{\tilde{W}}}{\mu} 
    - \frac{1}{6} \log\frac{m_H}{\mu}
    \notag\\
    &- \frac{1}{6} \sum_{i=1}^2 \sum_{p=1}^2
    \log\frac{m_{H_{pi}}}{\mu} 
    - \frac{1}{3} \sum_{i=1}^3 \sum_{p=1}^2
    \log\frac{m_{\tilde{H}_{piL}}}{\mu} 
    \notag\\
    &- \frac{1}{6}\log\frac{m_{H'}}{\mu} 
    - \frac{1}{3}\log\frac{m_{\tilde{H'}_{L}}}{\mu} 
    \notag\\
    &- \frac{1}{6}\log\frac{m_{\bar{H'}}}{\mu} 
    - \frac{1}{3}\log\frac{m_{\tilde{\bar{H'}}_{L}}}{\mu} 
    \Bigg]
    ,
  \label{eq:g2-threshold-program}
\end{align}
where we have added a finite counterterm which converts $g_2$ from
the \MSbar\ to the \DRbar\ scheme \cite{Martin:1993yx} and neglected
mixing.

\subsection{Yukawa couplings in the \ESSM}
\label{sec:calculation-of-yukawa-coupling-TCs}
Since the SM fermion masses are measured but not the Yukawa couplings,
we don't define the \ESSM\ Yukawa couplings in terms of threshold
corrections to the running SM Yukawa couplings.  Instead we directly
match them at the one-loop level to the measured SM fermion masses and
gauge couplings via
\begin{align}
   y_t^{\DRbar,\ESSMmath} &=
   \frac{g_2^{\DRbar,\ESSMmath} m_t^\text{on-shell,SM}}{\sqrt{2} m_W^\text{on-shell,SM} \sin{\beta}}
   \left(
     1 - \frac{\delta m_W}{m_W^\text{on-shell,SM}}
     + \frac{\delta m_t}{m_t^\text{on-shell,SM}}
   \right),\label{eq:yt-ESSM}\\
   y_b^{\DRbar,\ESSMmath} &=
   \frac{g_2^{\DRbar,\ESSMmath} m_b^\text{\DRbar,(5)}(m_Z)}{\sqrt{2} m_W^\text{on-shell,SM} \cos{\beta}}
   \left(
     1 - \frac{\delta m_W}{m_W^\text{on-shell,SM}}
     + \frac{\delta m_b - \delta m_b^{(5)} + \delta m_b^{\text{shift}}}{m_b^\text{\DRbar,(5)}(m_Z)}
   \right),\label{eq:yb-ESSM}\\
   y_\tau^{\DRbar,\ESSMmath} &=
   \frac{g_2^{\DRbar,\ESSMmath} m_\tau^\text{on-shell,SM}}{\sqrt{2} m_W^\text{on-shell,SM} \cos{\beta}}
   \left(
     1 - \frac{\delta m_W}{m_W^\text{on-shell,SM}}
     + \frac{\delta m_\tau}{m_\tau^\text{on-shell,SM}}
   \right).\label{eq:ytau-ESSM}
\end{align}
Here $m_t^\text{on-shell,SM}$, $m_\tau^\text{on-shell,SM}$ and
$m_W^\text{on-shell,SM}$ are Standard Model on-shell masses
\cite{Nakamura:2010zzi}.  To avoid large logarithms from the bottom
mass we use the $\DRbar$ value of $m_b$ in the 5-flavor QCD at $m_Z$,
$m_b^\text{\DRbar,(5)}(m_Z) = \unit{2.83}{\giga\electronvolt}$
\cite{Baer:2002ek} which we then shift to the scale $\mu$ where the
Yukawa couplings are evaluated at.  The counterterms in
Eqs.~\eqref{eq:yt-ESSM}--\eqref{eq:ytau-ESSM} are defined as
\begin{alignat}{2}
  \delta m_f &= \delta m_f^{\text{on-shell},\ESSMmath} - \delta m_f^{\DRbar,\ESSMmath}
  &&= \left.\rretilde \Sigma_f(m_f^2)\right|_\text{finite}
  \qquad (f = t, b, \tau),\\
  \delta m_b^{(5)} &= \delta m_b^{\text{on-shell,QCD(5)}} - \delta m_b^{\DRbar,(5)}
  &&= -\frac{4m_b^\text{\DRbar,(5)}(m_Z)}{3(4\pi)^2}g_3^2 \left(5+3\log\frac{\mu^2}{m_b^2}\right),\\
  \delta m_b^\text{shift} &= \beta_{m_b}^{\DRbar,(5)} \log\frac{\mu}{m_Z}
  &&= -2 m_b^\text{\DRbar,(5)}(m_Z) \frac{g_3^2}{4\pi^2} \log\frac{\mu}{m_Z}, \\
  \delta m_W &= \delta m_W^{\text{on-shell},\ESSMmath} - \delta m_W^{\DRbar,\ESSMmath}
  &&= \left.\rretilde \Pi_{WW,T}(m_W^2)\right|_\text{finite}
  .
\end{alignat}
The self-energies $\Sigma_f$ and $\Pi_{WW,T}$ are listed in Appendix
\ref{sec:ESSM-self-energies}.  It was checked that the divergences of
\eqref{eq:yt-ESSM}--\eqref{eq:ytau-ESSM}, see Appendix
\ref{sec:ESSM-counterterm-def}, are in agreement with the prediction
from the one-loop RGEs.

\section{Obtaining accurate spectra in the \cESSM}
\label{sec:CESSM}

\subsection{The \cESSM\ and its parameters}

The threshold corrections presented in the previous section improve
the precision of the high-scale--low-scale connection in the general
\ESSM. In the present paper we apply them to the constrained version
of the \ESSM\ (\cESSM). The \cESSM\ is defined by a universal scalar
mass $m_0$, gaugino mass $M_{1/2}$ and trilinear mass $A$ at the gauge
coupling unification scale $M_X$ \cite{Athron:2009bs,Athron:2009ue}.
Owing to the unification of matter and Higgs fields in complete GUT
multiplets these constraints are particularly well motivated. The soft
scalar masses from the survival Higgs (which appear in the incomplete
$(\dimrep{27})'$ and $(\dimrep{\overline{27}})'$ representations) sector, $m_{H'}$,
$m_{\bar{H'}}$ are not assumed to be unified with $m_0$ and the soft
bilinear $B\mu'$ is also unconstrained.

\begin{figure}[tbh]
  \centering
  \includegraphics[width=\textwidth]{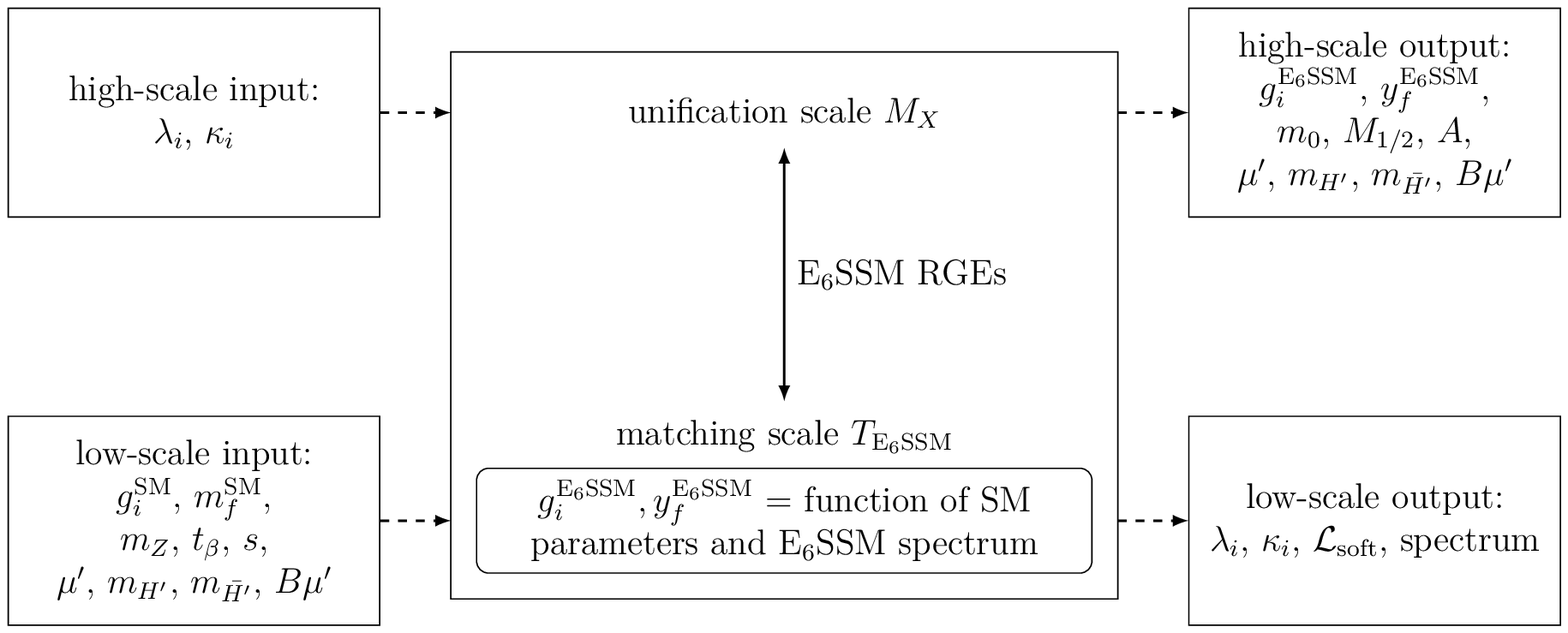}
  \caption{The structure of high- and low-scale input and output of
    the \cESSM.}
  \label{fig:essm-sm-matching}
\end{figure}

The \cESSM\ is an example of a highly predictive, non-minimal SUSY
model. It contains many new states at the TeV scale but has only few
free parameters. In theory, the \cESSM\ is fully determined by
specifying the gauge couplings, the superpotential parameters
$y_{t,b,\tau}$, $\lambda_i$ and $\kappa_i$ and the universal soft
parameters $m_0$, $M_{1/2}$, $A$ at the GUT scale as well as the
survival Higgs parameters. All high-scale parameters are shown in
the top row of boxes in \figref{fig:essm-sm-matching}.

In practice, the model predictions of course need to agree with known
SM constraints, in particular with the four known masses $m_Z$,
$m_{t,b,\tau}$ and the known low-energy gauge couplings. Furthermore,
it is useful to take the low-energy values of $\tan\beta=v_u/v_d$ and
$s=\sqrt{2}\langle S\rangle$ as free parameters of the model. In this
way, the six GUT-scale parameters $m_0$, $M_{1/2}$, $A$ and
$y_{t,b,\tau}$ are traded for the four known SM mass constraints and the
two low-scale input parameters $\tan\beta$ and $s$. Compared to the
familiar CMSSM, the \cESSM\ has no $\mu$ and $B\mu$ parameters,
which could be adjusted to fulfill SM constraints. Hence, the CMSSM
input parameters $m_0$, $M_{1/2}$ and $A$ are calculable in the \cESSM.

\figref{fig:essm-sm-matching} shows the resulting structure of input
and output of the \cESSM\ and the connection of the high and low
scales.  High-scale input parameters are the superpotential parameters
$\lambda_i(M_X)$ and $\kappa_i(M_X)$; low-scale input parameters are
$s(\Qfix)$, $\tan\beta(\Qfix)$ and the survival Higgs parameters
(defined at the matching scale $\Qmatch$), all defined in the \DRbar\
scheme.  The low-scale SM constraints fix the remaining parameters.
In previous studies $\lambda_i$, $\kappa_i$, $s$ and $\tan\beta$ were
sufficient to fix the spectrum.  Here, with the inclusion of threshold
corrections, we will be able to study the perturbation caused by the
survival Higgs sector as well.

As the Figure shows, $6+3+4=13$ high-scale parameters have been traded
against these $13$ low-scale input parameters, so this high- and
low-scale input determines the structure of the model completely and
at all scales.  The fixed scale $\Qfix$, where $s$ and $\tan\beta$ are
defined at, is set in Sec.~\ref{sec:results} to a value of the order
of the matching scale $\Qmatch$.

\subsection{Numerical procedure and the improved spectrum generator}
We extended the particle spectrum generator previously written for
\cite{Athron:2009bs}, which is partly based on \softsusy\ 2.0.5
\cite{Allanach:2001kg}.
As input the program gets a \cESSM\ parameter point specified by the
high- and low-scale input
\begin{align}
  \begin{split}
    \lambda_{1,2,3}(M_X), \kappa_{1,2,3}(M_X), s(\Qfix), \tan\beta(\Qfix),\\
    \mu'(\Qmatch), m_{H^\prime}(\Qmatch), m_{\bar{H}^\prime}(\Qmatch),
    B\mu'(\Qmatch),
  \end{split}
  \label{eq:program-input}
\end{align}
and the program knows about the SM constraints discussed above.
As illustrated in \figref{fig:essm-sm-matching}, RGEs and threshold
corrections connect high and low scales, and the aim is to find 
output parameters consistent with the
given input and then to compute the physical particle masses from the
obtained low-energy SUSY breaking parameters.  However, the threshold
corrections can only be computed once the full mass spectrum is
known. Hence, these steps need to be iterated until convergence is
reached, and in the first iteration the threshold corrections must be
ignored.

\begin{figure}[tbh]
  \centering
  \includegraphics[scale=0.75]{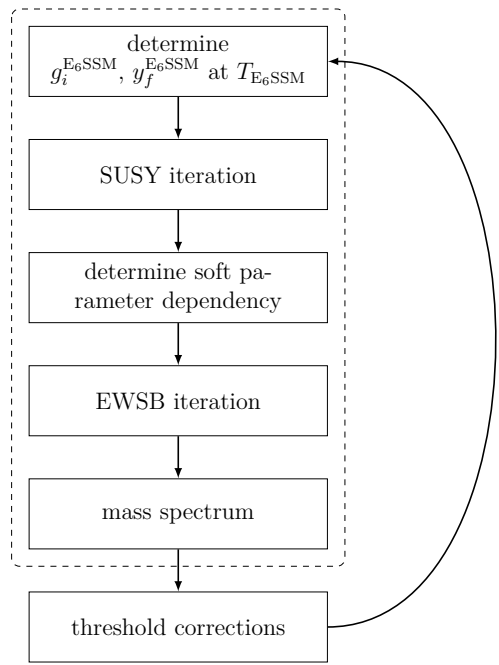}
  \caption{Program flow chart.  The dashed box marks the structure of
    the old spectrum generator, which had no threshold corrections.}
  \label{fig:program-flow}
\end{figure}

In the actual computation, we divide the basic strategy of each
iteration into five steps, see \figref{fig:program-flow}. The details
are as follows.
\begin{enumerate}
\item
  Determine the gauge and Yukawa couplings in the \ESSM\ at the
  threshold scale $\Qmatch$ from the known SM gauge couplings and
  masses, using SM RGEs and the threshold corrections.
  \begin{enumerate}
  \item Evolve the SM \MSbar\ gauge and Yukawa couplings from their
    known values at $m_Z$ \cite{Nakamura:2010zzi} to the
    intermediate matching scale $\Qmatch$ using SM RGEs.
  \item
    \label{enu:matching-step}
    Convert the SM \MSbar\ couplings to \ESSM\ \DRbar\ couplings using
    the individual particle threshold corrections discussed in Sec.\
    \ref{sec:thresholdcorrections}. This step is only
    possible once the physical particle masses are known, i.e.\ after
    the first iteration. Hence, in the first iteration, the threshold
    corrections are replaced by the trivial definitions
    \begin{align}
      y_f^\ESSMmath &:= \frac{y_f^\SMmath}{\cos\beta}
      \ (f=b,\tau) , &
      y_t^\ESSMmath &:= \frac{y_t^\SMmath}{\sin\beta}
      ,\\
      g_i^\text{\DRbar,\ESSMmath} &:= g_i^\text{\MSbar,\SMmath}
      \ (i=1,2,3) .
      \label{eq:trivial-matching-procedure}
    \end{align}
    Note that, in contrast to Ref.~\cite{Athron:2009bs}, we do not use
    an intermediate matching to the MSSM because the MSSM and the
    \ESSM\ spectra are typically mixed.
  \end{enumerate}
\item 
  Use RGEs to determine the high-scale gauge and Yukawa couplings
  and the low-scale values of the $\lambda_i$, $\kappa_i$. This step
  is completely independent of all soft SUSY breaking parameters
  because of the structure of the RGEs. 
  \begin{enumerate}
  \item Estimate low-energy values for the Yukawa couplings
    $\lambda_i(\Qmatch)$, $\kappa_i(\Qmatch)$. The necessity of this
    estimate is an additional complication not present e.g.\ in the
    CMSSM.  The difference is due to the fact that in the CMSSM all
    high-scale input parameters are soft breaking parameters, while
    here they are superpotential parameters.
  \item Evolve all \ESSM\ gauge and Yukawa couplings from $\Qmatch$ up
    to the unification scale $M_X$, defined as the scale where $g_1 =
    g_2$, using two-loop \ESSM\ RGEs.
  \item Set Yukawa couplings $\lambda_i(M_X)$, $\kappa_i(M_X)$ to
    program input values.
  \item Set $g_N := g_0 /\sqrt{40}$, where $g_0$ is defined by
    $g_1(M_X)$.
  \item Perform an iteration between $M_X$ and $\Qmatch$ to obtain
    values for the gauge and Yukawa couplings and $\lambda_i$,
    $\kappa_i$ until consistency is reached with low-energy boundary
    conditions 
    $y_f(\Qmatch)$, $g_i(\Qmatch)$ and high energy boundary conditions
    $\lambda_i(M_X)$, $\kappa_i(M_X)$.
  \end{enumerate}
\item With the gauge, Yukawa and $\lambda_i$, $\kappa_i$ couplings now
  known at all scales we find solutions for the soft breaking
  parameters. They are determined by the high-scale constraints of
  universal $m_0$, $M_{1/2}$, $A$, and by low-scale electroweak
  symmetry breaking (EWSB) conditions, i.e.\ consistency with the
  measured value of $m_Z$ and the input values for $\tan\beta$ and
  $s$. First, the dependency of the low-energy soft mass parameters on the
    GUT scale values $m^2_0$, $M_{1/2}$, $A_0$ is expressed in terms
    of the semi-analytical formulas
    \begin{align}
      m^2_i(t) &= a_i(t) m^2_0 + b_i(t) M^2_{1/2} + c_i(t) A_0 M_{1/2}
      + d_i(t) A^2_0 ,
      \\
      A_i(t) &= e_i(t) A_0 + f_i(t) M_{1/2},
      \\
      M_i(t) &= p_i(t) A_0 + q_i(t) M_{1/2},
    \end{align}
    where the coefficients $a_i(t),\dotsc, q_i(t)$ are calculated
    numerically at the scale $t=\log\Qmatch / M_X$.
\item Next, the EWSB conditions are used to fix the values of the
  soft parameters.
  \begin{enumerate}
  \item The obtained expressions for the low-energy soft parameters
    $m^2_i(t)$, $M_i(t)$, $A_i(t)$ are then combined with tree level
    EWSB conditions
    \begin{align}
      \frac{\partial V}{\partial v_1} = \frac{\partial V}{\partial v_2} =
      \frac{\partial V}{\partial s} = 0
      \label{eq:EWSB-conditions}
    \end{align}
    and the known values of $m_Z$, $\tan\beta$ and $s$ to form three
    quadratic equations in the soft masses $m_0$, $M_{1/2}$ and $A$,
    which can be reduced to a single quartic equation.
  \item This quartic equation is then solved numerically and the
    values used to determine the mass spectra.  Note that in principle
    there will be four solutions, though some or all may be
    complex. Therefore our routine deals with between zero and four
    sets of real solutions.
  \item Tadpole corrections can now be calculated and added to the
    EWSB conditions, and a solution of consistent EWSB for the leading
    one-loop effective potential is found iteratively.
  \end{enumerate}
\item The full \cESSM\ particle mass spectrum is now determined for
  each set of $\{m^2_0, M_{1/2}, A_0\}$ found to be consistent with
  both high scale and low scale boundary conditions.
\end{enumerate}

The final solution is found by iterating over all five steps until
convergent solutions are obtained.

It is important to note that in general EWSB is not guaranteed in the
\cESSM, i.e.\ solutions for $\{m^2_0, M_{1/2}, A_0\}$ from
Eq.~\eqref{eq:EWSB-conditions} are not always found.  But for
sufficiently large values of $\kappa_i$ the soft parameter
$m^2_S$ always gets negative at low energies, which triggers EWSB
\cite{Athron:2009bs}.

\section{Results}
\label{sec:results}

In this section we present the results obtained with our improved
\cESSM\ spectrum generator. We start by quantifying how the threshold
corrections stabilize the results for parameters like gauge couplings
and the mass spectrum. We will then extensively discuss the resulting,
more accurate mass spectrum. On the one hand we will obtain more
precise information on previously defined benchmark points, implying
that some are now excluded by LHC data. On the other hand we scan over
the parameter space and find large regions which are compatible with
LHC limits on SUSY particles and are consistent with the latest
discovery of the new boson which we associate with the lightest Higgs
boson in the model. Finally we investigate the impact of the survival
Higgs sector, which is possible due to the threshold corrections. It
has implications on gauge coupling unification and on the predictions
for the low-energy mass spectrum.  In subsequent studies we sometimes
make use of a test point, PP1, to illustrate effects, which is
defined by,
\begin{align}
  \begin{split}
    \tan\beta(\Qfix) = 35, \lambda_{1,2,3}(M_X) = \kappa_{1,2,3}(M_X) = 0.2,
    s(\Qfix) = \unit{10}{\tera\electronvolt},\\
    \mu'(\Qmatch) = m_{H^\prime}(\Qmatch) = m_{\bar{H}^\prime}(\Qmatch)
    = \unit{10}{\tera\electronvolt}, B\mu'(\Qmatch) = 0.
  \end{split}
  \label{eq:PP1-definition}
\end{align}
Furthermore we set $\Qfix=\unit{3}{\tera\electronvolt}$ for all the
following analyses.

\subsection{Matching scale dependency}\label{sec:matching-scale-dependency}
\begin{figure}[tbh!]
  \centering
  \includegraphics[width=0.8\textwidth]{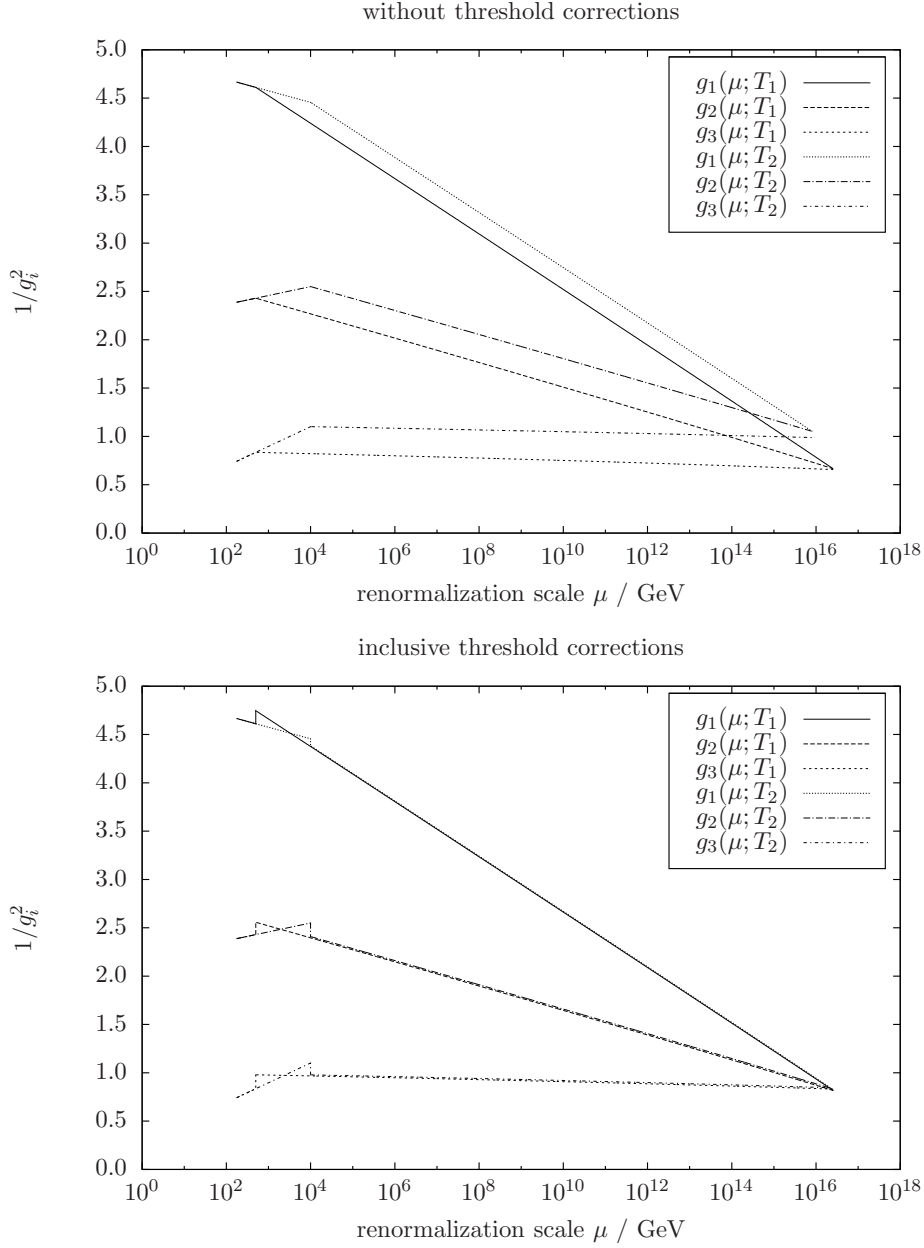}
  \caption{Running gauge couplings for parameter point PP1 for two
    different matching scales $T_1=\unit{500}{\giga\electronvolt}$ and
    $T_2=\unit{10}{\tera\electronvolt}$ with and without threshold
    corrections.}
  \label{fig:PP1-gauge-coupling-unification}
\end{figure}

In Sec.~\ref{sec:visualisation-of-TC} we presented
\figref{fig:visualisation-of-gauge-tc} to illustrate how the threshold
corrections should adjust the RG flow of the gauge couplings so that
the matching scale dependency is removed.  Now in
\figref{fig:PP1-gauge-coupling-unification} we present an explicit
demonstration of this effect in the \ESSM\ with the threshold
corrections we have calculated.  The gauge couplings are plotted as a
function of the renormalization scale for two different matching
scales, $T_1=\unit{500}{\giga\electronvolt}$ and
$T_2=\unit{10}{\tera\electronvolt}$.  If no threshold corrections are
used (upper plot) different matching scales yield different
predictions of $M_X$, which leads to a different phenomenology at the
TeV scale.  The inclusion of threshold corrections reduces this
unphysical behavior.  As shown in the lower plot, it leads to an
approximately matching-scale independent prediction of $M_X$.

\begin{figure}[tbh!]
  \centering
  \includegraphics[width=0.8\textwidth]{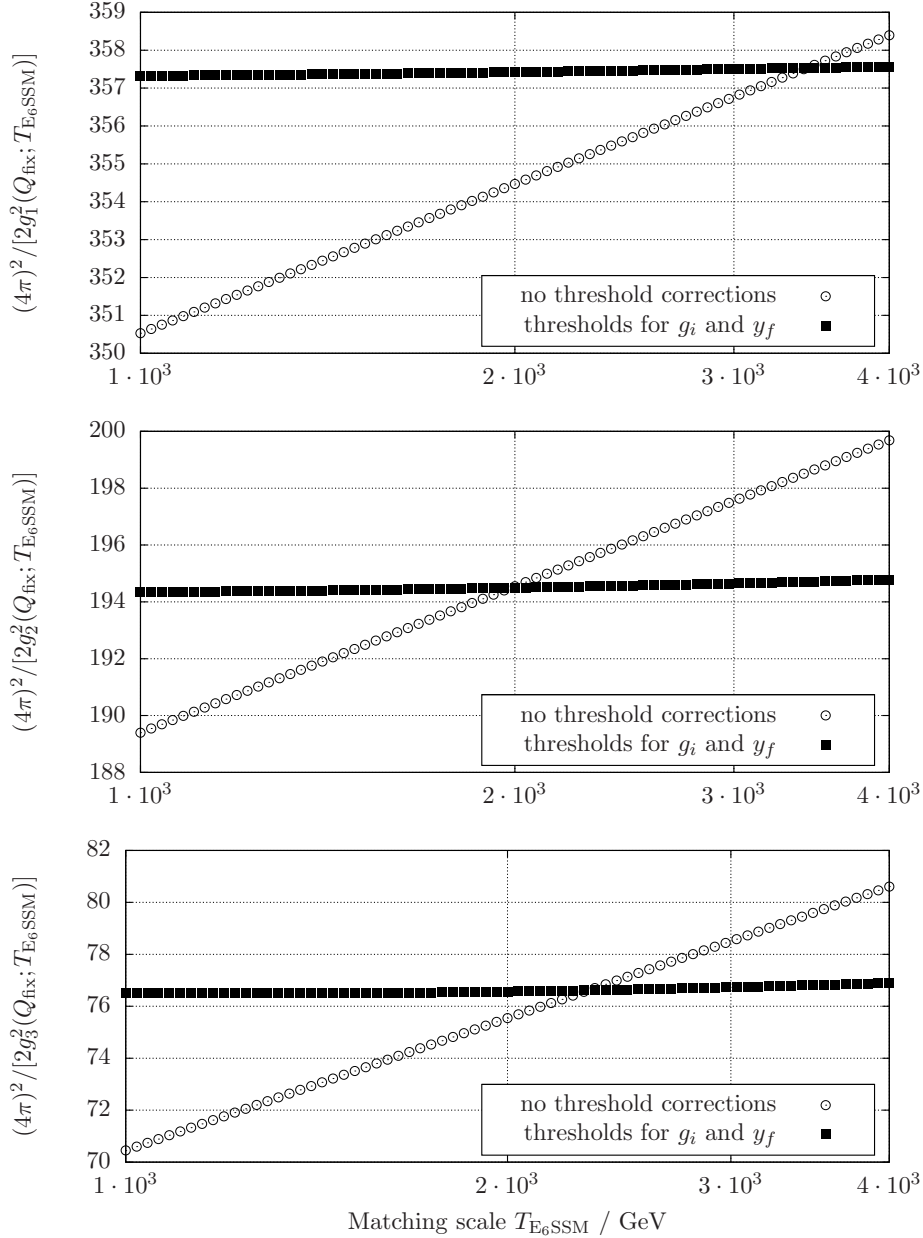}
  \caption{Dependency of the gauge couplings $g_i$ at
    $\Qfix=\unit{3}{\tera\electronvolt}$ on the matching scale
    $\Qmatch$ for parameter point PP1.  The circles show the behavior
    without threshold corrections and the squares with corrections for
    $g_i$ and $y_f$.}
  \label{fig:PP1-couplings-g-of-Q}
\end{figure}

This improved behavior should manifest itself in a reduced scale
dependence of model predictions at the TeV scale.
As a  test of this, \figref{fig:PP1-couplings-g-of-Q} shows the
\ESSM\ gauge couplings $g_i(\Qfix)$ at the fixed scale
$\Qfix=\unit{3}{\tera\electronvolt}$, as a function of the
matching scale $\Qmatch$ for parameter point PP1.  In the case of the
trivial matching relations \eqref{eq:trivial-matching-procedure} one
finds a clear (unphysical) dependence of $g_i(\Qfix)$ on
$\Qmatch$. As shown in Eq.~\eqref{eq:toymodel-threshold-test} the
slopes of $(4\pi)^2/[2g_i^2(\Qfix;\Qmatch)]$ are
given by the difference $\Delta\beta_i :=
\beta_i^{\ESSMmath}-\beta_i^{\SMmath}$ of the one-loop gauge coupling
beta functions, up to two-loop and iteration effects. 

In \tabref{tab:slopes}, columns 4 and 5, the numerical values of
$\Delta\beta_i$ as well as the slopes without threshold corrections
for PP1 are listed.  The values roughly coincide, thus confirming
Eq.~\eqref{eq:toymodel-threshold-test}.  When the threshold
corrections \eqref{eq:g1-threshold-program},
\eqref{eq:g2-threshold-program} and \eqref{eq:g3-threshold-program}
are used, the matching scale dependency reduces to about
$\unit{4}{\%}$ or less, resulting in the flattening of the curves
shown in \figref{fig:PP1-couplings-g-of-Q}.  The remaining matching
scale dependency as well as the difference between columns 4 and 5 are
due to still missing contributions of higher orders.

\begin{table}[tbh]
  \centering
  \begin{tabular}{lrrrrrr}
    \toprule
    & & & & \multicolumn{2}{c}{slopes of $(4\pi)^2/[2g_i^2(\Qfix;\Qmatch)]$} \\
    Coupling & $\beta^\ESSMmath_i$ & $\beta^\SMmath_i$ & $\Delta\beta_i$ & w/o thresh.\ & w/ thresh. \\
    \midrule
    $g_1$ & $9.6$ & $4.1$   & $5.5$  & $5.67$ & $0.18$ \\
    $g_2$ & $4$   & $-3.17$ & $7.17$ & $7.42$ & $0.32$ \\
    $g_3$ & $0$   & $-7$    & $7$    & $7.33$ & $0.29$ \\
    \midrule
    & & & & \multicolumn{2}{c}{slopes of $(4\pi)^2\log y_f(\Qfix;\Qmatch)$} \\
    & $\beta^\ESSMmath_f(\Qmatch)$ & $\beta^\SMmath_f(\Qmatch)$ & $-\Delta\beta_f$ & w/o thresh.\ & w/ thresh. \\
    \midrule
    $y_t$   & $-2.33$ & $-6.14$ & $-3.77$ & $-3.81$ & $0.77$ \\
    $y_b$   & $-4.66$ & $-8.18$ & $-4.92$ & $-4.95$ & $-0.20$\\
    $y_\tau$ & $-0.34$ & $0.77$  & $-0.29$ & $-0.21$ & $0.53$\\
    \bottomrule
  \end{tabular}
  \caption{Effect of the threshold corrections on the 
    dependency of the gauge and Yukawa couplings on $\Qmatch$ for parameter 
    point PP1.  
    The slopes in the last two columns are
    obtained by linear fits to the data in \figref{fig:PP1-couplings-g-of-Q} and
    \ref{fig:PP1-couplings-y-of-Q}.}
  \label{tab:slopes}
\end{table}

\begin{figure}[tbh!]
  \centering
  \includegraphics[width=0.8\textwidth]{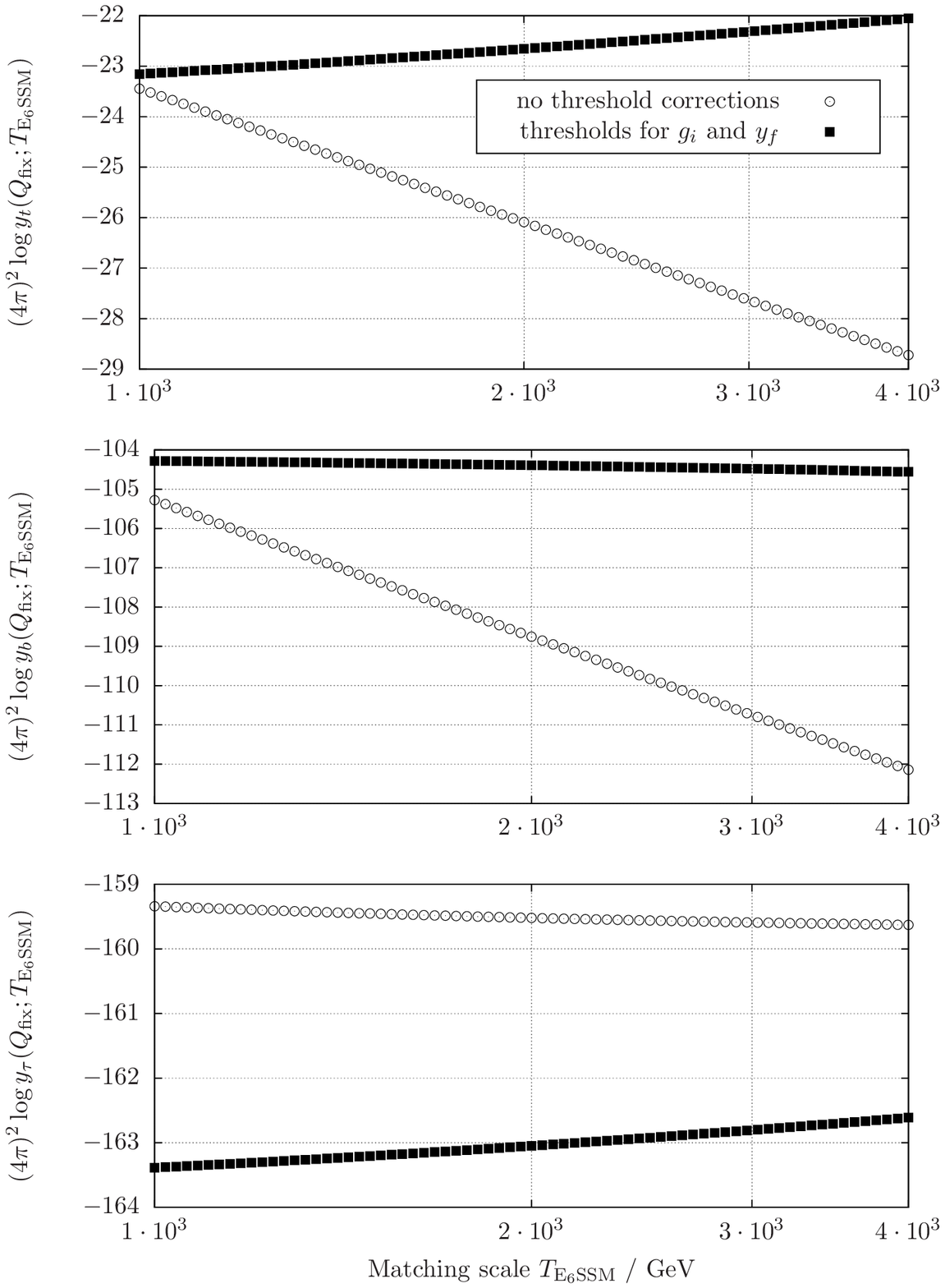}
  \caption{Dependency of the Yukawa couplings $y_f$ at
    $\Qfix=\unit{3}{\tera\electronvolt}$ on the matching scale
    $\Qmatch$ for parameter point PP1.  The circles show the behavior
    without threshold corrections and the squares with corrections for
    $g_i$ and $y_f$.}
  \label{fig:PP1-couplings-y-of-Q}
\end{figure}

\figref{fig:PP1-couplings-y-of-Q} shows the same analysis for the
Yukawa couplings $y_t$, $y_b$ and $y_\tau$.  Analogous to the gauge
couplings one expects the following slopes of the Yukawa couplings as
a function of the matching scale
\begin{align}
  \frac{\dd \log y_t(\Qfix;t)}{\dd t} &= -\frac{1}{(4\pi)^2}
  \left[\beta_t^\ESSMmath(t) - \beta_t^\SMmath(t) + \beta_{\tan\beta}(t) \cos^2\beta\right]
  \equiv -\frac{1}{(4\pi)^2} \Delta\beta_t
  ,\\
  \frac{\dd \log y_{b,\tau}(\Qfix;t)}{\dd t} &= -\frac{1}{(4\pi)^2}
  \left[\beta_{b,\tau}^\ESSMmath(t) - \beta_{b,\tau}^\SMmath(t) - \beta_{\tan\beta}(t) \sin^2\beta\right]
  \equiv -\frac{1}{(4\pi)^2} \Delta\beta_{b,\tau}
  .
\end{align}
Note that these relations also contain the $\beta$ function for
$\tan\beta$, defined in Eq.\ \eqref{eq:betatanbeta}.
In \tabref{tab:slopes}, columns 4 and 5, the numerical values of
$\Delta\beta_f$ as well as the slopes without threshold corrections
for PP1 are listed.  One finds an agreement with the predicted slopes
within a few percent.  When the threshold corrections from
Sec.~\ref{sec:calculation-of-gauge-coupling-TCs} and
\ref{sec:calculation-of-yukawa-coupling-TCs} are taken into account
one finds an overall reduction of the matching scale dependency 
effects. The remaining scale dependence is due to neglected
higher-order effects, including in particular
QED-logarithms of the small fermion masses in higher powers than taken
into account in the one-loop computation of $\delta m_f$ in
Eqs.~\eqref{eq:yt-ESSM}--\eqref{eq:ytau-ESSM}.

\subsection{Particle masses}
In softly broken SUSY models, like the \cESSM, where the
pattern of soft breaking masses is set 
substantially above the EW scale (e.g.~$M_X$), the low-energy \DRbar\
running masses are very sensitive to the renormalization group flow of
the gauge and Yukawa couplings.  This implies that the physical mass
spectrum strongly depends on the matching scale used for the gauge
and Yukawa couplings unless stabilized by the inclusion of threshold
corrections.  Here we show how the inclusion of the \cESSM\ threshold
corrections improves the prediction of the low-energy \DRbar\
soft masses and yields a mass spectrum where the dependence on the
unphysical matching scale is considerably reduced.

\begin{figure}[tbh]
  \centering
  \includegraphics[width=0.8\textwidth]{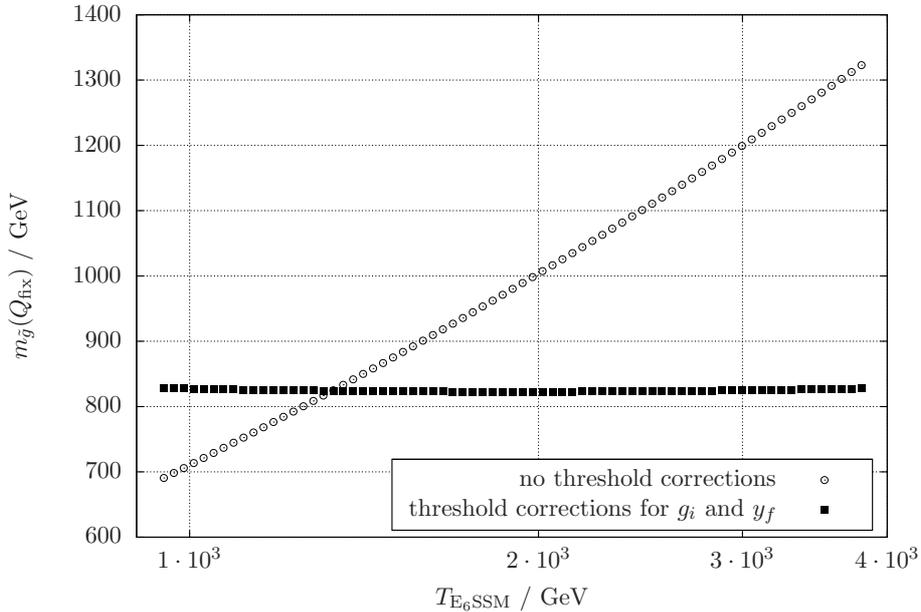}
  \caption{Dependency of the gluino mass at
    $\Qfix=\unit{3}{\tera\electronvolt}$ on the matching scale
    $\Qmatch$ for parameter point PP1.  The circles show the behavior
    without threshold corrections and the squares with corrections for
    $g_i$ and $y_f$.}
  \label{fig:PP1-GluinoMass-of-Q}
\end{figure}

\begin{figure}[tbh]
  \centering
  \includegraphics[width=0.8\textwidth]{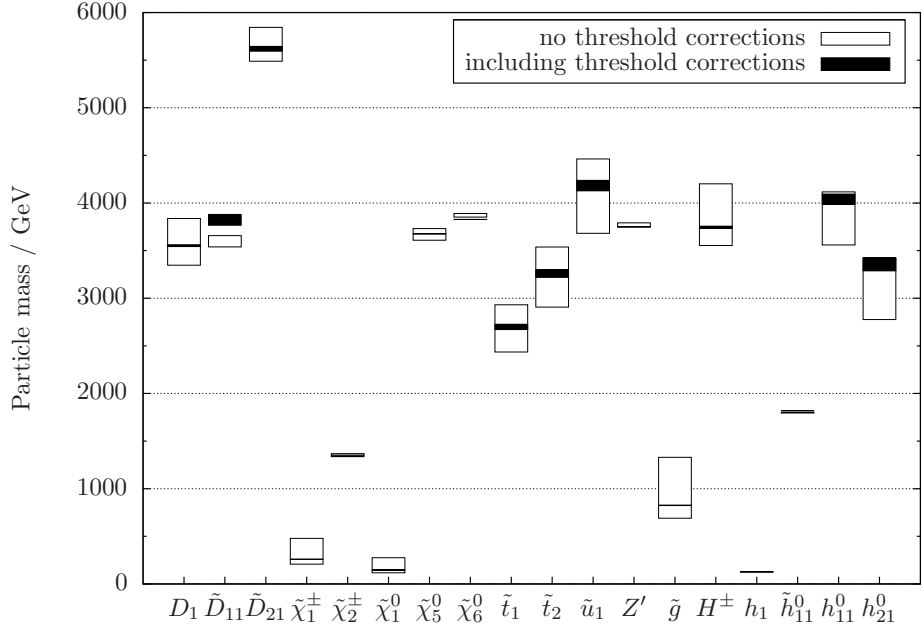}
  \caption{Particle spectra for parameter point PP1.  The white and
    the black boxes show the variation of the particle masses when the
    matching scale $\Qmatch$ is varied in the interval
    $[\frac{1}{2}\Qinter, 2\Qinter]$, where
    $\Qinter=\unit{1.9}{\tera\electronvolt}$ is the geometric average
    of all shown particle masses.  The black boxes show the error with
    threshold corrections and the white boxes without.}
  \label{fig:PP1-particle-spectrum}
\end{figure}

In \figsref{fig:PP1-GluinoMass-of-Q} and
\ref{fig:PP1-particle-spectrum} the dependency of the particle masses
on the matching scale for parameter point PP1 is shown.  The matching
scale is varied in the range $[\frac{1}{2}\Qinter, 2\Qinter]$, where
$\Qinter=\unit{1.9}{\tera\electronvolt}$ is the geometric average of
all particle masses shown in \figref{fig:PP1-particle-spectrum}.

\figref{fig:PP1-GluinoMass-of-Q} focuses on the gluino. The theory
uncertainty implied by the $\Qmatch$ dependence is $\unit{65}{\%}$ without
threshold corrections (the percentage value is defined as the full
variation of the particle mass divided by the mass at $\Qinter$).
This huge uncertainty is entirely due to the missing threshold
corrections, and including these reduces the uncertainty to only
$\unit{0.5}{\%}$.

\figref{fig:PP1-particle-spectrum} shows a subset of the generated
particle spectrum for the parameter point PP1.  The variation of the
particle masses is drawn with a white box in case of trivial matching
and with a black box in case of implemented threshold corrections. The
gluino, the lightest chargino and the lightest neutralino show the
biggest dependency with $\unit{65}{\%}$, $\unit{83}{\%}$ and
$\unit{85}{\%}$, respectively.
With the implemented threshold corrections, one finds a reduction of
the variation down to $0.1$--$\unit{4}{\%}$.  The biggest decrease is
found for the gluino, the lightest chargino and the lightest
neutralino whose remaining variation is of the order of
$\unit{0.5}{\%}$.  This is because the gluino mass is very sensitive
to $g_3$, as can be seen from the RGEs of the soft parameter $M_3$
\cite{Athron:2009bs}.  Note that the remaining variation of the
particle masses is due to 2-loop and iterative effects, which are of
the order of up to $\unit{4}{\%}$.

\begin{figure}[tbh]
  \centering
  \includegraphics[width=0.49\textwidth]{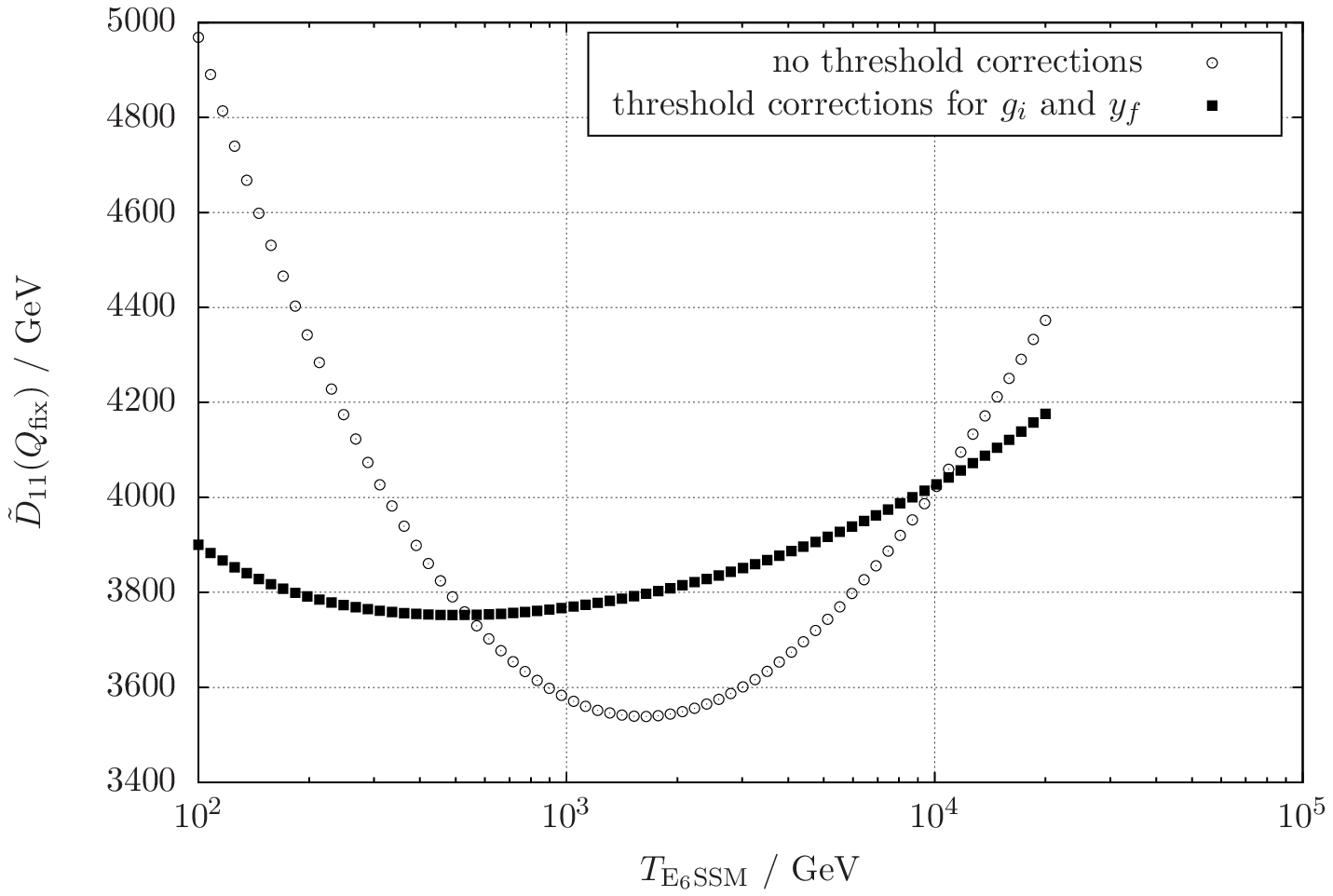}
  \includegraphics[width=0.49\textwidth]{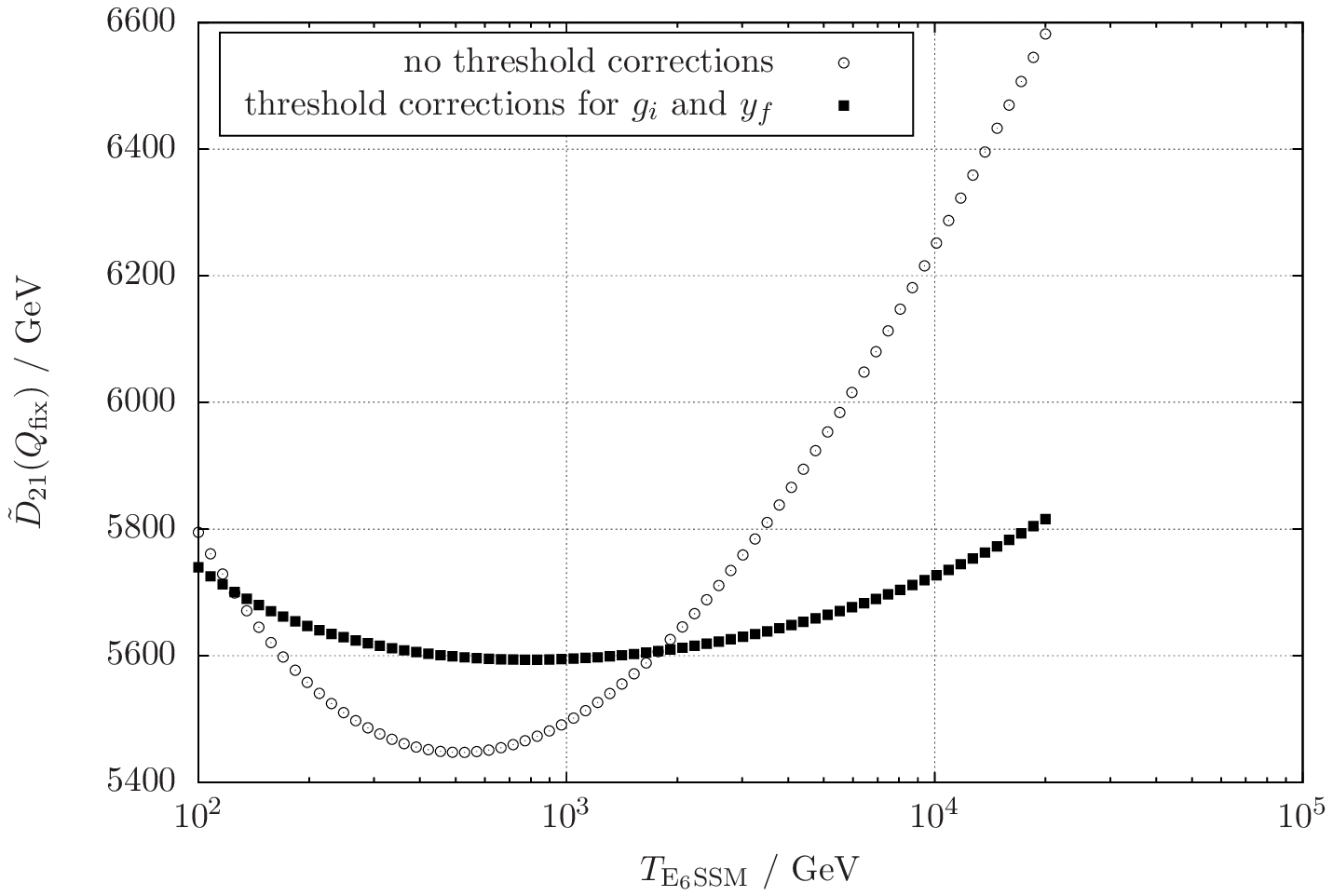}
  \caption{Dependency of the first generation scalar exotic masses at
    $\Qfix=\unit{3}{\tera\electronvolt}$ on the matching scale
    $\Qmatch$ for parameter point PP1.  The circles show the behavior
    without threshold corrections and the squares with corrections for
    $g_i$ and $y_f$.  In the interval $[\frac{1}{2}\Qinter, 2\Qinter]$
    the exotic mass $m_{\tilde{D}_{11}}$ without threshold corrections
    has a minimum which leads to poor error estimation, whereas
    $m_{\tilde{D}_{21}}$ is approximately linear within this
    interval.}
  \label{fig:PP1-scalar-exotics-of-Q}
\end{figure}

Note that the variation of the matching scale is not always a good
estimation of the theoretical uncertainty.  For example the error band
of $m_{\tilde{D}_{11}}$ without threshold corrections in
\figref{fig:PP1-particle-spectrum} is shifted when adding threshold
corrections but is not reduced in size.  As can be seen in
\figref{fig:PP1-scalar-exotics-of-Q} the reason for the shift is the
non-linear behavior of $m_{\tilde{D}_{11}}$ when $\Qmatch$ is varied.
In the interval $[\frac{1}{2}\Qinter, 2\Qinter]$ the exotic mass
$m_{\tilde{D}_{11}}$ without threshold corrections happens to have a minimum which
leads to an abnormally  narrow error band.  In contrast
$m_{\tilde{D}_{21}}$ shows the typical, approximately linear behavior
in $[\frac{1}{2}\Qinter,
2\Qinter]$ which leads to much better uncertainty estimation.

\subsection{Exploration of the \ESSM\ parameter space}
\subsubsection{Parameter space}
\label{sec:parameterspace}
We now turn from the specific benchmark point PP1 to a fuller
investigation of the \ESSM\ parameter space. Since our main focus is
the impact of threshold corrections we still restrict ourselves to a
two-dimensional slice of the $12$-dimensional parameter space. 
We choose to keep the Yukawa couplings of the $SU(5)$ $5$-plets
containing $\SuperField{H}_{1i}$, $\SuperField{H}_{2i}$,
$\SuperField{D}_i$ and $\SuperField{\bar{D}}_i$ the same and
generation-independent, thus leaving a single unified 
exotic Yukawa coupling, defined at the GUT scale $M_X$,
\begin{align}
  \lambda_1 = \lambda_2 = \lambda_3 =\kappa_1 =
  \kappa_2 = \kappa_3
  \label{lambdauniversality}
\end{align}
between those states and the third generation
singlet.  In addition we fix
\begin{align}
  s = \mu' = m_{H'} = m_{\bar{H'}} = \unit{10}{\tera\electronvolt},
  B\mu' = 0.\label{parameters}
\end{align}
The remaining input parameters are $\tan\beta$ and $\lambda_3$, and we
scan over them in the range
\begin{align}
  \tan\beta \in [2, 45],\qquad \lambda_3 \in [0, 3]. \label{scanrange}
\end{align}
Note that we are defining $\lambda_3$ at the GUT scale, and in this
respect the numerical value should not be compared with $\lambda$ in
the NMSSM which is often defined near the electroweak scale (EW).  By
defining $\lambda_3$ at the GUT scale we ensure automatically that it
is perturbative at all scales between the electroweak and the GUT
scale.\footnote{In the NMSSM it is well known that to ensure
  perturbativity $\lambda(\text{EW}) \lesssim 0.7$, while an analysis
  of this in the \ESSM\ \cite{King:2005jy} found that
  $\lambda_3(\text{EW}) \lesssim 0.85$ is required for perurbativity
  up to the GUT scale, though this limit depends on the values of the
  other exotic Yukawa couplings and $\tan \beta$.}

 For each choice of $(\tan\beta,\lambda_3)$, requiring electroweak
symmetry breaking simultaneously with 
\eqref{parameters} leads to up to four
solutions for $(m_0,M_{1/2})$, which we number consecutively.  In the
following $(m_0,M_{1/2})$-plots we will show all of them, where
solution 1 is preferred over solution 2 in the overlap region.  In
\figsref{fig:surv-scan-gaugeMx}--\ref{fig:surv-scan-mgluino-and-n1}
we will show only the first solution for simplicity.  

\begin{figure}[tbh!]
  \centering
  \includegraphics[width=0.8\textwidth]{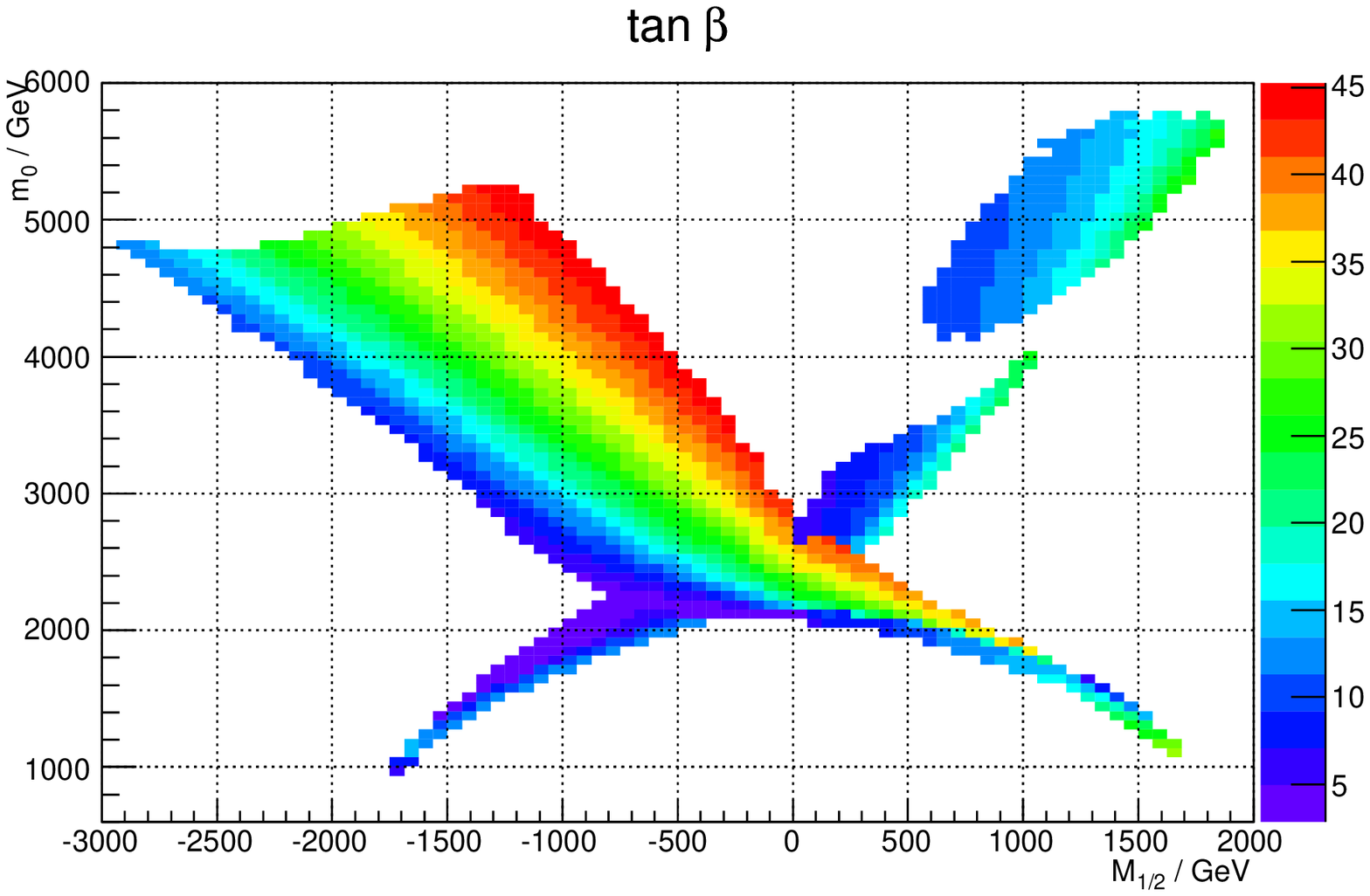}
  \includegraphics[width=0.8\textwidth]{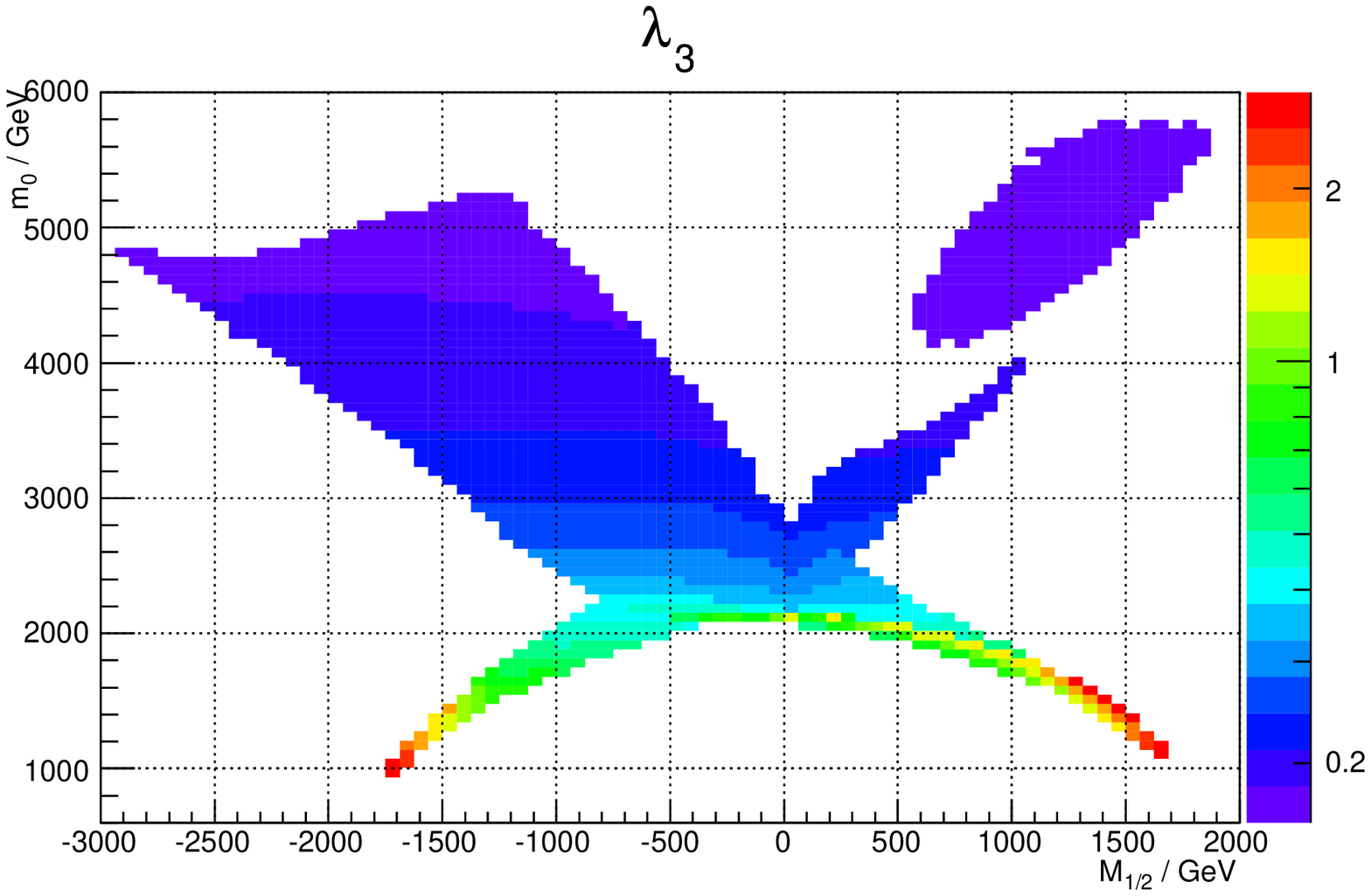}
  \caption{Illustration of the mapping
    $(\tan\beta,\lambda_3)\rightarrow(m_0,M_{1/2})$.  The top panel
    shows the input parameter $\tan\beta$ varying across the
    $(m_0,M_{1/2})$ plane (of output parameters). The bottom panel
    shows the same for the input parameter $\lambda_3$.}
  \label{fig:mapping-tanb-lam3-to-M12-m0}
\end{figure}

Since our input parameters $(\tan\beta,\lambda_3)$ have less direct
physical meaning than the output parameters $(m_0,M_{1/2})$, we
display many results in the $(m_0,M_{1/2})$ plane.  To better
understand the connection between these two sets we show in
\figref{fig:mapping-tanb-lam3-to-M12-m0} the mapping of
$(\tan\beta,\lambda_3)\rightarrow(m_0,M_{1/2})$.  In the top panel we
show as a color contour plot how $\tan\beta$ varies over the
$(m_0,M_{1/2})$ plane, while on the lower plot we do the same for
$\lambda_3$.  \figref{fig:mapping-tanb-lam3-to-M12-m0} thus allows to
read off the original input values for $(\tan\beta,\lambda_3)$ for any
given point in the subsequent plots.

The butterfly shape of \figref{fig:mapping-tanb-lam3-to-M12-m0} is due
to the superposition of the four solutions.  Most of the
$(m_0,M_{1/2})$ space is covered by relatively small $\lambda_3 <
0.4$, while larger values of lambda are concentrated in a narrow
region of the parameter space due to the renormalisation group flow
leading to a focusing on a significantly smaller range of $\lambda_3$
at the electroweak scale.  The $\tan\beta$ dependency depends on the
solution, but in much of the parameter space large values of $\tan
\beta$ are associated with large $m_0$ and small $M_{1/2}$.

\figref{fig:mapping-tanb-lam3-to-M12-m0} is not exactly symmetric in
$\pm M_{1/2}$ because $\lambda_3$ and therefore the effective
$\mu$-parameter is required to be positive in our analysis.

\subsubsection{Experimental constraints}
\label{exp-constraints}

At this point we briefly summarize the various experimental
constraints on the model. A more detailed discussion 
can be found in \cite{Athron:2012sq}.

Since the publication of the first studies of the
\cESSM~\cite{Athron:2009bs,Athron:2009ue} the Atlas and CMS experiments
have already placed strong constraints on supersymmetry from the first
$\approx \unit{5}{\femto\barn}^{-1}$ of data gathered from the LHC
running at $\sqrt{s} = \unit{7}{\tera\electronvolt}$.  With
$\unit{4.7}{\femto\barn}^{-1}$ of data Atlas have performed searches
for squarks and gluinos, by looking for jets plus missing transverse
momentum and possibly one isolated
lepton \cite{ATLAS_susy1,ATLAS_susy2,ATLAS_susy3} and presented the
constraints in a CMSSM interpretation.\footnote{They also presented
  constraints in a simplified model interpretation, however this is
  less relevant for our present purposes since the physical spectrum
  of our model, especially if the colored exotics are heavy, is far
  closer to that of the CMSSM than the simplified models.}  CMS
collaboration have also placed similar constraints using the razor
analysis \cite{Rogan:2010kb} with $\unit{4.4}{\femto\barn}^{-1}$ of
data \cite{CMS_razor} and another analysis with the transverse mass
variable \cite{CMS_MT2} using $\unit{4.73}{\femto\barn}^{-1}$ of
data. Very recently the ATLAS released updated results with
$\unit{5.8}{\femto\barn}^{-1}$ data from the $\sqrt{s} =
\unit{8}{\tera\electronvolt}$ run \cite{ATLAS-8TeV}.

It was estimated in \cite{Athron:2011wu,Athron:2012sq} that the effect
of these constraints on the \cESSM\ is to place a limit on the gluino
mass, close to $\unit{850}{\giga\electronvolt}$. Since the squarks
must always be heavier than the gluino, due to the RG
flow,\footnote{The soft scalar masses get large contributions
  (relative to the gauginos) from $M_{1/2}$, due to the
  renormalization group evolution, which is very different from the
  MSSM case.} squark limits are automatically satisfied.  However the
$8$ TeV data has increased the limits on the gluino, in the heavy
squarks region the limit varies between $\approx 900$--$1000$ GeV.

  However it is also of interest to point out that in
certain \ESSM\ scenarios (where the bino-like neutralino is not
stable, leading to longer cascade decays and an alternative LSP) the
gluino limits from these searches may not apply \cite{Belyaev:2012si}
resulting in different limits.

Furthermore, there are limits on the Higgs mass
\cite{ATLAS:2012ae,Chatrchyan:2012tx} from analyses using between
$4.6$ and $\unit{4.9}{\femto\barn}^{-1}$ of LHC data.  At
$\unit{95}{\%}$ confidence level Atlas exclude a standard model Higgs
in the ranges $110.0$--$\unit{117.5}{\giga\electronvolt}$,
$118.5$--$\unit{122.5}{\giga\electronvolt}$ and
$129$--$\unit{539}{\giga\electronvolt}$, while at the same confidence
level CMS exclude $127.5$--$\unit{600}{\giga\electronvolt}$.  This
leaves a narrow allowed window of
$117.5$--$\unit{118.5}{\giga\electronvolt}$ and a wider one of
$123.5$--$\unit{127.5}{\giga\electronvolt}$.

As this paper was being finalized the discovery of a new particle
consistent with a Higgs boson with a mass of around
$125$--$\unit{126}{\giga\electronvolt}$ was announced by ATLAS
\cite{:2012gk} and CMS \cite{:2012gu}.  CMS report a $5.0 \sigma$
excess at $\unit{125.5}{\giga\electronvolt}$ and quote a best fit mass
of $125.3 \pm 0.4 \text{(stat.)} \pm 0.5 \text{(syst.)}$ GeV while
ATLAS find a $6.0 \sigma$ excess at $126.5$, and using the two
channels with the highest mass resolution estimate the mass as $126.0
\pm 0.4 \text{(stat.)} \pm 0.4 \text{(syst.)}$ GeV.

The strongest current constraint on the $Z^\prime_N$ comes from a CMS
analysis using $\approx \unit{5}{\femto\barn}^{-1}$ of data
\cite{:2012it}, announced as this paper is being finalized and setting
a lower limit of $\unit{2.08}{\tera\electronvolt}$ on the mass of the
$Z'_N$ boson. However one should note that the lower limit assumes
there are now light exotics available for the $Z^\prime$ to decay
into.  If such are present then the branching ratio to leptons is
diluted and the limits are weaker.

Finally the model also contains exotic (colored and inert) fermions
and sfer\-mi\-ons. The inert states are weakly produced and we do not
anticipate very large new limits to be set by the early data from the
LHC, but the colored states should be produced abundantly and could be
tightly constrained.  However the existing analyses do not apply to
the fermionic diquarks or leptoquarks described in this model since
they carry half integer spin and are odd under their respective
discrete symmetries, decaying with missing energy.  Thus it is
currently unknown what limit the LHC has placed on their mass. Here we
will only consider cases where these exotics are heavy anyway but when
discussing existing benchmarks from the literature we will not apply
any new constraints on them.

The \ESSM\ also contains a number of dark matter candidates.  It is
possible that the correct relic density can be achieved entirely from
the inert sector neutralinos (admixtures of inert singlinos and inert
higgsinos) \cite{Hall:2009aj} however such a scenario is now in
conflict with XENON-100 limits \cite{Hall:2010ix} though if one allows
for a relic density which is too small to explain observations (and
therefore would require some additional contributions to dark matter)
consistency could be achieved.  Alternatively another possible
scenario described in Ref.~\cite{King:2012wg} is that there is a large
splitting between the inert states with one at a few keV and another a
few GeV, thus giving warm dark matter scenarios, which
match observation.  For cases studied the couplings which are assumed
to be negligible here are too large to avoid perturbing the RG
evolution and may be too large for consistency with the constrained
model generally.  However such scenarios have only recently been
proposed and the full parameters space far from explored so it is
simply unknown whether or not this could be applied with couplings
which would be consistent with the scenarios explored here.

A third possibility which is known to be consistent with the
constrained version of the model decouples the singlinos from the rest
of the spectrum, giving a new contribution to the effective number of
neutrinos (which is consistent with observation) \cite{Hall:2011zq}.
The bino is then the dark matter candidate, but the
scenario is still very distinct from typical models with bino
dominated dark matter because the inert Higgsinos play a crucial role in
enabling the correct relic density to be achieved so long as the
mass of the lightest inert neutralino is related to that of the
lightest neutralino by the condition \cite{Hall:2011zq},
\begin{align}
  \mu_{\tilde{H}_1} \approx |m_{\tilde\chi_1^0}| + \unit{10}{\giga\electronvolt}.
  \label{DM-cond}
\end{align}
In the scans we perform here we do not explicitly satisfy this
constraint (and instead our inert Higgsinos are typically far heavier)
as we assume a universality amongst all exotic Yukawa couplings.
Splitting one generation of all exotics or just inerts could solve
this problem.  Doing so would perturb the renormalization group
equations and distort the parameter space somewhat, though the
qualitative results and understanding of the impact that these
thresholds can have would not be altered.  Alternatively one can keep
in mind the ``keVins and GeVins'' scenarios though it is currently
unknown if this could result in the correct relic density or not.

\subsubsection{Change in masses from threshold corrections}
\begin{figure}[tbh!]
  \centering
  \includegraphics[width=0.8\textwidth]{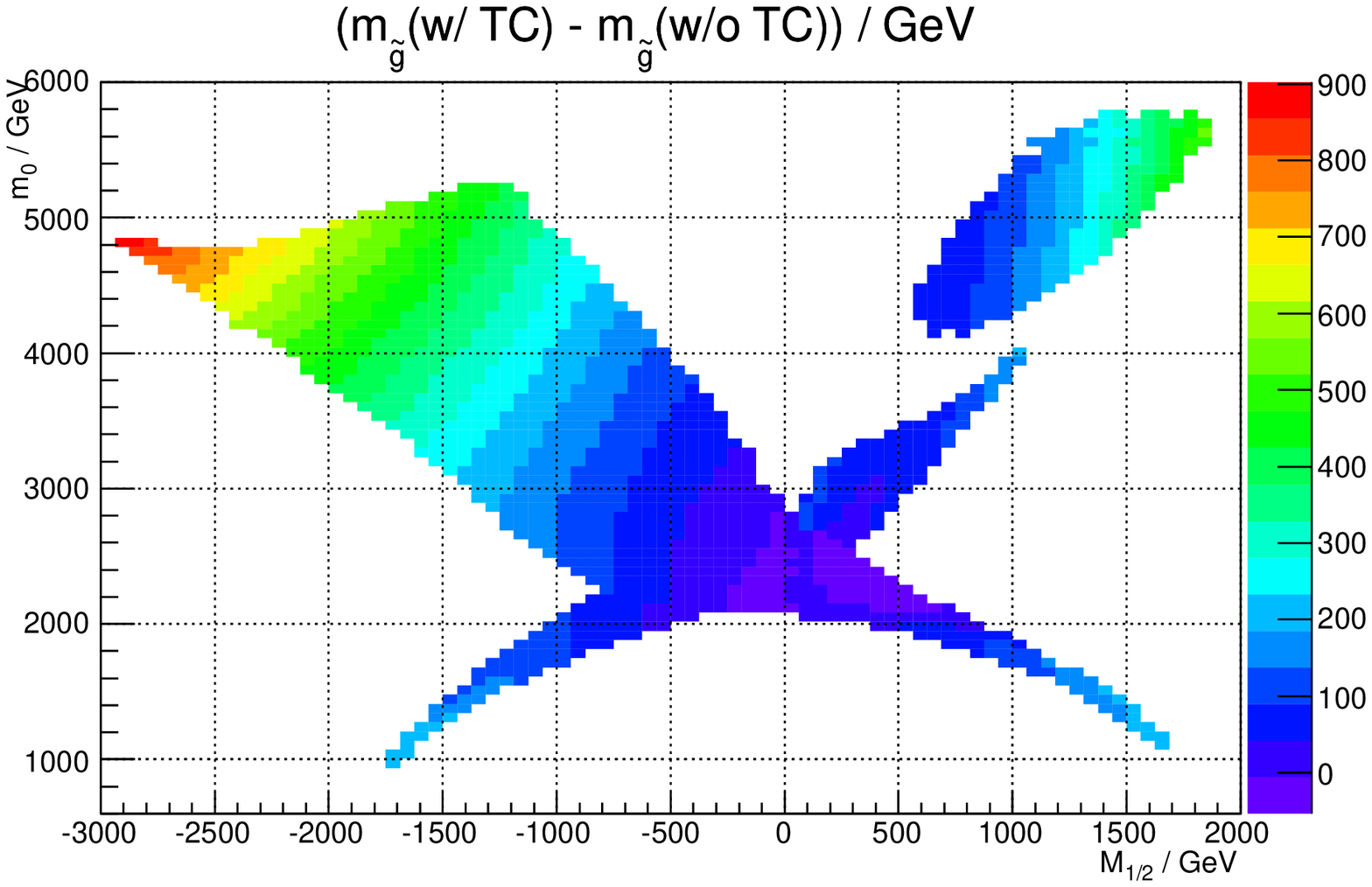}
  \includegraphics[width=0.8\textwidth]{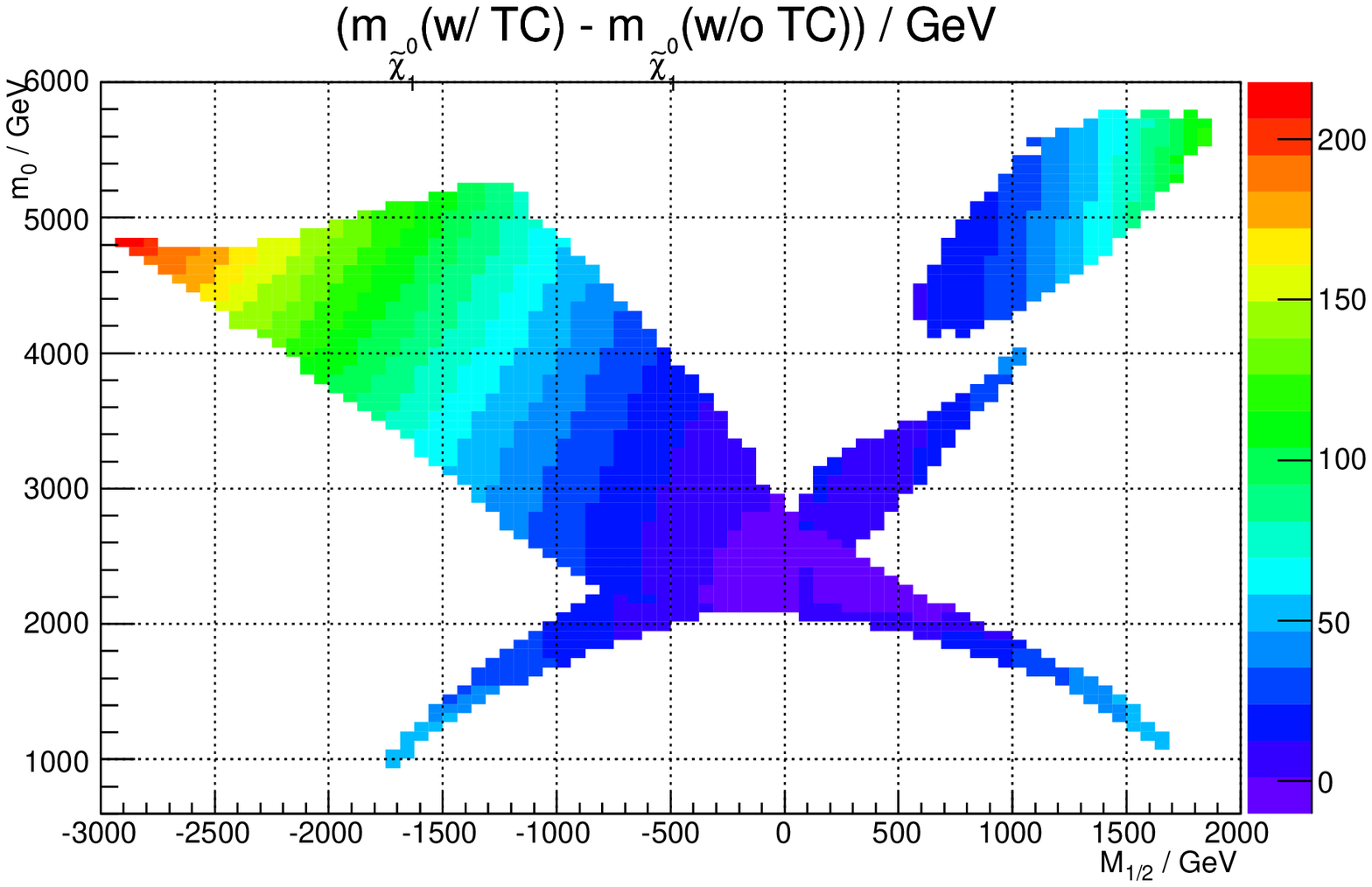}
  \caption{Change in the gluino and in the lightest neutralino mass
    when adding threshold corrections for $s = \mu' = m_{H'} =
    m_{\bar{H'}} = \unit{10}{\tera\electronvolt}$, $\tan\beta \in [2,
    45]$ and $\lambda_i(M_X) = \kappa_i(M_X) \in [0, 3]$.}
  \label{fig:gluino_neutralino_corrections}
\end{figure}

Now we look at the significance of the threshold effects throughout
the parameter space described in Sec.~\ref{sec:parameterspace}.  
Like for the point PP1, the largest corrections are
observed in the gaugino sector for states whose masses are set
by $M_{1/2}$.  \figref{fig:gluino_neutralino_corrections} shows the
shifts in mass for the gluino and lightest neutralino in the
$(m_0,M_{1/2})$-plane.\footnote{Note that these are the $m_0$
  and $M_{1/2}$ values obtained from the spectrum generator when
  thresholds are included.  Since $m_0$ and $M_{1/2}$ are outputs they
  are affected by the threshold corrections and so do not match the
  values for the point without threshold corrections.}  We compare
between results with trivial matching conditions to those with full
threshold corrections, where in both cases the matching scale used is
the ``optimal'' choice, where $\Qmatch$ is set to the geometric
average of the particle masses.

\begin{figure}[tbh!]
  \centering
  \includegraphics[width=0.8\textwidth]{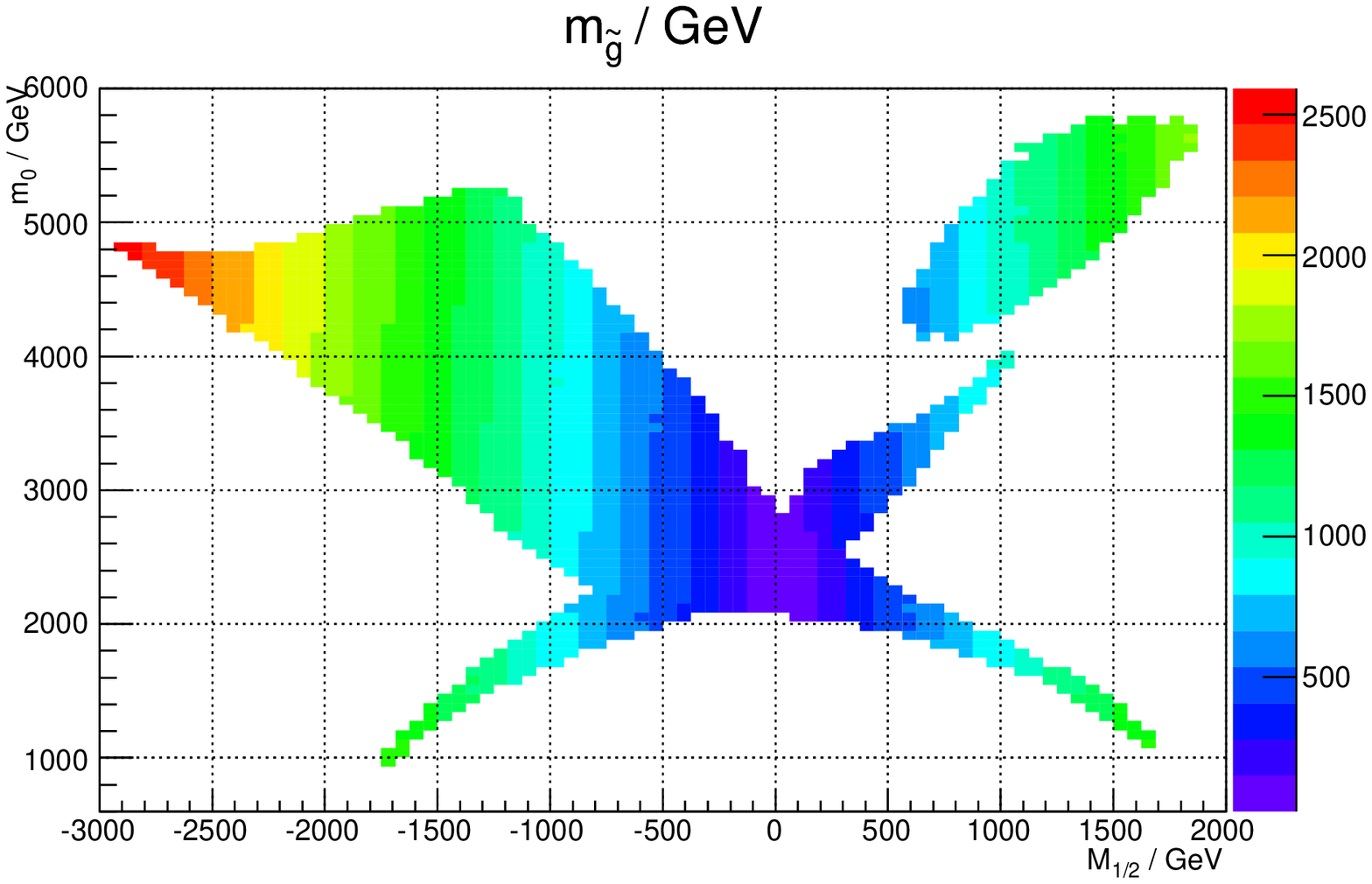}
  \includegraphics[width=0.8\textwidth]{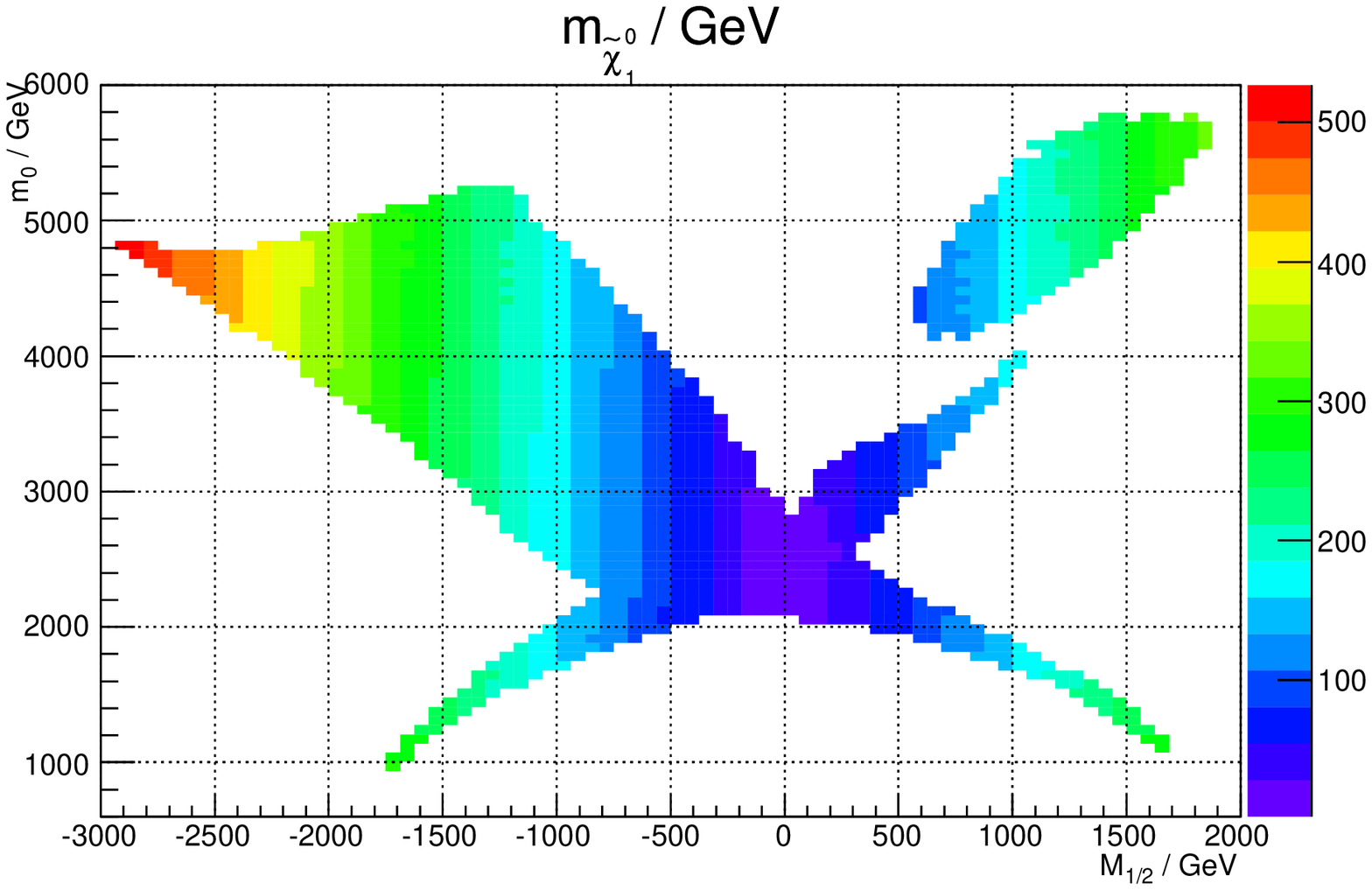}
  \caption{Gluino and in the lightest neutralino mass as a function of
    $M_{1/2}$ and $m_0$, where it was varied $\tan\beta\in[3,50]$,
    $\lambda_i=\kappa_i\in[0,3]$, $s=\unit{10}\tera\electronvolt$,
    $\mu'=m_{H'}=m_{\bar{H'}}=\unit{10}\tera\electronvolt$.}
  \label{fig:scan-M12-m0-mgluino}
\end{figure}

We show the absolute change in masses rather than relative change
because, when the masses are light due to a cancellation, the relative
corrections can be very large, distorting the plot. However it is
important to note that the gaugino masses though not shown can be
estimated since the low-energy soft gaugino masses $M_i$ (which
give the dominant contribution to $m_{\tilde{g}}$ and $m_{\tilde\chi_1^0}$)
are proportional $M_{1/2}$, with $m_{\tilde{g}} \sim 0.85 M_{1/2}$,
$m_{\tilde\chi_1^0} \sim 0.15 M_{1/2}$.  \figref{fig:scan-M12-m0-mgluino}
shows the actual gluino and neutralino masses after including
threshold corrections for comparison.

Once again we see substantial changes in both the gluino and
neutralino mass across a wide range of the parameter space with
corrections being small only in a very narrow region of the space.
Around the LHC limit on gluinos we can see there can be large
corrections of the order of several $\unit{100}{\giga\electronvolt}$
for the gluino mass, therefore these corrections can make a
significant impact in determining whether or not a point is ruled out
by LHC search constraints. The corrections to the gluino mass can be
positive or negative, depending on whether the masses appearing in the
logarithmic corrections are larger or smaller than the threshold
scale.\footnote{In this case the dominant corrections arise from
  corrections to $g_3$, see Eq.~\eqref{eq:g3-threshold-program}.}

The corrections to the neutralino are also substantial, varying from
just below zero to just over $200$ GeV.  They are distinctly
correlated with the corrections to the gluino, showing close to
identical variation, which is because the dominant effect is a shift
in $M_{1/2}$.

\begin{figure}[tbh!]
  \centering
  \includegraphics[width=0.8\textwidth]{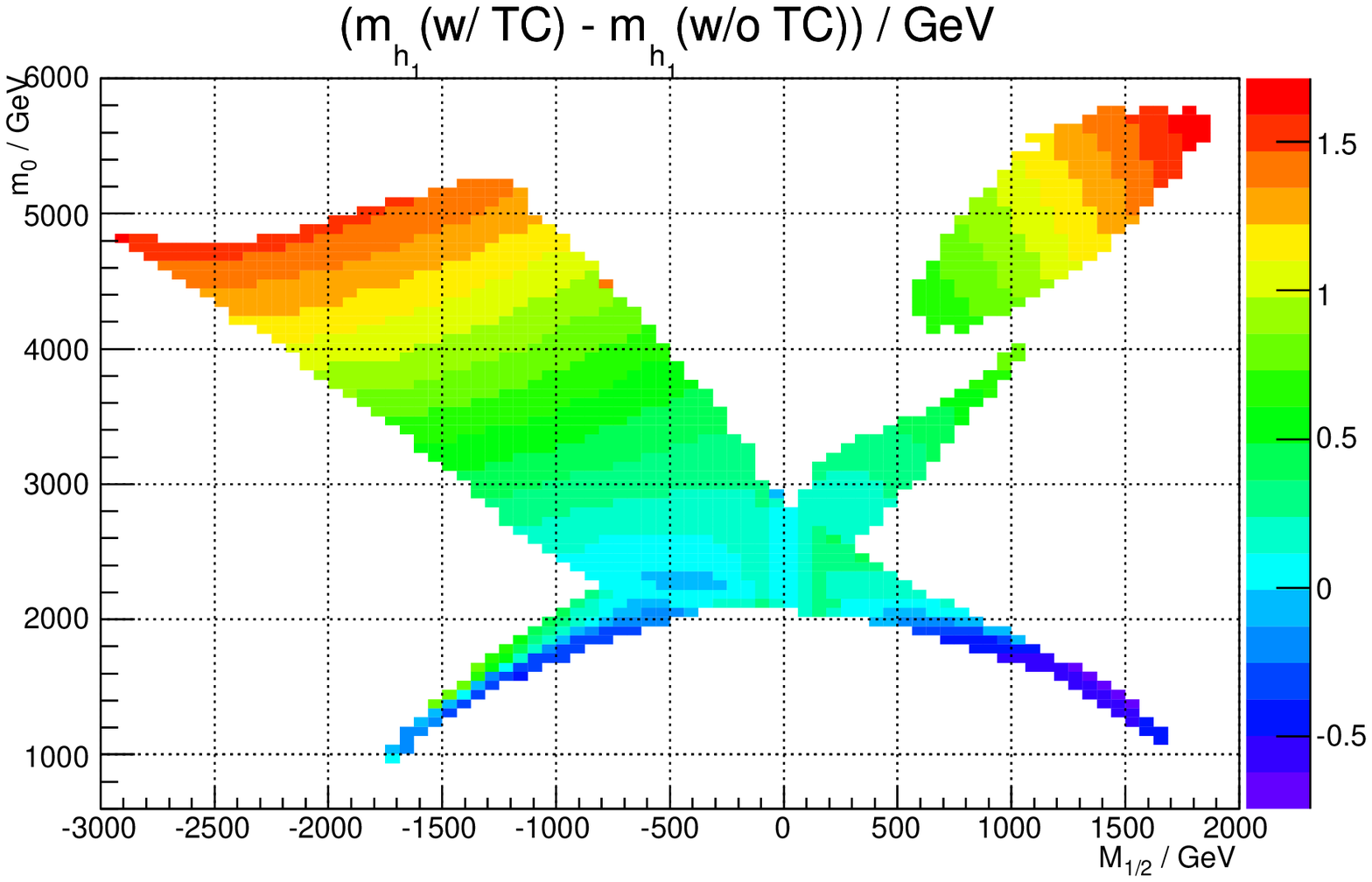}
  \caption{Change in the lightest CP even Higgs mass when adding
    threshold corrections for $s = \mu' = m_{H'} = m_{\bar{H'}} =
    \unit{10}{\tera\electronvolt}$, $\tan\beta \in [2, 45]$ and
    $\lambda_i(M_X) = \kappa_i(M_X) \in [0, 3]$.  We show only points
    where the change is less than \unit{5}{\giga\electronvolt}.}
  \label{fig:higgs_corrections}
\end{figure}

With the recent discovery of a Higgs-like boson \cite{:2012gk,:2012gu} with a
mass $\approx 125$--$126$ GeV and strong constraints
on a SM-like Higgs away from the mass of this new state, the lightest
Higgs mass is a crucial observable in constraining the parameter space
of SUSY models. The corrections to the lightest Higgs mass from
threshold effects are not expected to be very large since it's mass is
not set by soft mass parameters at tree level. However even small
corrections to the Higgs mass can significantly shift the $m_0$ and
$M_{1/2}$ values which a particular Higgs mass is compatible with and
this can dramatically alter how the Higgs mass measurement combines
with the constraints coming from squark and gluino searches.

\begin{figure}[tbh!]
  \centering
  \includegraphics[width=0.8\textwidth]{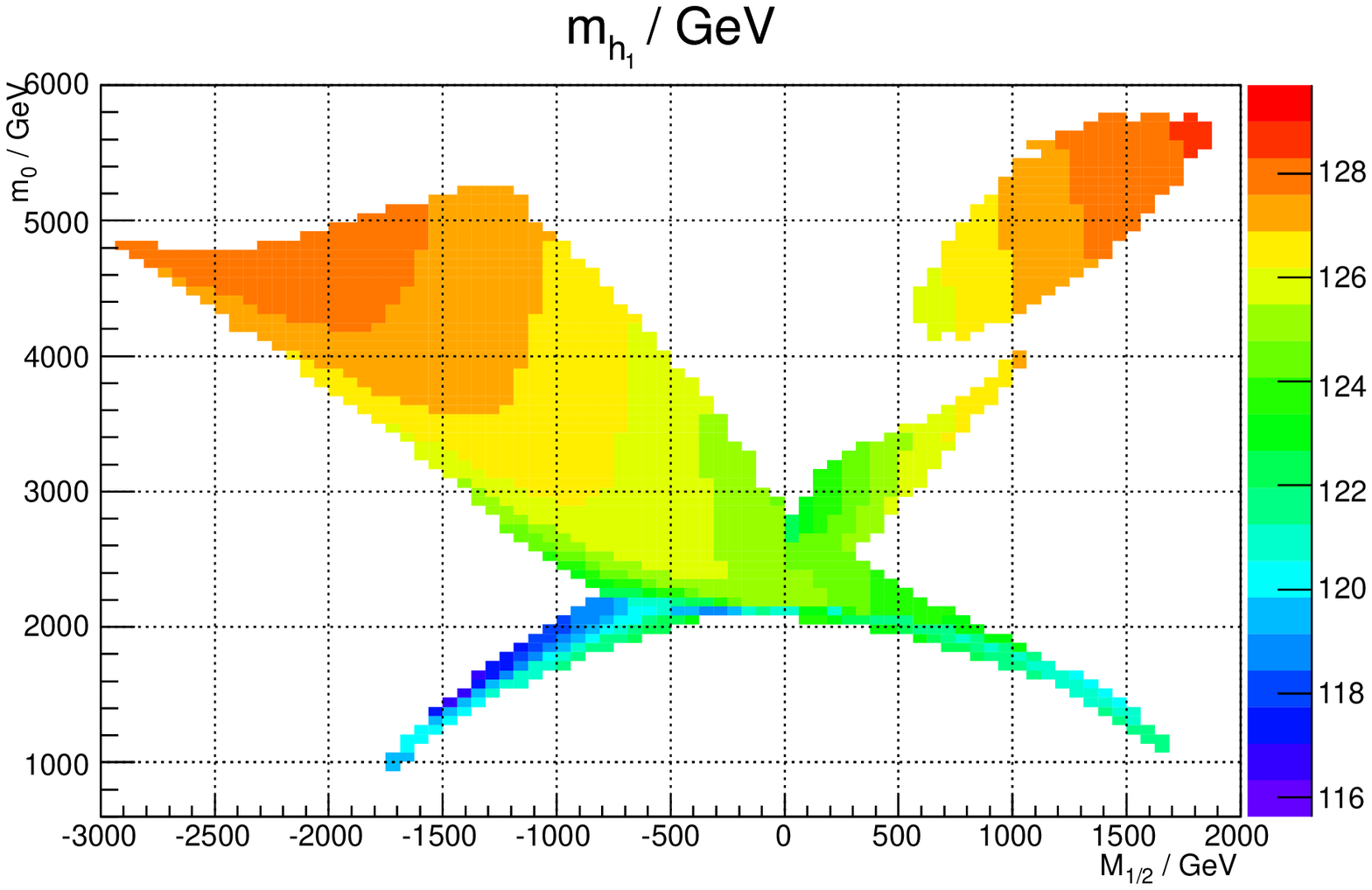}
  \caption{Lightest CP even Higgs mass $m_{h_1}$ as a function of
    $M_{1/2}$ and $m_0$, where it was varied $\tan\beta\in[3,50]$,
    $\lambda_i=\kappa_i\in[0,3]$, $s=\unit{10}\tera\electronvolt$,
    $\mu'=m_{H'}=m_{\bar{H'}}=\unit{10}\tera\electronvolt$.}
  \label{fig:scan-M12-m0-h1}
\end{figure}

In \figref{fig:higgs_corrections} the corrections to the Higgs mass
are shown across the $(m_0,M_{1/2})$-plane. The corrections vary between
around $-0.7$ GeV to $+1.7$ GeV. One can also see in
\figref{fig:scan-M12-m0-h1} the Higgs mass itself and note that
for the parameter set studied here the Higgs mass is consistent with
the new particle discovered by ATLAS and CMS in the large $m_0$ and
$M_{1/2}$ regions of the parameter space where the corrections from
the thresholds are largest.

\subsubsection{Allowed parameter space}\label{sec:allowed-parameter-space}
With the more reliable predictions at hand, we turn to understanding
the overall allowed region of parameter space.  Dominant experimental
constraints come from the gluino and the Higgs, see
\figsref{fig:scan-M12-m0-mgluino} and \ref{fig:scan-M12-m0-h1}.  We
can see that for our choice of $s$ the gluino is comfortably above the
LHC limit in a large volume the parameter space.  In the Higgs contour
plot \figref{fig:scan-M12-m0-h1} we also see a very large volume of
the parameter space where the light Higgs mass is within the narrow
window on it's allowed mass from the LHC discovery.  Nevertheless in
fact the combination of the two constraints creates very stringent
limits of this slice of the parameter space.

\begin{figure}[tbh!]
  \centering
  \includegraphics[width=0.8\textwidth]{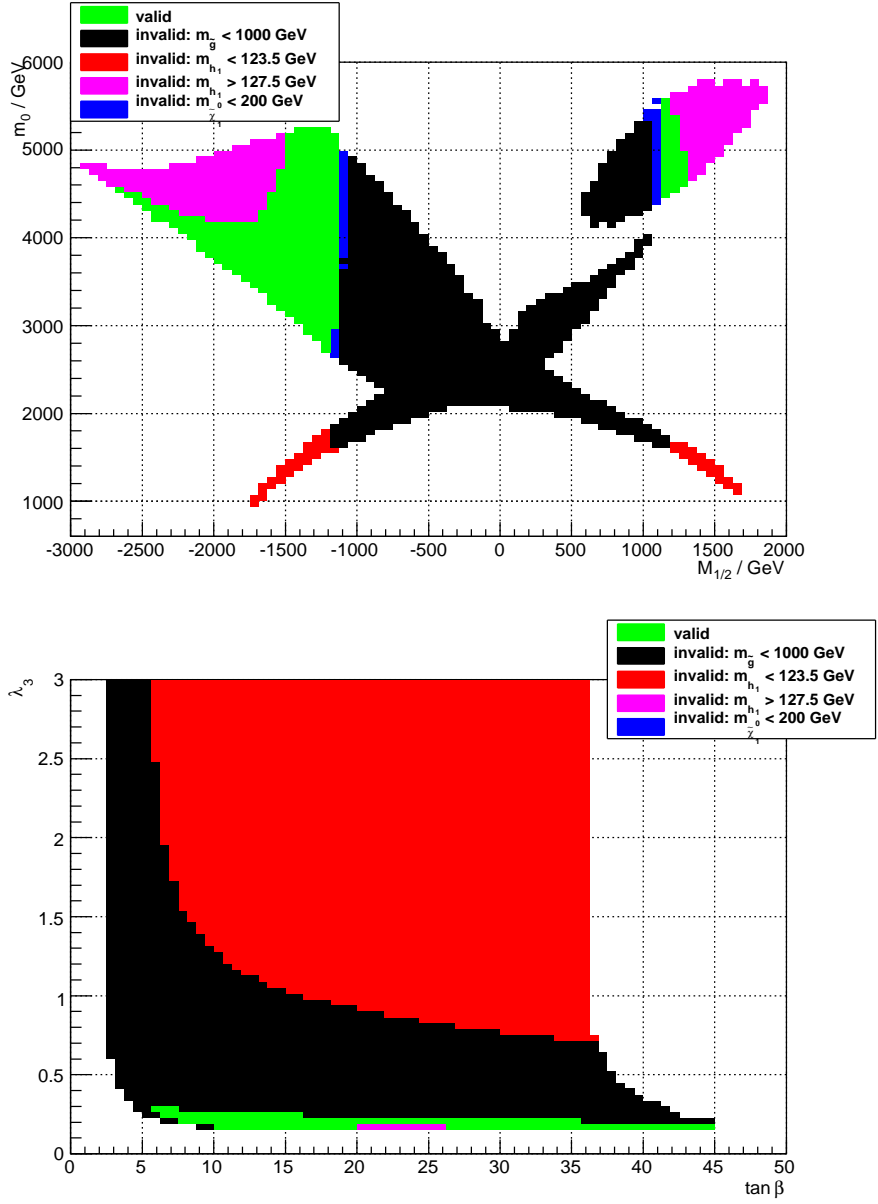}
  \caption{Exclusion plot of the \ESSM\ parameter space.}
  \label{fig:valid-parameter-space}
\end{figure}

To show the impact more precisely we now plot in
\figref{fig:valid-parameter-space} the valid and invalid parameter
space including all relevant constraints, both in terms of
$(\tan\beta,\lambda_3)$ and of $(m_0,M_{1/2})$.  While it is beyond
the scope of the current project to take account of full experimental
likelihoods or include precise $95$\% confidence limit contours, to
provide a guide as to the parameter space which evades LHC constraints we
define valid points as those satisfying,
\begin{align}
  \unit{123.5}{\giga\electronvolt} < m_{h_1} < \unit{127.5}{\giga\electronvolt},\\
  m_{\tilde{g}} > \unit{1}{\tera\electronvolt},\\
  m_{\tilde{t}_1} > \unit{300}{\giga\electronvolt},\\
  m_{\tilde{\chi}_1^0}, m_{\tilde{\chi}_1^\pm} > \unit{200}{\giga\electronvolt},\\
  m_{Z'} > \unit{2.1}{\tera\electronvolt},\\
  m_{\tilde{h}_{li}^0}, m_{\tilde{h}_i^+}, m_{h_{ij}^0}, m_{h_{ij}^\pm} > \unit{100}{\giga\electronvolt},
\end{align}
which are based on the discussion of Sec.~\ref{exp-constraints}.

  One finds that the lower bound on $m_{\tilde{g}}$ and the bounds on
  $m_{h_1}$ are most constraining.  This is because the gluino is
  driven lighter than the sfermions by the renormalization group flow
  with, $M_3\sim 0.7M_{1/2}\sim
  200$--$\unit{1400}{\giga\electronvolt}$, while for the Higgs we now
  have a very narrow range of allowed masses as a result of Higgs
  searches that ultimately lead to the recent discovery. Here we find
  the lightest Higgs mass is around $m_{h_1}\sim
  118$--$\unit{128}{\giga\electronvolt}$ \cite{Athron:2009bs}, as shown in
  \figref{fig:scan-M12-m0-h1}, including a substantial region of
  parameter space where the \cESSM\ predicts a lightest Higgs with a
  mass of $123.5$--$\unit{127.5}{\giga\electronvolt}$, consistent with
  the discovery.  

While the allowed region appears strongly restricted in 
$(\tan \beta,\lambda_3)$-space, it appears reasonably large
in the $(m_0,M_{1/2})$-plane.  The reason is the non-trivial mapping
between these two spaces, as illustrated by
\figref{fig:mapping-tanb-lam3-to-M12-m0}.
The valid green region in the $(\tan \beta,\lambda_3)$-space of
\figref{fig:valid-parameter-space} is mapped onto different wings of
the butterfly in the $(m_0,M_{1/2})$-plane.

\begin{figure}[tbh]
  \centering
  \includegraphics[width=0.8\textwidth]{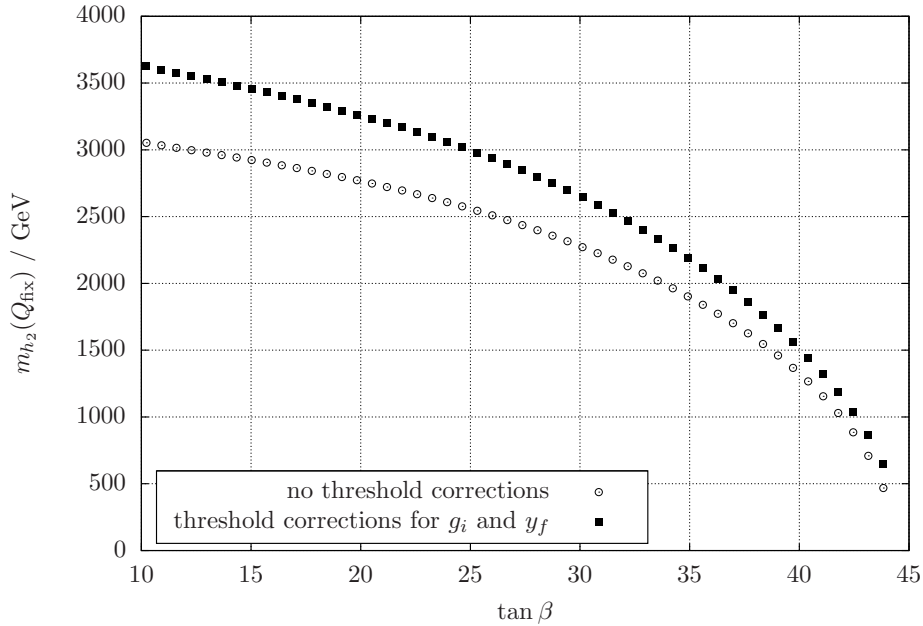}
  \caption{Second lightest CP even Higgs mass as a function of
    $\tan\beta$ for $s = \mu' = m_{H'} = m_{\bar{H'}} =
    \unit{10}{\tera\electronvolt}$ and $\lambda_i(M_X) = \kappa_i(M_X)
    = 0.2$}
  \label{fig:h2-of-tanb}
\end{figure}

\figref{fig:valid-parameter-space} also shows that even without imposing experimental constraints
there are upper and lower limits on all the parameters. Outside the
colored regions no simultaneous solution to electroweak symmetry
breaking and unification without tachyonic masses can be found. In
particular, there is an upper limit on $\tan \beta$ at around
$45$. Here the EWSB minima become unstable giving a tachyonic Higgs
mass, as shown in \figref{fig:h2-of-tanb} for both the uncorrected
Higgs mass and for the threshold corrected value, which converge in
this region. 
At such large $\tan \beta$ the bottom quark Yukawa coupling is much closer
to the top quark Yukawa coupling, and in such a situation
generating a large splitting between the Higgs masses required for
correct EWSB is difficult.  A simple tree-level condition for keeping
$m_A^2>0$ at large $\tan\beta$ can be derived in the MSSM
\cite{Carena:1994bv} and in the \ESSM,
\begin{align}
  \mbox{MSSM: } m_{H_d}^2 - m_{H_u}^2 &\gtrsim m_Z^2,\label{mmsmsplitting}\\
  \mbox{\ESSM: } m_{H_d}^2 - m_{H_u}^2 &\gtrsim m_Z^2 +
  \frac{g_N^2}{4}\Sigma_Q(N_{H_2} - N_{H_1}). \label{e6splitting}
\end{align}
The \ESSM\ condition is stricter since it contains an additional
positive term involving,\footnote{The term is positive so long as $s > v$, which
  is always the case due to limits on the $Z^\prime$ mass.}
\begin{align}
  \Sigma_Q = \frac{1}{2}(N_{H_1}v_1^2 + N_{H_2}v_2^2 + N_Ss^2).
\end{align}

\subsection{Dependency on the survival Higgs parameters}
The inclusion of threshold corrections not only reduces the unphysical
matching-scale dependence. It also leads to a physical dependence on
the survival Higgs sector.
Without thresholds the spectrum does not explicitly depend on the
survival Higgs masses as the renormalization group equations
decouple. With threshold corrections the survival
Higgs parameters
appear in the threshold corrections to the gauge couplings, and thus
affect both gauge coupling unification and the low-energy mass
spectrum. We will now study these two aspects in a simplified setting
by taking a single scale
\begin{align}
  m_\text{surv}\equiv \mu' = m_{H'} = m_{\bar{H'}} .
\end{align}

\begin{figure}[tbh]
  \centering
  \includegraphics[width=0.8\textwidth]{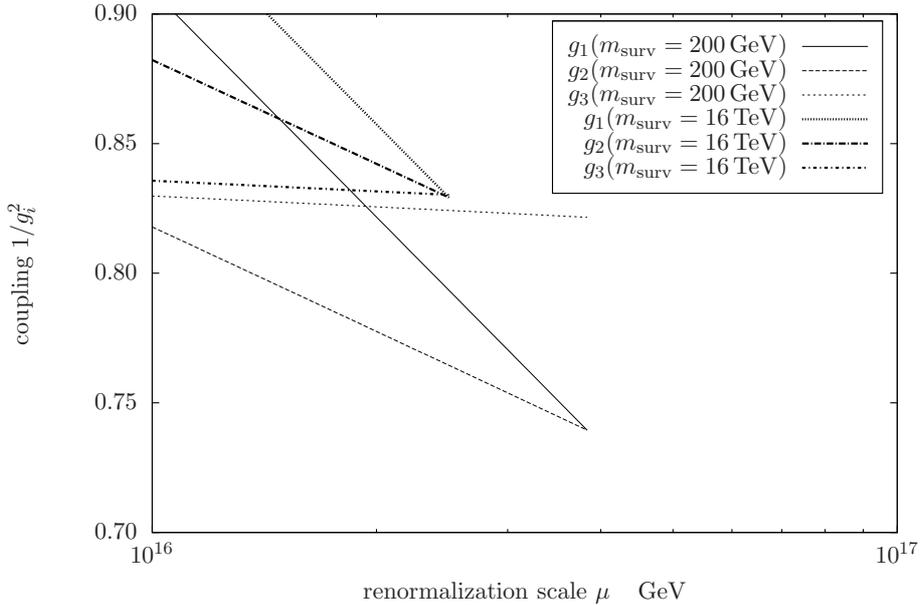}
  \caption{Splitting between $g_1$ and $g_3$ at $M_X$ for two
    different choices of $m_\text{surv}$ for PP1.}
  \label{fig:survival-unification}
\end{figure}

\begin{figure}[tbh]
  \centering
  \includegraphics[width=0.8\textwidth]{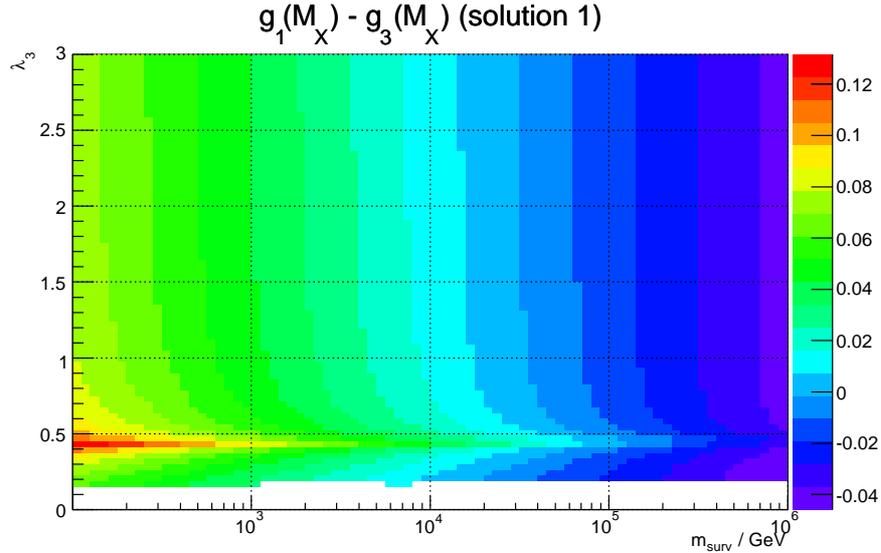}
  \caption{Splitting between $g_1$ and $g_3$ at $M_X$ in the
    $(\lambda_3, m_\text{surv})$ parameter space}
  \label{fig:surv-scan-gaugeMx}
\end{figure}

\figsref{fig:survival-unification} and \ref{fig:surv-scan-gaugeMx}
focus on gauge-coupling unification. We recall that in our approach
$M_X$ is defined by the intersection of $g_1$ and $g_2$, while $g_3$
is determined by its running from its low-energy measured
value. \figref{fig:survival-unification} shows that for the parameter
choice PP1, exact unification can be achieved by adjusting
$m_\text{surv}=\unit{200}{\giga\electronvolt}$, while e.g.\
$m_\text{surv}=\unit{16}\tera\electronvolt$ leads to a substantial
deviation between $g_3$ and $g_2$ at the GUT scale.

Of course in a full GUT model there will be GUT scale threshold
corrections. Nevertheless \figref{fig:survival-unification} makes
clear that requiring a precise splitting between the gauge couplings
from any given GUT threshold correction will fix the survival Higgs
scale.

\figref{fig:surv-scan-gaugeMx} shows gauge coupling (non-)unification
for the entire parameter space defined by
Eqs.~\eqref{lambdauniversality}--\eqref{scanrange} except that
$\tan\beta=10$ is fixed and $m_\text{surv}$ is varied in the range
\begin{align}
  m_\text{surv}
  \in [0.1, 1000]\;\tera\electronvolt .
\end{align}
Like in \figref{fig:survival-unification}, we find that the deviation
$(g_1-g_3)$ at $M_X$ depends substantially on the survival
Higgs masses, and for any given lambda there is a survival Higgs scale
which allows exact unification or any splitting required by GUT
thresholds between $0$ and $O(0.1)$.

\begin{figure}[tbh!]
  \centering
  \includegraphics[width=0.8\textwidth]{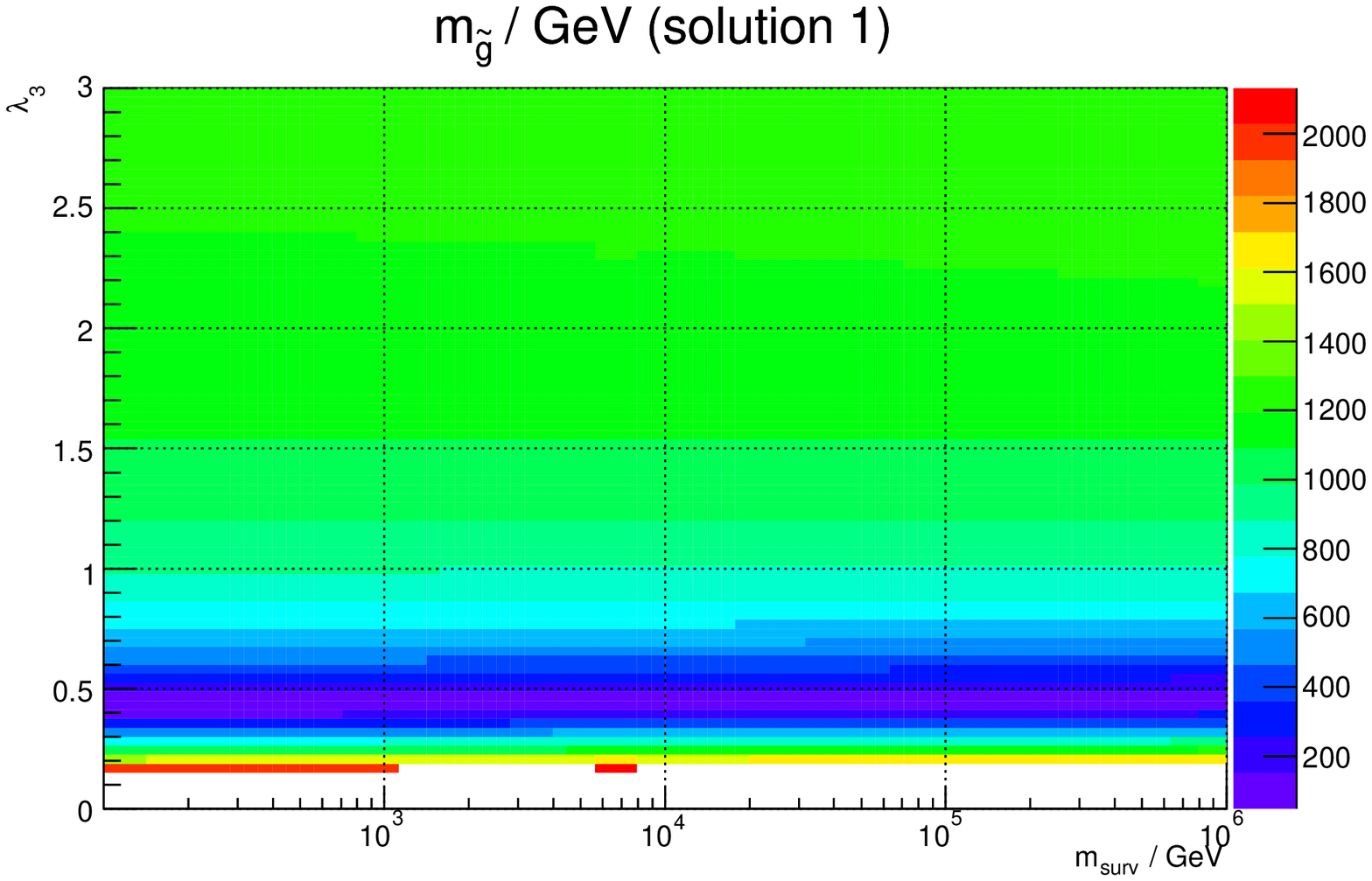}
  \includegraphics[width=0.8\textwidth]{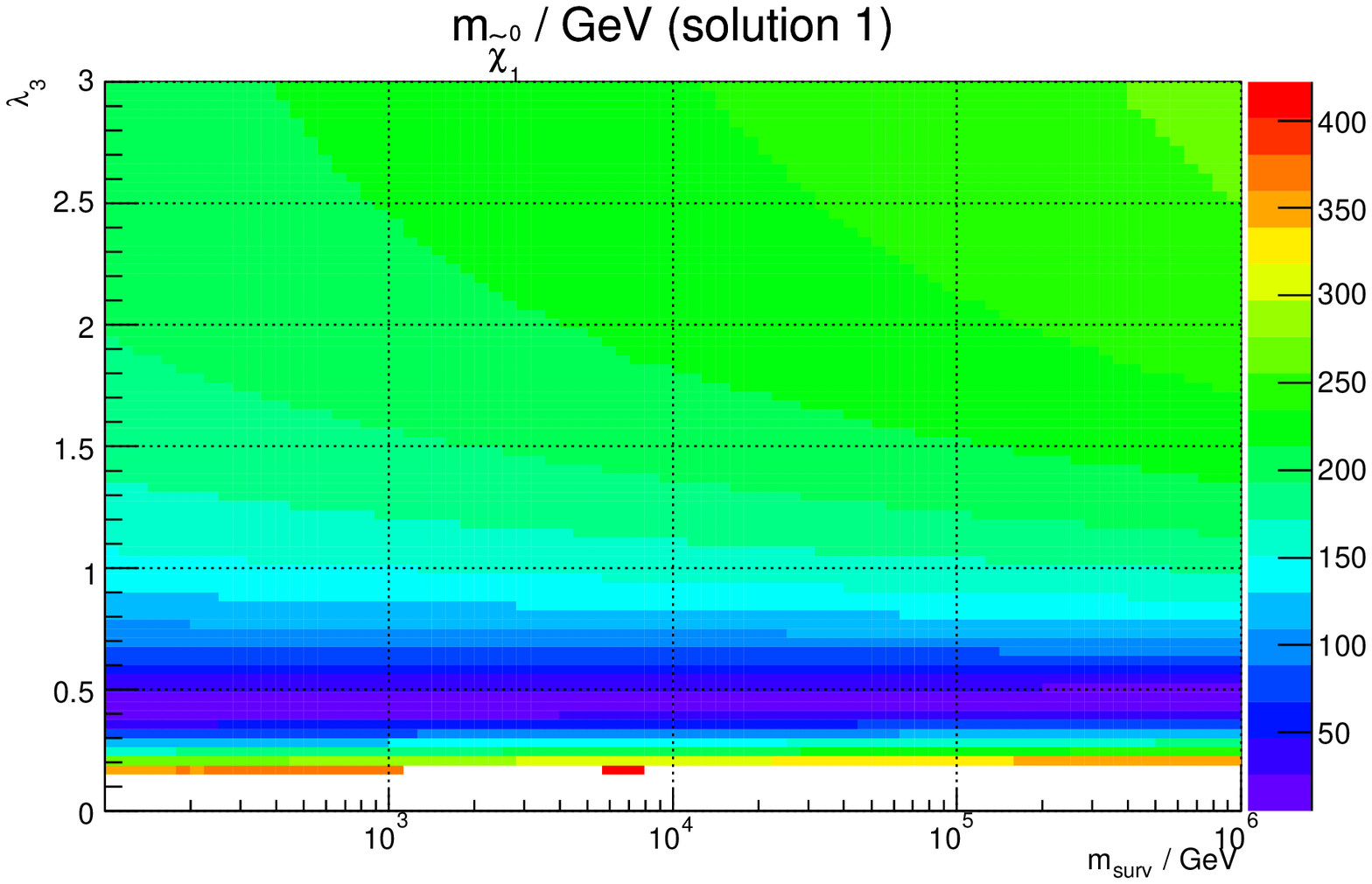}
  \caption{Gluino mass $m_{\tilde{g}}$ and mass of the lightest
    neutralino $m_{\tilde{\chi}^0_1}$ in the $(\lambda_3, m_\text{surv})$
    parameter space}
  \label{fig:surv-scan-mgluino-and-n1}
\end{figure}

\begin{figure}[tbh]
  \centering
  \includegraphics[width=0.8\textwidth]{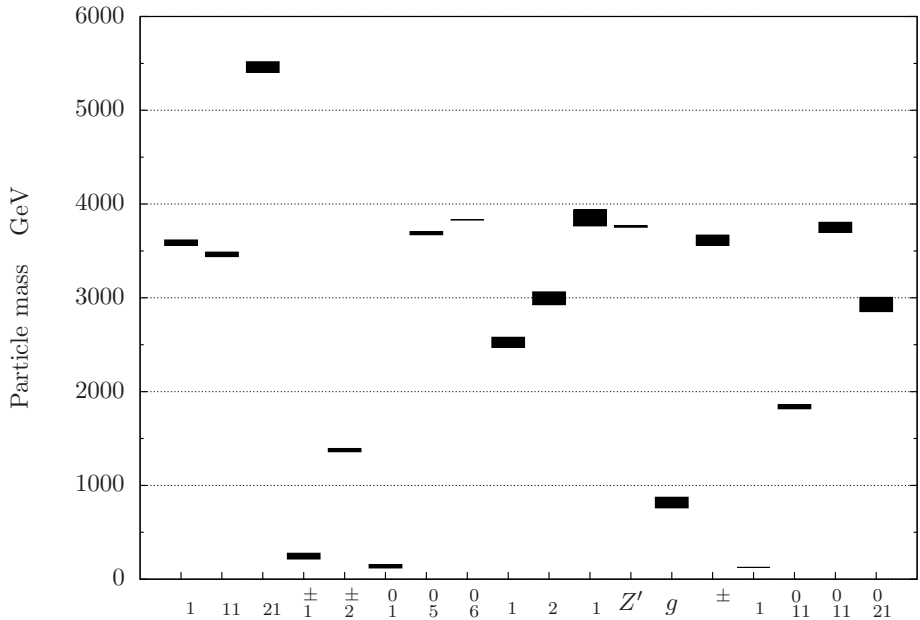}
  \caption{Dependence of the particle spectrum on the survival Higgs
    scale $m_\text{surv} = 0.1$--$\unit{10}{\tera\electronvolt}$
    for parameter point PP1.}
  \label{fig:PP1-gauge-coupling-unification-dependance-on-muprime}
\end{figure}

\figsref{fig:surv-scan-mgluino-and-n1} and
\ref{fig:PP1-gauge-coupling-unification-dependance-on-muprime} focus
on the influence of $m_\text{surv}$ on the mass spectrum.  In
\figref{fig:surv-scan-mgluino-and-n1} we plot the variation of the
gluino mass across the $(m_\text{surv}, \lambda_3)$-plane and see that
the dependence on the survival Higgs masses is extremely weak.  The
figure also shows that the lightest neutralino mass is affected
slightly more, in particular in regions where it is rather heavy
($\sim 250$--$\unit{300}{\giga\electronvolt}$).

Typically the survival Higgses and Higgsinos are too heavy to be
observed at the LHC, since they are only weakly produced and have the
couplings of a 4th generation lepton with current mass limit on the
mass of $\unit{100.8}{\giga\electronvolt}$.

For cases where they are not observable and where we do not constrain
the splitting between the gauge couplings at the GUT scale, we can
therefore consider them as an additional error in the theoretical
calculation. This is the approach we now take here.  To look at this
in detail we again turn to our test case PP1.

\figref{fig:PP1-gauge-coupling-unification-dependance-on-muprime}
shows the dependency of the particle spectrum on the new model
parameter $m_\text{surv}$ for PP1.  
It is instructive to compare the result to
\figref{fig:PP1-particle-spectrum}, where the sensitivity to the
matching scale $\Qmatch$ with and without threshold corrections is
shown. In the comparison one has to bear in mind that the
$m_\text{surv}$-dependence is physical and that we vary $m_\text{surv}$ in
a much larger interval than $\Qmatch$.

One finds a significant $m_\text{surv}$-dependence only for masses
which were also strongly 
sensitive to the choice of matching scale without threshold
corrections.  This is because a variation of $\Qmatch$ 
changes the size of all the logarithmic contributions to the threshold
corrections, while the variation of $m_\text{surv}$ changes a subset of
contributions to the thresholds. However, we also find that after
including threshold corrections the $m_\text{surv}$-dependence is
generally larger than the remaining matching-scale dependency. 

\subsection{Benchmarks in the literature}

As a final application of the threshold corrections presented here we
now study the impact they have on previously published benchmarks in
the model which have appeared in
\cite{Athron:2009bs,Athron:2009ue,Athron:2012sq}.  The
$\unit{1}{\tera\electronvolt}$ limit for the gluino which we require
to be consistent with Atlas and CMS searches is above the masses of
the all light benchmarks \cite{Athron:2009bs,Athron:2009ue}, while
heavy benchmarks \cite{Athron:2012sq} were chosen to be safe from
these limits. However since the threshold effects presented here have
a very large impact on the gluino mass, clearly they play an important
role in determining whether or not a particular \cESSM\ point is
excluded.  We will therefore study what impact the threshold effects
have on all these points and in particular whether it changes their
experimental status.\footnote{All but one of the light benchmarks have
  a $Z^\prime$ mass substantially below the limit discussed in Sec.\
  \ref{exp-constraints}. Light exotics, like singlinos, which are
  necessarily present in the \ESSM, will give a substantial
  reduction in the limit. This is probably not sufficient to evade the
  limits, but nonetheless testing against constraints from gluino
  searches provides another important limitation on the viability of
  these scenarios.}

The benchmarks in Refs.~\cite{Athron:2009bs,Athron:2009ue} were
selected to have very light gauginos or be close to the lower limit
on $m_0$ (for that choice of singlet VEV $s$), where one can get
lighter exotic sfermions.  Below this lower limit on $m_0$ the inert
Higgs scalars become tachyonic due to the large $U(1)_N$ $D$-terms
which give a negative contribution to the mass.

As a result of this the threshold corrections can push some points
into the region with tachyonic masses. Since we are applying the
threshold corrections iteratively this creates a problem. It may be
that such points in fact do not contain tachyonic solutions but merely
the first step in the iteration jumps to that region and subsequent
steps, if they could be performed, would lead to a self consistent
tachyon free solution.  In this case we try pushing the point into a
convergent iteration by adjusting the thresholds corrections in the
early steps.  If this does not work we then vary the survival Higgs
mass to see if this can lead to tachyon free solutions.

The impact on the gluino mass of these threshold corrections, for all
the points is shown in \tabref{tab:OldBMsGluinochange} for $\mu' =
m_{H^\prime} = m_{\bar{H}^\prime} = \unit{10}{\tera\electronvolt}$,
unless marked with $^a$ or  $^b$ where the survival Higgs masses
were varied until a self consistent solution could be found.

\begin{table}[tbp]
  \centering
  \begin{tabular}{lrrrrrr}
    \toprule
    $m_{\tilde{g}}$ / GeV   & BMA  & BMB  & BMC & BMD & BME &\\
    \midrule
    without thresholds   & $336$  & $330$  & $353$ & $327$ & $338$ &\\ 
    including thresholds & $224$  & $269^b$ & $260$ & $230$ & $203^b$ &\\
    \midrule
    $m_{\tilde{g}}$ / GeV   & BM1 & BM2  & BM3 & BM4 & BM5 & BM6\\
    \midrule
    without thresholds   & $350$ & $673$  & $362$ & $642$ & $338$ & $805$\\ 
    including thresholds & $322$ & $613^a$ & $275$ & $423^a$ & $190$ & $825$\\
    \midrule
    $m_{\tilde{g}}$ / GeV   & HBM1 & HBM2  & HBM3 & HBM4 & HBM5 \\
    \midrule
    without thresholds   & $984$ & $1352$  & $1659$ & $1129$ & $1001$ \\ 
    including thresholds & $1090$ & $1494$ & $1886$ & $827$  & $1067$\\
    \bottomrule
  \end{tabular}
  \caption{Comparison of the originally reported gluino mass and the gluino mass including threshold corrections for previously published \ESSM\ benchmarks, where we have chosen $\mu^\prime =m_{H^\prime} = m_{\bar{H}^\prime}= \unit{10}{\tera\electronvolt}$ for all points except those marked $^a$ where we used $\mu^\prime =m_{H^\prime} = m_{\bar{H}^\prime} = \unit{100}{\tera\electronvolt}$ to evade tachyonic problems with convergence and $^b$ where they were increased to $\unit{10^4}{\tera\electronvolt}$.}
  \label{tab:OldBMsGluinochange}
\end{table}

But while the survival Higgs masses can have
important effects, all light benchmarks are left with a
gluino mass substantially below the  limit.
 Therefore we can confirm that all of these points have been
excluded by the LHC even when threshold effects are taken account of
and the survival Higgs masses are varied without regard to maintaining
gauge coupling unification.

The situation is more optimistic for the new heavy benchmarks proposed
in \cite{Athron:2012sq}.  These benchmarks were chosen to satisfy the
latest LHC limits and also have the right relic density. For the
benchmarks HBM1--HBM3 there are significant changes to the gluino mass as
shown in the third column of \tabref{tab:OldBMsGluinochange}
pushing them further away from current constraints.  However beyond
this all qualitative features remain unchanged with the Higgs mass
staying in the Higgs signal range, though it is modified to
$\unit{125}{\giga\electronvolt}$ for HBM1 and HBM3.  The light inert
Higgsinos, fixed to be light to achieve the correct relic abundance
density, remain light and can be tuned carefully to match the relic
abundance density measurement exactly without perturbing the rest of
the spectrum. HBM2 and HBM3 also have light exotic quarks and these
also stay light, with their masses reduced by approximately $\unit{4}{\%}$ and
$\unit{6}{\%}$ respectively.  Therefore the essential features of these
benchmarks are not changed by the inclusion of threshold effects.

However for HBM4 and HBM5 the situation is more dramatic.  These
points have a very large singlet VEV, $\unit{50}{\tera\electronvolt}$
and $\unit{100}{\tera\electronvolt}$ respectively and in such cases
while light gauginos can always be obtained, this is only achievable
for a very narrow range of the input exotic Yukawa couplings.
Consequentially when one varies these Yukawa couplings the masses can
change a lot and the modification in these values (at low energies)
from threshold corrections can result in very large changes to the
soft masses.  For HBM4 although a gluino mass of the same order is
maintained the sign of $M_{1/2}$ is actually changed, while for HBM5 a
positive $M_{1/2}$ is obtained but the stability of the result is not
clear as scale variations lead to huge changes in the mass.

However these large changes in the physical spectrum are really an
artifact of the choices for input/output parameters.  One can instead
try to match the values of $m_0$, $M_{1/2}$ and then observe changes
in the exotic Yukawa couplings.  This would typically lead to a
similar spectrum, though $\lambda_3$ affects the light Higgs mass and
therefore the light Higgs mass is shifted.

\section{Summary and conclusions}
In this paper we have improved the prediction of low-scale quantities
from high-scale parameters in the Exceptional Supersymmetric Standard
Model (\ESSM).  Full sparticle threshold corrections to the gauge
couplings have been calculated by matching the \ESSM\ to the
SM. Similarly, the low-scale \DRbar\ \ESSM\ Yukawa couplings have been
calculated at the loop level directly from mass measurements of the SM
fermions and $m_W$. Full expressions for these corrections are
provided, and we have implemented them into an improved spectrum
generator for the constrained \ESSM.

Using the spectrum generator, we have studied the impact of these
corrections in detail. Both for the test point PP1 and in a parameter
scan we found a dramatic reduction in the scale dependency of many
masses, in particular of the gluino mass which is important
for setting limits on the model from collider constraints. Also
charginos, neutralinos, squarks and exotics receive large
corrections. Even the smaller threshold effects for the light Higgs
boson mass have a big impact
on the allowed regions of parameter space due to our knowledge of the
allowed values of the light Higgs mass, given recent LHC results on
Higgs searches and the discovery of a new particle.

The new spectrum generator with the implemented corrections allows us to draw firmer conclusions about
the allowed parameter space of the model. In line with previous
results, we found a substantial region of parameter space where the
mass of lightest Higgs is compatible with the new boson discovered at
the LHC and the limits from collider searches for squark, gluinos and
$Z^\prime$ bosons, are all satisfied. However the threshold corrections
imply a significant change in the high-scale parameters where this is
achieved and can alter the way these constraints combine.

An interesting consequence of the threshold corrections is a dependency of the mass
spectrum on the survival Higgs fields, which are included in the model
to assist gauge coupling unification. In previous studies these states
simply decoupled from the rest of the spectrum.  With the full sparticle threshold
corrections included, we found that in most of the spectrum the
dependency of the spectrum is rather weak.  However if the dependence
on the unknown survival Higgs masses is viewed as a theoretical
uncertainty, this uncertainty is larger than that from the remaining
scale variation, though much smaller than the uncertainty of earlier
studies neglecting threshold corrections.

If the survival Higgs bosons are taken seriously as physical fields,
the threshold corrections allow to fix their masses by the requirement
of gauge coupling unification. Indeed we found that the
survival Higgs masses can always be chosen such that gauge coupling
unification is valid, up to hypothetical GUT threshold
corrections, with required survival Higgs masses  in the multi-TeV
region.

Finally we looked at how the threshold corrections affect benchmark
points that have previously appeared in the literature.  Many
benchmarks proposed in earlier papers appeared to be ruled out by
gluino searches. We confirmed all of these to be ruled out once the
threshold corrections are included.  The benchmarks proposed more
recently to be consistent with experiment receive significant
corrections. Nonetheless these benchmarks are still experimentally
viable after including the threshold corrections. For two benchmarks,
HBM4 and HBM5, the stability of the predictions is poor due to fine
tuning of input parameters required to get the very hierarchical
spectrum these benchmarks illustrate.

In conclusion, our study shows that it is possible and valuable to
take into account higher-order corrections to the
high-scale--low-scale connection even in complicated, non-minimal
supersymmetric models such as the \ESSM. The qualitative features of
the \cESSM\ itself have not been changed by the threshold
corrections. Still, the model is theoretically attractive, predictive
and viable. As more LHC data comes in, distinguishing between
different supersymmetric models and hypotheses of GUT-scale physics
becomes more relevant, and precise spectrum generators like the one
presented here will be helpful.

\section*{Acknowledgements}

The work of PA is supported by the ARC Centre of Excellence for
Particle Physics at the Terascale, the work of AV is
funded by the DFG Graduate College 1504, and we acknowledge financial
support by the German Ministry for Science and Education.
PA would like to thank S.~F.~King,
A.~Merle, D.~J.~Miller, M.~Sch\"onherr and Patrik Svantesson for
helpful comments and discussions regarding this work.

\appendix
\numberwithin{equation}{section}

\newpage

\section{\ESSM\ covariant derivative and GUT relations}
The $E_6$ covariant derivative reads
\begin{align}
  D_\mu = \partial_\mu 
  + \mathi g_0 \tilde{A}_\mu^{\bar{a}} \tilde{T}^{\bar{a}}
  \qquad \qquad
  (\bar{a} = 1,\dotsc, 78),
\end{align}
with a single gauge coupling $g_0$, where $\tilde{T}^{\bar{a}}$ are
the generators and $\tilde{A}_\mu^{\bar{a}}$ are gauge fields in the
adjoint representation $(\dimrep{78})$ of the $E_6$.  This covariant
derivative can be decomposed in terms of the \SUc, \SUL, \UY\ and \UN\
sub-groups
\begin{align}
    D_\mu = \partial_\mu
  + \mathi g_3 T^a G_\mu^a
  + \mathi g_2 \frac{\vec{\tau}}{2} \cdot \vec{W}_\mu
  + \mathi g_Y \frac{Y}{2} B_\mu
  + \mathi g_N \frac{N}{2} Z'_\mu
  + \dotsb
  \label{eq:covarDerivDecomp}
\end{align}
All generators in \eqref{eq:covarDerivDecomp} are normalized such that
the quantum numbers are the same as in \tabref{tab:ESSMparticles}.
This, together with the condition $\Tr\tilde{T^a}\tilde{T^a} =
\delta^{ab}/2$, yields a GUT relation between the couplings
\begin{align}
  g_3 = g_2 = g_1 = g_1' =: g_0 ,
\end{align}
where
\begin{align}
  g_1 &:= \sqrt{\frac{5}{3}}\; g_Y , &
  g_1' &:= \sqrt{40}\; g_N .
\end{align}
At the electroweak scale the generator $Q$ of the unbroken \Uem\ is
then given by
\begin{align}
  Q = \frac{\tau_3}{2} + \frac{Y}{2} .
\end{align}

\section{\ESSM\ mass eigenstates}
\subsection{Higgs sector}
We write the Higgs bosons in the \ESSM\ as
\begin{align}
  H_{pi} =
  \begin{pmatrix}
    H_{pi}^1 \\ H_{pi}^2
  \end{pmatrix},\qquad
  S_i,
  \label{eq:higgs-doublett-notation}
\end{align}
where $i=1,2,3$ is the generation index and $p=1,2$ denotes the down
($p=1$) and up ($p=2$) Higgs bosons.  When the $SU(2)_L\times U(1)_Y\times
U(1)_N$ is broken to $U(1)_\text{em.}$ the third generation Higgs
bosons get a vacuum expectation value
\begin{align}
  H_{13} &=
  \begin{pmatrix}
    H_{13}^0 \\ H_{13}^-
  \end{pmatrix}
  \rightarrow
  \begin{pmatrix}
    \frac{v_1}{\sqrt{2}} + \rre H_{13}^0 + i\iim H_{13}^0 \\
    H_{13}^-
  \end{pmatrix},
  \\
  H_{23} &=
  \begin{pmatrix}
    H_{23}^+ \\ H_{23}^0
  \end{pmatrix}
  \rightarrow
  \begin{pmatrix}
    H_{23}^+ \\
    \frac{v_2}{\sqrt{2}} + \rre H_{23}^0 + i\iim H_{23}^0
  \end{pmatrix}
  ,\\
  S_3 &\rightarrow \frac{s}{\sqrt{2}} + \rre S_3 + i\iim S_3 .
\end{align}
Furthermore we define
\begin{align}
  \tan\beta := \frac{v_2}{v_1} ,
  \qquad
  \tan\phi := \frac{v}{2s}\sin 2\beta ,
  \qquad
  \mu_{\text{eff},i} := \frac{\lambda_i s}{\sqrt{2}}.
\end{align}
From the real parts of $H^0_{13}$, $H^0_{23}$ and $S_3$ we construct
three CP even Higgs bosons.  The diagonalization of the CP even mass
matrix is done in two steps.  At first we transform $\rre H^0_{13}$,
$\rre H^0_{23}$ and $\rre S_3$ into intermediate states
$(\mathsf{h}_1, \mathsf{h}_2, \mathsf{h}_3)$ via
\begin{align}
  \begin{pmatrix}
    \rre H^0_{13} \\ \rre H^0_{23} \\ \rre S_3
  \end{pmatrix}
  =
  U_\text{MSSM} \frac{1}{\sqrt{2}}
  \begin{pmatrix}
    \mathsf{h}_1 \\ \mathsf{h}_2 \\ \mathsf{h}_3
  \end{pmatrix}
  + \frac{1}{\sqrt{2}}
  \begin{pmatrix}
    v_1 \\ v_2 \\ s
  \end{pmatrix},
\end{align}
where the mixing matrix $U_\text{MSSM}$ has the form
\begin{align}
  U_\text{MSSM} =
  \begin{pmatrix}
    \cos\beta & - \sin\beta & 0 \\
    \sin\beta & \cos\beta   & 0 \\
    0         & 0           & 1
  \end{pmatrix}.
\end{align}
In the basis of $\mathsf{h} = (\mathsf{h}_1, \mathsf{h}_2,
\mathsf{h}_3)^T$ the Lagrangian for the CP even Higgs masses reads
\begin{align}
  \Lagr_\text{even} = -\frac{1}{2} \mathsf{h}^T M_\text{MSSM}
  \mathsf{h} ,
\end{align}
where the matrix $M_\text{MSSM}$ is non-diagonal in general.  Note,
that the above transformation is analogous to the MSSM, where the
mixing angle in $U_\text{MSSM}$ is $\beta$.  In a second step we
diagonalize the mass matrix $M_\text{MSSM}$ by the unitary matrix
$U_{E_6}$.  The resulting CP even Higgs mass eigenstates are labeled
$h = (h_1, h_2, h_3)^T$.  The diagonalization transformation reads
\begin{align}
  h &= U_{E_6} \mathsf{h}, &
  M_{E_6} &= U_{E_6}^* M_\text{MSSM} U_{E_6}^\dagger ,
\end{align}
where $M_{E_6}$ is diagonal.  From the gauge eigenstates $\mathsf{A} =
(\iim H^0_{13}, \iim H^0_{23}, \iim S_3)^T$ we construct three CP odd
Higgs boson mass eigenstates $A=(A_1,A_2,A_3)^T\equiv(G^0,G',A^0)^T$
via
\begin{align}
  \frac{1}{\sqrt{2}} A  &:= U_A \mathsf{A}, &
  M_A &= U_A^* M U_A^\dagger,
\end{align}
where $M_A$ is diagonal.  The mixing matrix $U_A$ is parametrized as
\begin{align}
  U_A =
  \begin{pmatrix}
     \cos\beta         & -\sin\beta         & 0        \\
    -\sin\beta\sin\phi & -\cos\beta\sin\phi & \cos\phi \\
     \sin\beta\cos\phi &  \cos\beta\cos\phi & \sin\phi
  \end{pmatrix}.
\end{align}
The charged Higgs and goldstone bosons $(H^\pm_i) = (G^\pm ~ H^\pm)^T$
are constructed from the gauge eigenstates $H_{i3}^\pm$ via
\begin{align}
  H^\pm_i &= U^\pm_{ij} H^\pm_{j3}
  \quad (i,j=1,2),
  & &\text{where} &
  U^\pm &=
  \begin{pmatrix}
    \cos\beta & -\sin\beta \\
    \sin\beta & \cos\beta
  \end{pmatrix}.
\end{align}

\subsection{Inert Higgs sector}\label{sec:inert-Higgs-sector}
The first two generations $(i=1,2;~p=1,2)$ Higgs doublets in
Eq.~\eqref{eq:higgs-doublett-notation} are called inert Higgs bosons.
For each generation $i=1,2$ we mix the fields $H_{1i}^{0}$,
$H_{2i}^{0*}$ to mass eigenstates $h_{ik}^{0}$ with an
unitary matrix $U_\text{inert}^{0i}$ via
\begin{align}
  h_{ik}^{0} = (U_\text{inert}^{0i})_{kl}
  \begin{pmatrix}
    H_{1i}^{0} \\ H_{2i}^{0*}
  \end{pmatrix}_l ,
  \qquad
  U_\text{inert}^{0i} =
  \begin{pmatrix}
    \cos\theta_i^0 & \sin\theta_i^0 \\
    -\sin\theta_i^0 & \cos\theta_i^0
  \end{pmatrix}.
\end{align}
Here $k,l=1,2$ enumerates the mass eigenstates and we neglect
inter-generation mixing.  Furthermore for each generation $i=1,2$ we
mix the fields $H_{1i}^{-}$, $H_{2i}^{+*}$ to mass eigenstates
$h_{ik}^{-}$ with an unitary matrix $U_\text{inert}^{\pm i}$ via
\begin{align}
  h_{ik}^{-} = (U_\text{inert}^{\pm i})_{kl}
  \begin{pmatrix}
    H_{1i}^{-} \\ H_{2i}^{+*}
  \end{pmatrix}_l ,
  \qquad
  U_\text{inert}^{\pm i} =
  \begin{pmatrix}
    \cos\theta_i^\pm & \sin\theta_i^\pm \\
    -\sin\theta_i^\pm & \cos\theta_i^\pm
  \end{pmatrix}.
\end{align}
Here $k,l=1,2$ enumerates the mass eigenstates and we neglect
inter-generation mixing.  Furthermore for each generation $i=1,2$ we
mix the fields $\tilde{H}_{1iL}^0$, $\tilde{H}_{2iL}^0$ to mass
eigenstates $\psi_{li}^0$ with an unitary matrix $Z$ via
\begin{align}
  \psi_{li}^0 = Z_{ln}
  \begin{pmatrix}
    \tilde{H}_{1iL}^0 \\ \tilde{H}_{2iL}^0
  \end{pmatrix}_n ,
  \qquad
  Z = \frac{1}{\sqrt{2}}
  \begin{pmatrix}
    1 & 1 \\ -1 & 1
  \end{pmatrix}.
\end{align}
The Majorana mass eigenstates are then defined as
\begin{align}
  \tilde{h}_{li}^0 =
  \begin{pmatrix}
    \psi_{li}^0 \\ \overline{\psi_{li}^0}^T
  \end{pmatrix}.
\end{align}
For each generation $i=1,2$ we combine the fields $\tilde{H}_{1iL}^-$,
$\overline{\tilde{H}_{2iL}^+}^T$ to mass eigenstates
$\tilde{h}_{i}^-$ via
\begin{gather}
  \tilde{h}_{i}^- =
  \begin{pmatrix}
    \tilde{H}_{1iL}^- \\ \overline{\tilde{H}_{2iL}^+}^T
  \end{pmatrix}.
\end{gather}

\subsection{Survival Higgs sector}
We mix the neutral survival Higgs bosons ${H'}^{0}$, $\bar{H'}^{0*}$
to mass eigenstates ${h'}_{k}^{0}$ with an unitary matrix
$U_\text{surv}^{0}$ via
\begin{align}
  {h'}_{k}^{0} = (U_\text{surv}^{0})_{kl}
  \begin{pmatrix}
    {H'}^{0} \\ \bar{H'}^{0*}
  \end{pmatrix}_l,
  \qquad
  U_\text{surv}^{0} =
  \begin{pmatrix}
    \cos\theta'^0 & \sin\theta'^0 \\
    -\sin\theta'^0 & \cos\theta'^0
  \end{pmatrix}.
\end{align}
We mix the charged survival Higgs bosons ${H'}^{-}$, $\bar{H'}^{+*}$
to mass eigenstates ${h'}_k^{-}$ with an unitary matrix
$U_\text{surv}^{\pm}$ via
\begin{align}
  {h'}_{k}^{-} = (U_\text{surv}^{\pm})_{kl}
  \begin{pmatrix}
    {H'}^{-} \\ \bar{H'}^{+*}
  \end{pmatrix}_l,
  \qquad
  U_\text{surv}^{\pm} =
  \begin{pmatrix}
    \cos\theta'^\pm & \sin\theta'^\pm \\
    -\sin\theta'^\pm & \cos\theta'^\pm
  \end{pmatrix}.
\end{align}
The survival higgsinos obey the same mixing as the neutral and charged
inert higgsinos in Sec.~\ref{sec:inert-Higgs-sector}.  We write the mass
eigenstates as $\tilde{h}^{\prime\pm}$ and $\tilde{h}_i^{\prime 0}$
($i=1,2$).

\section{\ESSM\ self-energies}\label{sec:ESSM-self-energies}
\subsection{$W$ boson}

The $W^\pm$ boson 1PI correlation function is decomposed into a
transverse and longitudinal part as follows
\begin{align}
  \Gamma_{W_\mu^+W_\nu^-}(p) = - g^{\mu\nu}(p^2 - m_W^2)
  - \left( g^{\mu\nu} - \frac{p^\mu p^\nu}{p^2} \right) \Pi_{WW,T}(p^2)
  - \frac{p^\mu p^\nu}{p^2} \Pi_{WW,L}(p^2),
\end{align}
where the transverse part is in the \ESSM\ given by
\begin{align}
  \frac{(4\pi)^2}{g_2^2} \Pi_{WW,T}^\ESSMmath(p^2) &=
  \frac{1}{4} \sum_{i=1}^3 \sum_{k=1}^2 \left\{
    \left(U_{ik}^{A*}\right)^2 A_0(m_{A_i}) + \left(U_{ik}^{E_6*}\right)^2 A_0(m_{h_i})
  \right\}\notag\\
  &\phantom{{}=} + \frac{1}{2} A_0(m_{H^\pm}) + \frac{1}{2} A_0(m_{G^\pm})
  + m_W^2 \sum_{i=1}^3 \left(U_{i1}^{E_6*}\right)^2 B_0(p^2, m_{h_i}, m_W)\notag\\
  &\phantom{{}=} - \sum_{i=1}^3 \sum_{j=1}^2 \left|
    U_{i1}^A U_{j1}^\pm + U_{i2}^A U_{j2}^\pm
  \right|^2 B_{22}(p^2, m_{A_i}, m_{H_j^\pm})\notag\\
  &\phantom{{}=} - \sum_{i=1}^3 \sum_{j=1}^2 \left|
    \sum_{k=1}^3 U_{ik}^{E_6} \left(U_{1k}^\MSSMmath U_{j1}^\pm - U_{2k}^\MSSMmath U_{j2}^\pm\right)
  \right|^2 B_{22}(p^2, m_{A_i}, m_{H_j^\pm})
  \notag\\&\phantom{{}=}
  +\sum_{i,l=1}^2 \bigg\{
  \frac{1}{2} \left( |Z_{l1}|^2 + |Z_{l2}|^2 \right) H(p^2, m_{{\tilde{h}}_i^+}, m_{{\tilde{h}}_{li}^0})
  \notag\\
  &\phantom{{}= \sum_{i,l=1}^2\bigg\{}
  - 2 Z_{l1} Z_{l2} m_{{\tilde{h}}_i^+} m_{{\tilde{h}}_{li}^0} B_0(p^2, m_{{\tilde{h}}_i^+}, m_{{\tilde{h}}_{li}^0})
  \bigg\}\notag\\
  &\phantom{{}=} -2 \sum_{i=1}^2 \Big\{
  \cos^2(\theta_i^0 + \theta_i^\pm) \left(
    \widetilde{B}_{22}(p^2, m_{h_{i1}^\pm}, m_{h_{i1}^0}) + \widetilde{B}_{22}(p^2, m_{h_{i2}^\pm}, m_{h_{i2}^0})
  \right)\notag\\
  &\phantom{{}= -2 \sum_{i=1}^2 \Big\{} + \sin^2(\theta_i^0 + \theta_i^\pm) \left(
    \widetilde{B}_{22}(p^2, m_{h_{i1}^\pm}, m_{h_{i2}^0}) + \widetilde{B}_{22}(p^2, m_{h_{i2}^\pm}, m_{h_{i1}^0})
  \right)
  \Big\}
  \notag\\&\phantom{{}=}
  +\sum_{l=1}^2 \bigg\{
  \frac{1}{2} \left( |Z'_{l1}|^2 + |Z'_{l2}|^2 \right) H(p^2, m_{\tilde{h}^{\prime +}}, m_{\tilde{h}_l^{\prime 0}})
  \notag\\
  &\phantom{{}= \sum_{l=1}^2\bigg\{}
  - 2 Z'_{l1} Z'_{l2} m_{\tilde{h}^{\prime +}} m_{\tilde{h}_l^{\prime 0}} B_0(p^2, m_{\tilde{h}^{\prime +}}, m_{\tilde{h}_l^{\prime 0}})
  \bigg\}\notag\\
  &\phantom{{}=} -2 \sum_{i=1}^2 \Big\{
  \cos^2(\theta'^0 + \theta'^\pm) \left(
    \widetilde{B}_{22}(p^2, m_{{h'}_{1}^\pm}, m_{{h'}_{1}^0}) + \widetilde{B}_{22}(p^2, m_{{h'}_{2}^\pm}, m_{{h'}_{2}^0})
  \right)\notag\\
  &\phantom{{}= -2 \sum_{i=1}^2 \Big\{} + \sin^2(\theta'^0 + \theta'^\pm) \left(
    \widetilde{B}_{22}(p^2, m_{{h'}_{1}^\pm}, m_{{h'}_{2}^0}) + \widetilde{B}_{22}(p^2, m_{{h'}_{2}^\pm}, m_{{h'}_{1}^0})
  \right)
  \Big\}
  \notag\\&\phantom{{}=}
  +4 m_W^2 \frac{g_N^2}{g_2^2}
  \bigg\{\left(\frac{N_{H_{13}}}{2} + \frac{N_{H_{23}}}{2}\right)^2
  \sin^2\beta \cos^2\beta B_0(p^2, m_{H^\pm}, m_{Z'})
  \notag\\
  &\phantom{{}=4m_W^2\frac{g_N^2}{g_2^2}\bigg\{}+ \left(\frac{N_{H_{23}}}{2}\sin^2\beta
    - \frac{N_{H_{13}}}{2}\cos^2\beta\right)^2 B_0(p^2, m_{G^\pm}, m_{Z'})
  \bigg\}
  \notag\\&\phantom{{}=}
  - s_W^2 \left(8 \widetilde{B}_{22}(p^2, m_W, 0) + 4 p^2 B_0(p^2, m_W, 0)\right)
  \notag\\&\phantom{{}=}
  - \left\{ (4p^2 + m_Z^2 + m_W^2) c_W^2  - m_Z^2 s_W^4 \right\} B_0(p^2, m_Z, m_W)
  \notag\\
  &\phantom{{}=}- 8 c_W^2 \widetilde{B}_{22}(p^2, m_Z, m_W)
  \notag\\&\phantom{{}=}
  +\sum_{f_u/f_d} \left\{\frac{1}{2} N_c^f H(p^2, m_u, m_d)
    - \sum_{i,j=1}^2 2 N_c^f w_{fij}^2 \widetilde{B}_{22}(p^2, m_{\tilde{u}_i}, m_{\tilde{d}_j})
  \right\}
  \notag\\&\phantom{{}=}
  +\frac{1}{g_2^2} \sum_{i=1}^6 \sum_{j=1}^2 \bigg\{
  f_{ijW} H(p^2, m_{\tilde{\chi}^0_i}, m_{\tilde{\chi}^+_j})
  + 2 g_{ijW} m_{\tilde{\chi}^0_i} m_{\tilde{\chi}^+_j}
  B_0(p^2, m_{\tilde{\chi}^0_i}, m_{\tilde{\chi}^+_j})
  \bigg\}
  \label{eq:W-SE}
\end{align}
Analogous to \cite{Pierce:1996zz} the summation $\sum_{f_u/f_d}$ is
over quark and lepton doublets and
\begin{align}
  (w_{fij}) =
  \begin{pmatrix}
    c_u c_d & c_u s_d \\ s_u c_d & s_u s_d
  \end{pmatrix} .
\end{align}
The neutralino--chargino--W-boson couplings are given by
\begin{align}
  f_{ijW} &= |a_{\tilde\chi_i^0\tilde\chi_j^+W}|^2 + |b_{\tilde\chi_i^0\tilde\chi_j^+W}|^2, &
  g_{ijW} &= 2\rre\left(b^*_{\tilde\chi_i^0\tilde\chi_j^+W} a_{\tilde\chi_i^0\tilde\chi_j^+W}\right),
\end{align}
where the Feynman rule for the neutralino--chargino--$W_\mu$ vertex is
written as $-i\gamma_\mu(a{\cal P}_L+b{\cal P}_R)$.  The nonzero
couplings in the \ESSM\ are the same as in the MSSM
\begin{align}
  a_{\tilde\psi_2^0\tilde\psi_1^+W} &= b_{\tilde\psi_2^0\tilde\psi_1^+W} = -g_2, &
  a_{\tilde\psi_4^0\tilde\psi_2^+W} &= -b_{\tilde\psi_3^0\tilde\psi_2^+W} = \frac{g_2}{\sqrt2}.
\end{align}
The couplings to mass eigenstates for an incoming neutralino
$\tilde\chi_i^0$ are
\begin{align}
  a_{\tilde\chi_i^0\tilde\chi_j^+W} &= N_{ik}^*V_{jl} a_{\tilde\psi_{k}^0\tilde\psi_{l}^+W}, &
  b_{\tilde\chi_i^0\tilde\chi_j^+W} &= N_{ik} U_{jl}^* b_{\tilde\psi_{k}^0\tilde\psi_{l}^+W},
\end{align}
while for an incoming chargino $\tilde\chi_j^+$ the couplings read
\begin{align}
  a_{\tilde\chi_i^0\tilde\chi_j^+W} &= N_{ik} V_{jl}^* a_{\tilde\psi_{k}^0\tilde\psi_{l}^+W}, &
  b_{\tilde\chi_i^0\tilde\chi_j^+W} &= N_{ik}^* U_{jl} b_{\tilde\psi_{k}^0\tilde\psi_{l}^+W}.
\end{align}

\subsection{Fermions}
We decompose the fermion 1PI correlation function as
\begin{align}
  \label{eq:Gamma-ff-def}
  \Gamma_{f\bar{f}}(p) = \slashed{p} \left(
    P_L\Gamma_{f\bar{f}}^L(p^2)
    + P_R\Gamma_{f\bar{f}}^R(p^2)
  \right)
  + P_L\Gamma_{f\bar{f}}^l(p^2)
  + P_R\Gamma_{f\bar{f}}^r(p^2)
\end{align}
and then define the fermion self-energy to be
\begin{align}
  \label{eq:sigmaf-def}
  \Sigma_f(p^2) := \frac{1}{2} \left\{
  m_f \left[ \Gamma_{f\bar{f}}^L(p^2) + \Gamma_{f\bar{f}}^R(p^2) \right]
  + \Gamma_{f\bar{f}}^l(p^2) + \Gamma_{f\bar{f}}^r(p^2)
  \right\}.
\end{align}
In the \ESSM\ it is given by
\begin{align}
  (4\pi)^2 \frac{\Sigma_t(p^2)}{m_t} &= \frac{4g_3^2}{3}
  \Biggl\{B_1(p^2,m_{\tilde g},m_{\tilde t_1})+B_1(p^2,m_{\tilde
    g},m_{\tilde t_2}) -\biggl(5+3\ln\frac{\mu^2}{m^2_t}\biggr)
  \notag\\
  &\phantom{{}=}\qquad -\sin(2\theta_t)\frac{m_{\tilde{g}}}{m_t}\left(
    B_0(p^2,m_{\tilde{g}},m_{\tilde{t}_1})- B_0(p^2,m_{\tilde{g}},m_{\tilde{t}_2})\right)\Biggr\}
  \notag\\
  &\phantom{{}=}+\frac{y_t^2}{2}
  \sum_{i=1}^3 \bigg\{
  \underline{A}_{ti}^2 \left[B_1(p^2,m_t,m_{h_i}) + B_0(p^2,m_t,m_{h_i})\right]
  \notag\\
  &\phantom{{}=}\phantom{+\frac{1}{2}y_t^2\sum_{i=1}^3 \bigg\{}
  +\underline{B}_{ti}^2 \left[B_1(p^2,m_t,m_{A_i}) - B_0(p^2,m_t,m_{A_i})\right]
  \bigg\}
  \notag\\
  &\phantom{{}=}+\frac{1}{2}
  \biggl[(y^2_b s^2_\beta + y^2_t c^2_\beta )
  B_1(p^2,m_b,m_{H^+}) +(g_2^2 +y_b^2c_\beta^2+y^2_t
  s^2_\beta) B_1(p^2,m_b,m_W)  \biggr]
  \notag\\
  &\phantom{{}=}+y_b^2 c_\beta^2
  \biggl[B_0(p^2,m_b,m_{H^+})-B_0(p^2,m_b,m_{W})\biggr] -(ee_t)^2
  \biggl(5+3\ln\frac{\mu^2}{m^2_t}\biggr)
  \notag\\
  &\phantom{{}=}+\frac{g_2^2}{c_W^2}\biggl[\biggl(g_{t_L}^2+ g_{t_R}^2\biggr) B_1(p^2, m_t,m_Z) +
  4g_{t_L}g_{t_R} B_0(p^2,m_t,m_Z)\biggr] 
  \notag\\
  &\phantom{{}=}+\frac{1}{2}\sum_{i=1}^6\sum_{j=1}^2\biggl[f_{it\tilde t_j}
  B_1(p^2,m_{\tilde\chi^0_i},m_{\tilde t_j}) + g_{it\tilde
    t_j}\frac{m_{\tilde\chi^0_i}}{m_t} B_0(p^2,m_{\tilde\chi^0_i},m_{\tilde
    t_j})\biggr]
  \notag\\
  &\phantom{{}=}+\frac{1}{2}\sum_{i,j=1}^2\biggl[f_{it\tilde b_j}
  B_1(p^2,m_{\tilde\chi^+_i},m_{\tilde b_j}) + g_{it\tilde
    b_j}\frac{m_{\tilde\chi^+_i}}{m_t} B_0(p^2,m_{\tilde\chi^+_i},m_{\tilde
    b_j})\biggr]
  \notag\\
  &\phantom{{}=}+ g_N^2 \left[
    \left(\left(\frac{N_{t_L}}{2}\right)^2
      + \left(\frac{N_{t_R}}{2}\right)^2\right) B_1(p^2, m_t, m_{Z'})
    + N_{t_L} N_{t_R} B_0(p^2, m_t, m_{Z'})
  \right]
  \label{eq:fermion-SE}
\end{align}
The matrix elements $\underline{A}_{fi}$ and $\underline{B}_{fi}$ are
defined as
\begin{align}
  \underline{A}_{fi} &=
  \begin{cases}
    (U_\text{MSSM})_{2k} (U_{E_6})_{ik}^* & \text{if $f$ is up-type},\\
    (U_\text{MSSM})_{1k} (U_{E_6})_{ik}^* & \text{if $f$ is down-type},
  \end{cases}
  \\
  \underline{B}_{fi} &=
  \begin{cases}
    (U_A)_{i2}^* & \text{if $f$ is up-type},\\
    (U_A)_{i1}^* & \text{if $f$ is down-type}.
  \end{cases}
\end{align}
In analogy to \cite{Pierce:1996zz} the Feynman rules for the
$\tilde\chi_if\tilde f_j$ couplings are written as $-i(a{\cal
  P}_L+b{\cal P}_R)$ and we have defined
\begin{align}
  f_{if\tilde f_j} &= |a_{\tilde\chi_if\tilde f_j}|^2 +
  |b_{\tilde\chi_if\tilde f_j}|^2, & g_{if\tilde f_j} &=
  2\rre(b^*_{\tilde\chi_if\tilde f_j}a_{\tilde\chi_if\tilde f_j}).
\end{align}
In the gauge eigenstate basis $\tilde\psi^0,\tilde\psi^+$ one has
\begin{align}
  a_{\tilde\psi_1^0f\tilde f_R} &= \frac{g_Y}{\sqrt{2}}Y_{f_R},
  \qquad b_{\tilde\psi_1^0f\tilde f_L} = \frac{g_Y}{\sqrt{2}}Y_{f_L}
  \\
  b_{\tilde\psi_2^0f\tilde f_L} &=\sqrt{2}g_2\tau_3^{f_L},
  \qquad a_{\tilde\psi_1^+d\tilde u_L} = b_{\tilde\psi_1^+u\tilde d_L} = g_2,
  \\
  a_{\tilde\psi_3^0d\tilde d_L} &= b_{\tilde\psi_3^0d\tilde d_R} = -b_{\tilde\psi_2^+d\tilde u_L}
  =-b_{\tilde\psi_2^+u\tilde d_R} = y_d,
  \\
  a_{\tilde\psi_4^0u\tilde u_L} &= b_{\tilde\psi_4^0u\tilde u_R} =-a_{\tilde\psi_2^+u\tilde d_L}
  =-a_{\tilde\psi_2^+d\tilde u_R} =y_u
  ,
  \\
  a_{\tilde\psi_6^0 f\tilde f_R} &= \frac{g_N}{\sqrt{2}} N_{f_R},\qquad
  b_{\tilde\psi_6^0 f\tilde f_L} = \frac{g_N}{\sqrt{2}} N_{f_L},\qquad
\end{align}
where the quantum numbers $Y_f/2$, $N_f/2$ and $\tau_3^f$ are listed
in the \tabref{tab:ESSMparticles}.  The couplings to the mass
eigenstates $\tilde\chi_i^0$ and $\tilde\chi_i^+$ are obtained by the
rotations
\begin{align}
  a_{\tilde\chi_i^0f\tilde f } &= N_{ij}^*a_{\tilde\psi_j^0f\tilde f}, &
  b_{\tilde\chi_i^0f\tilde f } &= N_{ij} b_{\tilde\psi_j^0f\tilde f},\\
  a_{\tilde\chi_i^+f\tilde f'} &= V_{ij}^*a_{\tilde\psi_j^+f\tilde f'}, &
  b_{\tilde\chi_i^+f\tilde f'} &= U_{ij} b_{\tilde\psi_j^+f\tilde f'}.
\end{align}
To obtain the couplings to the sfermion mass eigenstates one rotates
these couplings (both $a$- and $b$-type) by the sfermion mixing
matrix,
\begin{align}
  \begin{pmatrix}
    a_{\tilde\chi f\tilde f'_1} \\ a_{\tilde\chi f\tilde f'_2}
  \end{pmatrix}
  =
  \begin{pmatrix}
    c_{f'} & s_{f'} \\ -s_{f'} & c_{f'}
  \end{pmatrix}
  \begin{pmatrix}
    a_{\tilde\chi f\tilde f'_L} \\ a_{\tilde\chi f\tilde f'_R}
  \end{pmatrix}.
\end{align}
The self-energies for the other fermions are obtained from $\Sigma_t$
by the substitutions
\begin{align}
  \Sigma_\tau(p^2) &= \left.\Sigma_t(p^2)\right|_{t\rightarrow b, g_3=0}
  \label{eq:tau-SE},\\
  \Sigma_b(p^2) &= \left.\Sigma_t(p^2)\right|_{t\rightarrow b, c_\beta\leftrightarrow s_\beta}
  \label{eq:bottom-SE}.
\end{align}
In Eqs.~\eqref{eq:W-SE} and \eqref{eq:fermion-SE} we use the loop
functions as defined in \cite{Pierce:1996zz}.

\section{\ESSM\ counterterms} \label{sec:ESSM-counterterm-def}

If not otherwise stated we renormalize the $W^\pm$ boson and the SM
fermions in the on-shell scheme.  The corresponding on-shell and
$\DRbar$ counterterms are
\begin{align}
  \delta m_W^{2,\text{on-shell}} &= \rretilde \Pi_{WW,T}(m_W^2)
  \label{eq:deta-MW-def}
  , & \delta m_W^{2,\DRbar} &= \left.\rretilde \Pi_{WW,T}(m_W^2)\right|_{\Delta}
  \\
  \delta m_f^{\text{on-shell}} &= \rretilde \Sigma_f(m_f^2)
  \label{eq:deta-mf-def}
  , & \delta m_f^{\DRbar} &= \left.\rretilde \Sigma_f(m_f^2)\right|_{\Delta}
\end{align}
where the self-energies $\Pi_{WW,T}$ and $\Sigma_f$ are given in
Eqs.~\eqref{eq:W-SE}, \eqref{eq:fermion-SE}, \eqref{eq:tau-SE}
and \eqref{eq:bottom-SE}.  Using Eqs.~\eqref{eq:W-SE} and
\eqref{eq:fermion-SE} one can derive the following divergences for
the $\delta m_W $ and $\delta m_f$ counterterms in the \ESSM\
\begin{align}
  \left.\frac{\delta m_W}{m_W}\right|_\Delta &=
  \frac{\Delta}{(4\pi)^2}
  \Bigg\{
  \frac{11}{2} g_2^2
  + \frac{1}{2} g_Y^2
  + 2 g_N^2 \left[\left(\frac{N_{H_{13}}}{2}\right)^2 \cos^2\beta 
    + \left(\frac{N_{H_{23}}}{2}\right)^2 \sin^2\beta\right]
  \notag\\
  &\phantom{{}=\frac{\Delta}{(4\pi)^2}\Bigg\{}
  - \lambda_3^2
  -3 y_t^2 \sin^2\beta
  -3 y_b^2 \cos^2\beta
  - y_\tau^2 \cos^2\beta
  \Bigg\}
  ,\label{eq:delta-MW-div}
  \\
  \left.\frac{\delta m_{f_u}}{m_{f_u}}\right|_\Delta &=
  \frac{\Delta}{(4\pi)^2}
  \left\{
    - 2 g_3^2 C_2(r(f_u))_{\SUc}
    - \frac{3}{4} g_2^2
    - g_N^2 \left[\left(\frac{N_{L,f_u}}{2}\right)^2 + \left(\frac{N_{R,f_u}}{2}\right)^2\right]
  \right.
  \notag\\
  &\phantom{{}=\frac{\Delta}{(4\pi)^2}\Bigg\{}
  \left.
    - g_Y^2 \left[\left(\frac{Y_{L,f_u}}{2}\right)^2 + \left(\frac{Y_{R,f_u}}{2}\right)^2\right]
    + \frac{3}{2} y_{f_u}^2
    + \frac{1}{2} y_{f_d}^2
  \right\}
  ,\label{eq:delta-mfu-div}
  \\
  \left.\frac{\delta m_{f_d}}{m_{f_d}}\right|_\Delta &=
  \frac{\Delta}{(4\pi)^2}
  \left\{
    - 2 g_3^2 C_2(r(f_d))_{\SUc}
    - \frac{3}{4} g_2^2
    - g_N^2 \left[\left(\frac{N_{L,f_d}}{2}\right)^2 + \left(\frac{N_{R,f_d}}{2}\right)^2\right]
  \right.
  \notag\\
  &\phantom{{}=\frac{\Delta}{(4\pi)^2}\Bigg\{}
  \left.
    - g_Y^2 \left[\left(\frac{Y_{L,f_d}}{2}\right)^2 + \left(\frac{Y_{R,f_d}}{2}\right)^2\right]
    + \frac{3}{2} y_{f_d}^2
    + \frac{1}{2} y_{f_u}^2
  \right\}.
  \label{eq:delta-mfd-div}
\end{align}
Here $C_2(r)_{SU(N)}$ is a representation invariant of the
representation $r$ of $SU(N)$ and defined by
\begin{align}
  C_2(N)_{SU(N)} &= \frac{N^2-1}{2N} & &\text{(fundamental representation $N$)} ,\\
  C_2(G)_{SU(N)} &= N & &\text{(adjoint representation $G$)} .
\end{align}
Especially we have $C_2(r_{f_u})_{\SUc} = C_2(r_{f_d})_{\SUc} = 4/3$.
Furthermore it follows from the one-loop $\beta$ functions
\eqref{eq:beta-3}--\eqref{eq:beta-1prime} that the divergences of $\delta
g_i$ in the SM and the \ESSM\ are given by
\begin{align}
  \label{eq:delta-g2-div}
  \left.\frac{\delta g_i}{g_i}\right|_\Delta &=
  \frac{\Delta}{(4\pi)^2} \frac{\beta_i}{2} g_i^2,\\
  \beta_3^\ESSMmath &= 0 , & \beta_3^\SMmath &= -7
  , \label{eq:beta-3}\\
  \beta_2^\ESSMmath &= 4 , & \beta_2^\SMmath &= -\frac{19}{6}
  , \label{eq:beta-2}\\
  \beta_1^\ESSMmath &= \frac{3}{5}\beta_Y^\ESSMmath = \frac{48}{5}, &
  \beta_1^\SMmath &= \frac{3}{5}\beta_Y^\SMmath = \frac{41}{10}
  , \label{eq:beta-1}\\
 {\beta'_1}^\ESSMmath &= \frac{1}{40}\beta_N^\ESSMmath = \frac{47}{5}
  . \label{eq:beta-1prime}
\end{align}
The divergence of $\delta\tan\beta$ in the \ESSM\ is
\begin{align}
  \label{eq:delta-tanb-div}
  \left.\frac{\delta \tan\beta}{\tan\beta}\right|_\Delta =
  \frac{\Delta}{(4\pi)^2}
  \frac{\beta_{\tan\beta}}{2},
\end{align}
from which we obtain the one-loop RGE for $\tan\beta$, 
\begin{align}
  \label{eq:betatanbeta}
  \frac{\dd \tan\beta}{\dd t} &= \frac{\tan\beta}{(4\pi)^2} \beta_{\tan\beta},\\
  \beta_{\tan\beta} &= 2\left\{
    \frac{3}{2} y_{b}^2
    + \frac{1}{2} y_{\tau}^2
    - \frac{3}{2} y_{t}^2
    - g_N^2 \left[\left(\frac{N_{H_{13}}}{2}\right)^2 - \left(\frac{N_{H_{23}}}{2}\right)^2\right]
  \right\},
\end{align}
needed so that we consistently input $\tan \beta$, defined at a fixed
scale, when we vary the matching scale where the Yukawas are
calculated.

\end{document}